\documentclass[11pt,a4paper]{article}
\usepackage{jcappub}
\usepackage{multirow}
\usepackage{graphicx}
\usepackage{amsmath,amsfonts,amsthm}
\usepackage{bm}
\usepackage{array}
\usepackage{chngpage}

\newcommand{\ie}{i.e.}
\newcommand{\eg}{e.g.}

\newcommand{\bi}{\begin{itemize}}
\newcommand{\ei}{\end{itemize}}
\newcommand{\ben}{\begin{eqnarray}}
\newcommand{\een}{\end{eqnarray}}
\newcommand{\nn}{\nonumber}

\newcommand{\Msun}{{\rm M}_\odot}

\newcommand{\dd}{{\rm d}}

\newcommand{\mmin}{m_{\rm min}}

\newcommand{\citefig}[1]{Fig.~\ref{#1}}

\newcommand{\citesec}[1]{Sec.~\ref{#1}}
\newcommand{\citeapp}[1]{App.~\ref{#1}}
\newcommand{\citeeq}[1]{Eq.~(\ref{#1})}
\newcommand{\citeref}[1]{Ref.~\cite{#1}}
\newcommand{\citerefs}[1]{Refs.~\cite{#1}}

\begin{document}

\subheader{IFT-UAM/CSIC-XXX, LUPM-22-007, ULB-TH/22-05, LAPTH-015/22}

\title{Classification of gamma-ray targets for velocity-dependent and subhalo-boosted dark-matter annihilation
}

\author[a]{Thomas Lacroix,}
\author[b]{Gaétan Facchinetti,}
\author[a]{Judit Pérez-Romero,}
\author[c]{Martin Stref,}
\author[d]{Julien Lavalle,}
\author[e]{David Maurin,}
\author[a]{and Miguel A.~S\'{a}nchez-Conde}
\affiliation[a]{Instituto de F\'isica Te\'orica (IFT) UAM-CSIC, Departamento de F\'isica Te\'orica (DFT), Universidad Aut\'onoma de Madrid, Calle Nicol\'as Cabrera, 13-15, 28049 Madrid, Spain}
\affiliation[b]{Service de Physique Th\'eorique, CP225, Universit\'e Libre de Bruxelles, Boulevard du Triomphe, B-1050 Brussels, Belgium}
\affiliation[c]{Laboratoire d'Annecy-le-Vieux de
Physique Th\'eorique (LAPTh), Universit\'e Grenoble Alpes, Universit\'e Savoie Mont Blanc, CNRS, F-74000 Annecy, France}
\affiliation[d]{Laboratoire Univers et Particules de Montpellier (LUPM),\\
Universit\'e de Montpellier \& CNRS, Place Eug\`ene Bataillon, 34095 Montpellier Cedex 05, France}
\affiliation[e]{Laboratoire de Physique Subatomique \& Cosmologie (LPSC), Universit\'e Grenoble Alpes, CNRS/IN2P3, 53 avenue des Martyrs, 38026 Grenoble, France}

\emailAdd{thomas.lacroix@uam.es}
\emailAdd{gaetan.facchinetti@ulb.be}
\emailAdd{judit.perez@uam.es}
\emailAdd{martin.stref@univ-smb.fr}
\emailAdd{lavalle@in2p3.fr}
\emailAdd{david.maurin@lpsc.in2p3.fr}
\emailAdd{miguel.sanchezconde@uam.es}

\abstract{Gamma-ray observations have long been used to constrain the properties of dark matter (DM), with a strong focus on weakly interacting massive particles annihilating through velocity-independent processes.
However, in the absence of clear-cut observational evidence for the simplest candidates, the interest of the community in more complex DM scenarios involving a velocity-dependent cross-section has been growing steadily over the past few years.
We present the first systematic study of velocity-dependent DM annihilation (in particular $p$-wave annihilation and Sommerfeld enhancement) in a variety of astrophysical objects, not only including the well-studied Milky Way dwarf satellite galaxies, but nearby dwarf irregular galaxies and local galaxy clusters as well. Particular attention is given to the interplay between velocity dependence and DM halo substructure. Uncertainties related to halo mass, phase-space and substructure modelling are also discussed in this velocity-dependent context. 
We show that, for $s$-wave annihilation, extremely large subhalo boost factors are to be expected, up to $10^{11}$ in clusters and up to $10^6-10^7$ in dwarf galaxies where subhalos are usually assumed not to play an important role. Boost factors for $p$-wave annihilation are smaller but can still reach $10^3$ in clusters. 
The angular extension of the DM signal is also significantly impacted, with e.g. the cluster typical emission radius increasing by a factor of order 10 in the $s$-wave case. 
We also compute the signal contrast of the objects in our sample with respect to annihilation happening in the Milky Way halo. Overall, we find that the hierarchy between the brightest considered targets depends on the specific details of the assumed particle-physics model. }


\maketitle

\section{Introduction}
\label{sec:intro}

With the absence of robust non-gravitational evidence for dark matter (DM), astrophysical observations remain a prime avenue to find DM; the latter can be made of exotic particles, macroscopic compact objects, or be a result of an incomplete understanding of gravity. The crucial role of astrophysics has already proven essential in the quest for particle self-annihilating DM candidates, thanks to multi-wavelength and multi-messenger constraints on the properties of the underlying DM particle candidates \cite{LavalleEtAl2012,BringmannEtAl2012c,Gaskins2016,DiMauroEtAl2021}. In particular, $\gamma$-ray searches, traditionally used to look for signatures of the annihilation of weakly interacting massive particles (WIMPs), are a very powerful tool to explore broader classes of particle DM candidates.

In the past decades, significant efforts have been devoted to the ``vanilla'' WIMP DM scenario, in which the DM particles annihilate through an $s$-wave process with a thermal cross section of $\sim 3 \times 10^{-26}\, \rm cm^{3}\, s^{-1}$. The absence of a firm detection in the 10-100 GeV range by the \textit{Fermi} Large Area Telescope (\textit{Fermi} LAT) on board the NASA Fermi satellite~\citep{LAT_projections,fermi_instrument_paper} has led to a broadening of search strategies in recent years. First of all, the thermal $s$-wave scenario can only be robustly excluded in a small range of masses \cite{Leane2018}, while the (multi-)TeV range is still essentially unconstrained despite the close scrutiny of Cherenkov $\gamma$-ray telescopes from the ground \cite{2021arXiv211101198D}.\footnote{The upcoming Cherenkov Telescope Array (CTA), that will lead the field of $\gamma$-ray DM searches in the near future, is expected to provide competitive DM constraints in the TeV regime soon \cite{2013APh....43....3A,CTA_GC_2020}.} Moreover, the bulk of the effort in $\gamma$-ray DM searches has been focused on canonical velocity-independent ($s$-wave) DM annihilations, while velocity dependence of the annihilation cross section can appear in many theoretical contexts \cite{AbdallahEtAl2015}. This case leads to a much richer phenomenology, upon which we build in this work. 

Present-day WIMP annihilations may be naturally suppressed if the $s$-wave partial wave contribution is negligible and annihilations result from a $p$-wave process instead, i.e., $ \sigma v_{\rm rel} \propto v_{\rm rel}^{2}$. This arises, for instance, if the interaction between WIMPs and Standard Model particles is mediated by a scalar. Due to the small DM-induced $\gamma$-ray fluxes expected for $p$-wave annihilations, constraints focusing on dwarf spheroidal satellite galaxies (dSphs) are fairly weak and do not reach the level of the thermal $p$-wave annihilation cross section \cite{ZhaoEtAl2016,ZhaoEtAl2018,BoddyEtAl2020}. Moreover,
the standard paradigm has moved from scenarios involving one DM candidate to more complex dark sectors in which DM annihilation can be mediated, for instance, by new light states \cite{Arkani-HamedEtAl2009,PospelovEtAl2008}. In these scenarios, the exchange of light mediators induces a long-range interaction between the DM particles. This modifies the short-range annihilation cross section, enhancing it for an attractive interaction \cite{Sommerfeld1931,Hisano2003,HisanoEtAl2005,Profumo2005,CirelliEtAl2007,MarchRussellEtAl2008}. This effect is referred to as the Sommerfeld enhancement, and leads to a specific velocity dependence of the annihilation cross section; the latter boils down, in some regions of parameter space, to inverse powers of the relative velocity (i.e., enhanced annihilation at low velocities). 

Another crucial aspect of the calculation is the DM distribution in subhalos, which has long been recognized to play a very important part in predictions of annihilation signatures \cite{SilkEtAl1993,BergstroemEtAl1999a,CalcaneoRoldanEtAl2000,BerezinskyEtAl2003,StoehrEtAl2003,LavalleEtAl2007,KuhlenEtAl2008}. The presence of subhalos enhances $\gamma$-ray fluxes in the outskirts of their host halos, boosting the total signal. However, this boost is strongly dependent on the host halo mass and the structural properties of its subhalos; see for instance \cite{PieriEtAl2011,NezriEtAl2012,Sanchez-CondeEtAl2014,BonnivardEtAl2016,MolineEtAl2017,StrefEtAl2017} for boost calculated for velocity-independent annihilation fluxes. The calculation of the boost is further complicated for velocity-dependent annihilations, as discussed in \cite{Arkani-HamedEtAl2009,LattanziAndSilk2009,Bovy2009,KuhlenMadauSilk2009,KamionkowskiEtAl2010}. In this work, we go beyond previous results and provide a systematic study of the substructure boosts resulting from the interplay of both host halo mass and subhalo structural properties.

Overall, many studies have focused on refining the modelling of velocity-independent annihilations for $\gamma$-ray DM searches, considering and ranking potential astrophysical targets inside the Galaxy (Galactic centre, subhalos, and dSphs) and outside (nearby galaxies and galaxy clusters, diffuse extragalactic emission). These studies have helped to refine pointing and analysis strategies for ground-based Cherenkov telescopes and spaceborne $\gamma$-ray instruments. In contrast, velocity-dependent annihilations \cite{Arkani-HamedEtAl2009,LattanziAndSilk2009,RobertsonAndZentner2009,EssigEtAl2010,FerrerEtAl2013} have been less systematically considered. Recent studies have mostly focused on dSphs \cite{EssigEtAl2010,ZhaoEtAl2016,BoddyEtAl2017,BergstromEtAl2018,PetacEtAl2018,ZhaoEtAl2018,BoddyEtAl2020,AndoAndIshiwata2021,BaxterEtAl2021}, with fewer studies on other targets (Galactic centre \cite{BoddyEtAl2018,JohnsonEtAl2019}, Milky-Way \cite{Petac2020,BoardEtAl2021}, subhalos \cite{LuEtAl2018,RunburgEtAl2021}, galaxy clusters \cite{AbramowskiEtAl2012}, and diffuse extragalactic emission \cite{CampbellEtAl2010,CampbellAndDutta2011}).\footnote{Complementary to $\gamma$-ray searches, velocity-dependent DM annihilation has also been studied in a cosmological context, see \eg~\cite{ZavalaEtAl2009,HannestadEt2010,HisanoEtAl2011}.} For this reason, it is timely to perform a more systematic study and comparison for a wider variety of target objects and particle physics models, in particular for $p$-wave annihilation.

This paper is dedicated to detailed predictions of the astrophysical factors that determine expected DM-induced $\gamma$-ray fluxes, as well as a thorough discussion of the associated theoretical uncertainties; connecting these predictions to actual $\gamma$-ray data analysis and discussing the full implications in terms of DM models is left for a follow-up study.
For our purpose, we consider several targets, avoiding those with too diffuse or too extended expected signals. These targets are taken among three families, namely dSphs, dwarf irregular galaxies (dIrrs) -- these ones included for the first time in a velocity-dependent study -- and galaxy clusters. Our goal is to address how the intra- and inter-family ranking is impacted by the particle-physics model considered. In the process, we make several improvements with respect to previous calculations. The main novelty is that, for all targets, we consider the subhalo boost obtained from a self-consistent semi-analytic model, i.e.,~reconstructing the velocity dependence at all scales. This is done via the reconstruction of the phase-space distribution function from the mass modelling. We study in particular the complex interplay between velocity dependence and substructure boost, and the implications in terms of ranking of the various targets. A more in-depth and analytical study of the boost factor is presented in a companion paper \cite{CompanionPaper}.

The paper is organized as follows. In \citesec{sec:theoretical_ingredients}, we introduce the various ingredients needed to perform the calculation of the generalised astrophysical factor of the flux, the so-called $J$-factor (velocity-weighted DM squared density integrated over the phase-space distribution): (i) we recall the regimes where the Sommerfeld enhancement can occur, in particular we focus on the $s$- and $p$-wave cases; (ii) we describe how we invert the mass model to obtain the phase-space DM distribution entering the $J$-factor calculations; (iii) we discuss the calculation of the boost factors in the context of generalised $J$-factors. 
In \citesec{sec:mass_modelling}, we detail the mass modelling of specific targets, selected among the sample of known dSph galaxies, dIrr galaxies, and galaxy clusters.
In \citesec{sec:uncertainties_phase_space_smooth}, we present our $J$-factor results for the host halos alone (i.e., no substructures), highlighting the two  main  uncertainties at this stage of the calculation, namely the DM density profile and phase-space modelling. Then, in \citesec{sec:subhalo_boost_and_classification}, we detail how the targets of interest are boosted by the presence of subhalos and how this may affect the intensity and the spatial morphology of the DM signal. We also provide a ranking of these targets (in terms of their generalised $J$-factor) and highlight the potential of galaxy clusters for $p$-wave annihilation. Finally, we conclude and discuss the next steps of our work in \citesec{sec:conclusion}.
To ease the reading, we postpone to the appendix the discussion of uncertainties related to phase-space modelling (\citeapp{app:phase_space_modelling}), the details of the subhalo model (\citeapp{app:SL17}), the expressions used to perform the numerical computation of the subhalo boost factor (\citeapp{app:subhalo_boost}),
and a discussion of the signal contrast between our targets and the Milky Way (MW) DM annihilation foreground (\citeapp{app:MWcontrast}).

\section{Velocity-dependent annihilation: theoretical ingredients}
\label{sec:theoretical_ingredients}

\subsection{Dark matter annihilation and self-interaction: Sommerfeld effect}
\label{ssec:sommerfeld}

In this section, we provide a brief review of the impact of DM self-interaction on the physics relevant to $\gamma$-ray searches. We consider the phenomenological scenario in which DM particles self-interact through the exchange of a light mediator.
In the absence of such interactions, the annihilation cross section can be computed perturbatively from the physics of the short-range annihilation processes. 
However, a light mediator leads to a long-range interaction which can distort the wave function of the corresponding two-body system in a non-perturbative way, leading to Sommerfeld enhancement\footnote{We restrict ourselves to symmetric DM with attractive interactions, for which the Sommerfeld factor is effectively an enhancement factor.} of the annihilation cross section in the non-relativistic regime \cite{Sommerfeld1931,Hisano2003,HisanoEtAl2005}. 
Since the effect appears in the non-relativistic limit, the enhancement can be computed by solving the Schr\"{o}dinger equation for the scattering of two DM particles. The radial part of the wave function $R_{\ell}(r)=\chi_{\ell}(r)/r$, for the partial wave with angular momentum $\ell$, solves 
\begin{equation}
    \label{eq:radial_Schroedinger}
    \left(-\frac{\hbar^2}{m_{\chi}}\,\partial^2_{r} + \frac{\hbar^2\,\ell(\ell+1)}{m_{\chi}\,r^2} + V(r)\right)\,\chi_{\ell}(r) = \frac{(\hbar\,k)^2}{2\,m_{\chi}}\,\chi_{\ell}(r)\,,
\end{equation}
where $m_{\chi}$ is the DM mass, $V$ is the interaction potential and $k=m_{\chi}v/\hbar$ is the wave vector at infinity of one of the incoming DM particles in the centre-of-mass frame. 

Equation~\eqref{eq:radial_Schroedinger} is solved with the boundary conditions that the interaction only leads to outgoing spherical waves at infinity, and with $R_{\ell}(r) \propto r^{\ell}$ as $r \rightarrow 0$. The Sommerfeld enhancement factor reads \cite{Cassel2010,Iengo2009,Slatyer2010}
\begin{equation}
    {\cal S}_{\ell} = \left| \dfrac{(2 \ell + 1)!!\, \chi_{\ell}^{(\ell+1)}(0)}{(\ell+1)!\, k^{\ell+1}} \right|^{2}\,.
\end{equation}
The factor $S_\ell$ multiplies the corresponding term in the partial wave expansion of the annihilation cross-section 
\begin{equation}
    \sigma v_{\rm rel} = {\cal S}_s\,\sigma_0\,c + {\cal S}_p\,\sigma_1\,c\,\left(\frac{v_{\rm rel}}{c}\right)^2 + \mathcal{O}\left(\left(\frac{v_{\rm rel}}{c}\right)^4\right)\,,
\end{equation}
where ${\cal S}_s$ is the $s$-wave factor ($\ell=0$), ${\cal S}_p$ is the $p$-wave factor ($\ell=1$), etc. To make an explicit computation, we must fix the interaction potential.
The relevant one for an attractive interaction through a massive mediator is the Yukawa potential
\begin{equation}
V_{\rm Y}(r) = -\alpha_{\rm D}\, \dfrac{\mathrm{e}^{-m_{\phi}r}}{r}\,,
\end{equation}
with $m_\phi$ the mediator mass and $\alpha_{\rm D}$ the dark fine-structure constant. Unfortunately, solutions of the Schr\"{o}dinger equation for the Yukawa potential are only known numerically. 
Luckily, analytical expressions are known for the closely related Hulth\'{e}n potential
\begin{equation}
V_{\rm H}(r) = -\alpha_{\rm D}\,\dfrac{m_{\ast}\, \mathrm{e}^{-m_{\ast}r}}{1-\mathrm{e}^{-m_{\ast}r}}\,.
\end{equation}
Fixing $m_{\ast} = (\pi^{2}/6)m_{\phi}$ makes the Hulth\'{e}n analytical solution close to the Yukawa numerical solution \cite{Cassel2010}, therefore we use the former in the following.
From the Schr\"{o}dinger equation, one sees that the enhancement factor is only a function of two dimensionless parameters
\begin{equation}
\epsilon_{v} \equiv \dfrac{v}{\alpha_{\rm D}\,c} \ \ \ \mathrm{and} \ \ \ \epsilon_{\phi} \equiv \dfrac{m_{\phi}}{\alpha_{\rm D}\,m_{\chi}}\,.
\end{equation}

In this paper, we focus on $s$-wave and $p$-wave annihilation processes, corresponding to $\ell = 0$ and $\ell = 1$, respectively. This already covers a broad variety of underlying particle-physics models. For an $s$-wave annihilation process, the enhancement factor can be written as \cite{Slatyer2010}
\begin{equation}
\label{eq:Sommerfeld_enhancement_s_wave}
{\cal S}_{s}(v) \approx \begin{cases} \ \dfrac{\pi}{\epsilon_{v}} \dfrac{\sinh\left( \dfrac{2\pi\epsilon_{v}}{\epsilon_{\phi}^{*}} \right)}{\cosh \left( \dfrac{2\pi\epsilon_{v}}{\epsilon_{\phi}^{*}} \right) - \cos\left( 2\pi\sqrt{\dfrac{1}{\epsilon_{\phi}^{*}} - \dfrac{\epsilon_{v}^{2}}{\epsilon_{\phi}^{*2}}} \right)} & {\rm if}\ \ \ \epsilon_{v} \leqslant \sqrt{\epsilon_{\phi}^{*}} \\

\ \dfrac{\pi/\epsilon_{v}}{1 - {\rm e}^{-\pi/\epsilon_{v}}} & {\rm \ \ \ otherwise\,,}
\end{cases}
\end{equation}
where $\epsilon_{\phi}^{*} \equiv (\pi^{2}/6)\epsilon_{\phi}$. 
It should be noted that the expression of the Sommerfeld factor in the first line of Eq.~\eqref{eq:Sommerfeld_enhancement_s_wave} --- which is the standard result for the Hulth\'{e}n potential in the literature --- is only valid when both $\epsilon_{\phi}, \epsilon_{v} \ll 1$, \ie,~in the regime of large enhancements, as discussed in Ref.~\cite{Slatyer2010}. When $\epsilon_{v} \gg \sqrt{\epsilon_{\phi}}$, we recover the standard solution for the Coulomb potential corresponding to a massless mediator. 
For $p$-wave annihilation, the enhancement factor is
\begin{equation}
\label{eq:Sommerfeld_enhancement_p_wave}
{\cal S}_{p}(v) = \dfrac{\left( 1-\epsilon_{\phi}^{*}\right)^{2} + 4\, \epsilon_{v}^{2}}{\epsilon_{\phi}^{*2} + 4\, \epsilon_{v}^{2}} \times {\cal S}_{s}(v)\,.
\end{equation}
Different regimes arise according to the values of  $\epsilon_{v}$ and $\epsilon_{\phi}$, which encode the dependence of the Sommerfeld effect on the relative velocity of the DM particles and the masses of the DM candidate and light mediator:
\begin{itemize}
\item At large velocities, $\epsilon_{v} \gg 1$, there is no enhancement: ${\cal S}_{s} \approx 1$ and ${\cal S}_{p} \approx 1$;
\item In the intermediate regime, for which $\epsilon_{\phi} \ll \epsilon_{v} \ll 1$, we have ${\cal S}_{s} \approx \pi/\epsilon_{v} \propto 1/v$ and ${\cal S}_{p} \approx \pi/(4\epsilon_{v}^{3}) \propto 1/v^{3}$. This contains the regime in which the interaction potential tends to a Coulomb potential (for $\epsilon_{v} \gg \sqrt{\epsilon_{\phi}}$) but spans a broader range of values of $\epsilon_{v}$; 
\item The regime of small velocities, \ie,~$\epsilon_{v} \ll \epsilon_{\phi} \ll 1$, corresponds to the saturation regime of the Sommerfeld effect which is almost independent of the velocity of the DM particles, except at a series of resonances --- namely $\epsilon_{\phi} = 6/(\pi^{2} n^{2})$ with $n$ is an integer --- for which ${\cal S}_{s} \approx 1/(n^{2}\epsilon_{v}^{2}) \propto 1/v^{2}$ and ${\cal S}_{p} \approx \left( n^{2} -1 \right)^{2}/(n^{2}\epsilon_{v}^{2}) \propto 1/v^{2}$;
\item Finally when the mediator is heavy, \ie,~$\epsilon_{\phi} \gg 1$, there is again no enhancement, so ${\cal S}_{s} \approx 1$ and ${\cal S}_{p} \approx 1$.
\end{itemize}
The analytic solution typically reproduces the numerical result within 10\%, except close to resonances where larger differences arise since the analytic resonances for the Hulth\'{e}n potential are slightly offset from the ones obtained for the Yukawa potential, as discussed for instance in Refs.~\cite{FengEtAl2010,BoddyEtAl2017}. However, for the purpose of this work the features of the solution, especially the resonances, are sufficiently well accounted for by the analytic solution. For numerical calculations, we consider a benchmark value of $\alpha_{\rm D} = 10^{-2}$, but generalised $J$-factors can be easily rescaled, and we provide the scalings whenever relevant.

It should be noted that long-range interaction can lead to the formation of unstable bound states which modify the annihilation cross-section and are not taken into account in the equations above. In the resonant regime, we regularise the resonances by performing the replacement $v\rightarrow v+\alpha_{\rm D}^4$ which accounts for the finite lifetime of these intermediate states \cite{HisanoEtAl2005,FengEtAl2010,BlumEtAl2016}. Bound states can also form in the Coulomb regime where they might significantly change the overall cross-section, however the velocity dependence is left unchanged \cite{PetrakiEtAl2015,PetrakiEtAl2016} hence we choose to ignore this effect.

\subsection{Generalized $J$-factors and phase-space modelling}

\label{ssec:phase-space}

The DM-induced $\gamma$-ray flux integrated over a sky region of solid angle $\Delta \Omega$ reads\footnote{To ease the comparison with the majority of previous works in the literature, we do not include in the definition of the $J$-factor the $1/(4\pi)$ pre-factor. The latter appears in the derivation of an intensity from a volume emissivity, which is accounted for in Eq.~\eqref{eq:flux} here. As a result, the $J$-factors given in this work are expressed in $\rm GeV^{2}\, cm^{-5}\, sr$.}
\begin{equation}
\label{eq:flux}
\dfrac{\mathrm{d}\Phi_{\gamma}}{\mathrm{d}E_{\gamma}} = \dfrac{1}{4 \pi} \dfrac{(\sigma v_{\rm rel})_{0}}{\eta m_{\chi}^{2}}\dfrac{\mathrm{d}N}{\mathrm{d}E_{\gamma}} J_{\rm S} (\Delta \Omega)\,,
\end{equation}
where $\mathrm{d}N/\mathrm{d}E_{\gamma}$ is the $\gamma$-ray spectrum per annihilation, $\eta = 2$ for self-conjugate DM ($\eta = 4$ for non-self-conjugate DM), and the astrophysical factor $J_{\rm S}$ encodes the information on the DM spatial and velocity distribution. 

\paragraph{Generalised $J$-factors.}
We introduce the following notation
\ben
\overline{\cal S} =
\begin{cases}
  {\cal S}_s\left(\dfrac{v_{\rm rel}}{2}\right) &\text{(for $s$-wave annihilation)}\\
  \left( \dfrac{v_{\rm rel}}{c} \right)^{2}\,{\cal S}_p\left(\dfrac{v_{\rm rel}}{2}\right)\;\;&\text{(for $p$-wave annihilation)}
\end{cases}
\een
to treat the $s$- and $p$-wave annihilations on equal footing. This allows to write 
\begin{eqnarray}
\label{eq:J_S}
\!\!J_{\mathrm{S}} (\Delta \Omega) &\!=\!& \int_{\Delta \Omega} \!\!\! \mathrm{d}\Omega \int_{\rm l.o.s.} \! \mathrm{d}s  \int \! \mathrm{d}^{3}\vec{v}_{1} \int \! \mathrm{d}^{3}\vec{v}_{2} \, f(r(s,\Omega),\Vec{v}_{1}) \, f(r(s,\Omega),\Vec{v}_{2})\, \overline{{\cal S}}\! \left(\dfrac{v_{\rm rel}}{2}\right)\,, 
\end{eqnarray}
where $s$ the line-of-sight (l.o.s.) coordinate, $\Omega$ the solid angle, $\vec{v}_{\rm rel} = \Vec{v}_{2} - \Vec{v}_{1}$ is the relative velocity with $v_{\rm rel} = |\vec{v}_{\rm rel}|$ and $f(r,\Vec{v})$ is the phase-space distribution function (PSDF) of the DM (assuming spherical symmetry), normalised to the total mass of the gravitational system of interest, such that the DM density $\rho_\chi$ at galactocentric radius $r$ is
\begin{equation}
\rho_\chi(r) = \int  f(r,\Vec{v})\, \mathrm{d}^{3}\vec{v}\,.    \label{eq:rho_from_f}
\end{equation}
Equation~\eqref{eq:J_S} is referred to as the generalised $J$-factor. As the name indicates, it is a generalisation of the standard $J$-factor relevant for $s$-wave and $p$-wave annihilation without Sommerfeld enhancement (recovered for $\mathcal{S}_{s}=\mathcal{S}_{p}=1$). For the sake of clarity, we will denote $J_{\mathrm{S},s}$ the generalised $J$-factor associated to the $s$-wave and $J_{\mathrm{S},p}$ the one associated to the $p$-wave.

Assuming spherical symmetry of the DM halo, the integral over the solid angle becomes an integral over the angular distance $\psi$ from the centre of the object, with $\mathrm{d}\Omega = 2 \pi \sin \psi \, \mathrm{d} \psi$ and $r(s,\Omega) \equiv r(s,\psi) = \sqrt{s^{2} + D^{2} - 2 s D \cos \psi}$, where $D$ is the distance from the observer to the centre of the object. In the following, we perform the integral over an angular size $\theta_{\rm int}$ that depends on the target and can also depend on the $\gamma$-ray detection technique. For instance, most dSphs are observed as point-like by \textit{Fermi}-LAT, whereas galaxy clusters are extended targets, so we take $\theta_{\rm int} = 0.5^{\circ}$ for dSphs and $\theta_{\rm int} \approx R_{\rm 200}/D$ for clusters, with $R_{\rm 200}$ the virial radius.

In practice, Eq.~\eqref{eq:J_S} can be rewritten in terms of a $J$-factor for an effective squared density profile as 
\begin{eqnarray}
J_{\mathrm{S}} (\theta_{\rm int}) &=& 2 \pi \int_{0}^{\theta_{\rm int}} \! \mathrm{d}\psi \, \sin \psi \int \! \mathrm{d}s \, \left\langle \overline{{\cal S}}\! \left(\dfrac{v_{\rm rel}}{2}\right) \right\rangle \! (r(s,\psi)) \, \rho_\chi^{2}(r(s,\psi))\,,
\label{eq:J_S_fov}
\end{eqnarray}
where the average of an observable $\mathcal{O}(v_{\rm rel})$ that depends on the relative velocity is given by 
\begin{eqnarray}
\left\langle \mathcal{O}(v_{\rm rel}) \right\rangle\! (r) = \frac{1}{\rho_\chi^2(r)}\int \mathrm{d}^3\vec{v}_{\rm rel}\,\mathcal{O}(v_{\rm rel})\int\mathrm{d}^3\vec{v}_{\rm c}\,f(r,\vec{v}_{1}) \, f(r,\vec{v}_{2})\,,
\label{eq:average_O_vrel}
\end{eqnarray}
with $\vec{v}_{\rm c}=(\vec{v}_1+\vec{v}_2)/2$ the centre-of-mass velocity.
It should be noted that the profile is truncated at the tidal radius for dSphs, or (conventionally) at the virial radius for the other objects, with no contribution to the line-of-sight integral outside that radius.

\paragraph{Phase-space modelling.}
The main results of this work are based on the Eddington formalism, which provides the full PSDF $f(\vec{r},\vec{v})$ of a given component of a system in dynamical equilibrium associated with a given density-potential pair. More specifically, under the assumptions of maximal symmetries, namely spherical symmetry of the system, and an isotropic velocity tensor, the PSDF $f(r,\vec{v})$ can be written as a function of the relative energy ${\cal E} = \Psi(r) - v^{2}/2$ only, $f(r,\vec{v}) \equiv f({\cal E})$, where $\Psi$ is the total gravitational potential of the system. In that case, \citeeq{eq:rho_from_f} can be uniquely inverted, leading to the well-known Eddington formula \cite{Eddington1916,BinneyTremaine2008}:
\begin{equation}
\label{Eddington_formula}
f(\mathcal{E}) = \dfrac{1}{\sqrt{8}\pi^{2}} \left[ \dfrac{1}{\sqrt{\mathcal{E}}} \left( \dfrac{\mathrm{d}\rho_\chi}{\mathrm{d}\Psi} \right)_{\Psi=0} + \int_{0}^{\mathcal{E}} \! \dfrac{\mathrm{d}^{2}\rho_\chi}{\mathrm{d}\Psi^{2}} \, \dfrac{\mathrm{d}\Psi}{\sqrt{\mathcal{E} - \Psi}}    \right]\,,   
\end{equation}
where the first term between brackets is related to the radial boundary of the system. We disregard this term in the following, since we model the system as infinite as far as the PSDF is concerned when computing $J$-factors, considering that the latter are not sensitive to the very outer parts of the system. 

In \citeapp{app:phase_space_modelling} we provide a brief overview of the prediction methods for the PSDF $f(r,\vec{v})$ of DM particles from first principles that we use in this study. In particular, the Eddington formalism can be extended to anisotropic PSDFs under some specific assumptions. We use this extended formalism to quantify the uncertainty on generalised $J$-factors from the modelling of the PSDF itself, which we discuss in \citeapp{app:phase_space_modelling}. In the main text, we restrict the presentation to the Eddington method which provides a very good approximation.
Technical details regarding the semi-analytic derivation of averages over the relative velocity distribution, \citeeq{eq:average_O_vrel}, for various assumptions on the anisotropy of the velocity distribution, can be found in Ref.~\cite{LacroixEtAl2018}.

\subsection{Host halo and  subhalos: generalized boost factor}

\label{ssec:subhalo_boost}

DM subhalos, which are characteristic of any self-annihilating CDM particle scenario, are expected to boost the gamma-ray signals that would be predicted assuming DM is smoothly distributed in target halos \cite{SilkEtAl1993,BergstroemEtAl1999a}. We account for this boost factor by means of the analytical subhalo population model developed in \citeref{StrefEtAl2017} (SL17 henceforth) --- see also \citerefs{Huetten2019,StrefEtAl2019,FacchinettiEtAl2020,FacchinettiEtAl2022}. Considering subhalos is particularly important when annihilation is velocity-dependent because the internal velocity dispersion in these objects is much smaller than that of the host halo.
Indeed, from dimensional arguments and assuming virial equilibrium, the velocity dispersion should scale as $m^{1/3}$ where $m$ is the virial halo mass.
This is quite relevant for Sommerfeld-enhanced processes, which depend on inverse powers of the velocity, leading to a potentially strong enhancement when the subhalo mass range extends down to very small masses. Interestingly, heavy WIMPs beyond $\sim$10 TeV are naturally subject to Sommerfeld effects \cite{Hisano2003,CirelliEtAl2007}, and also lead to subhalo virial masses as small as $\sim 10^{-12}\,\Msun$ \cite{Bringmann2009}.

\paragraph{Modelling the host halo and subhalos.}
The SL17 model assumes that the host halo can be described by a spherically symmetric and smoothed total DM density profile, $\rho_{\rm host}(r)$, comprising a genuine smooth component $\rho_{\rm sm}(r)$ and a subhalo component made of individual objects (DM inhomogeneities), but globally described by an average density profile $\rho_{\rm sub}(r)$. These components are simply bound to obey the relation
\ben
\label{eq:rhohost}
\rho_{\rm host}(r) = \rho_{\rm sm}(r) + \rho_{\rm sub}(r) \geqslant 0 \, ,
\een
where $r$ is the radial distance to the host's centre. The density profile $\rho_{\rm host}$ can in principle be constrained both theoretically and observationally, in particular its shape (cuspy or cored, external tail, etc.), and both its global and internal properties (virial and/or tidal mass, concentration, etc.). For all of the host halos studied in this paper, we do use observationally constrained density profiles for $\rho_{\rm host}$ (see next section), and the smooth halo component $\rho_{\rm sm}$ is obtained by subtracting $\rho_{\rm sub}$ from $\rho_{\rm host}$, see \citeeq{eq:rhohost}. The average density profile of subhalos, $\rho_{\rm sub}$, is calculated from the SL17 model. It can be expressed in terms of a continuous number density of subhalos, $n_{\rm sub}$, depending on the virial\footnote{We use the conventional definitions, where the index ``200'' indicates that quantities are defined with respect to some virial radius $R_{\rm 200}$ (or $x_{\rm 200}$ for subhalos) over which the average density of a halo is 200 times the critical density at redshift zero. The actual extension of a halo is not necessarily its virial radius; for subhalos, the physical extension is taken to be the {\em tidal} radius, $r_{\rm t}\leqslant R_{\rm 200}$.} mass $m=M_{\rm 200}$ and concentration $c=c_{\rm 200}$ of the subhalos, and their radial position $r$ in the host halo:
\ben
\label{eq:defrhosub}
\rho_{\rm sub}(r) &=& \int \dd m \int \dd c\,m_{\rm t}(m,c,r)\,\left\{\frac{\dd^5N_{\rm sub}(m,c,r)}{\dd^3\vec{r}\,\dd m\,\dd c}\equiv N_{\rm tot}\,\frac{\dd^5{\cal P}_{\rm sub}(m,c,r)}{\dd^3\vec{r}\,\dd m\,\dd c}\right\}\\
&=& n_{\rm sub}(r)\times \langle m_{\rm t}(m,c)\rangle_{(m,c)}(r) \,,\nn
\een
where $m_{\rm t}\neq m$, the physical tidal (not virial) mass of subhalos, critically depends on the position $r$. Tidal stripping effects are sourced by all gravitational components of the host, leading in particular to a calculated total number $N_{\rm tot}$ of surviving subhalos. Moreover, although the spatial dependence of the mass and concentration PDFs were initially set homogeneous (from cosmological considerations), the tidal effects make the calculated probabilistic parameter phase space $\dd{\cal P}_{\rm sub}(m,c,r)$ fully intricate and non-separable; the latter is normalised to unity over the whole halo phase-space volume (position, mass, and concentration parameters). A short presentation of the SL17 model with more technical details is given in \citeapp{app:SL17}.

\paragraph{Generalized $J$-factor for subhalos.}

From \citeeq{eq:defrhosub}, we can write 
\ben
\label{eq:rho2sub}
\underline{\rho^2_{\rm sub}}(r) &\equiv& N_{\rm tot}\,\rho_\circledast^2\,\int \dd m \int \dd c \,
\xi_{\rm t}(m,c,r)\, \frac{\dd^5{\cal P}_{\rm sub}(m,c,r)}{\dd^3\vec{r}\,\dd m\,\dd c}\\
&=& n_{\rm sub}(r)\,\rho_\circledast^2\,\langle \xi_{\rm t}(m,c)\rangle_{(m,c)}(r)
\,,\nn
\een
where we defined the subhalo effective tidal annihilation volume\footnote{This is the volume a subhalo would have in order to sustain its own annihilation rate if it had an arbitrary constant DM density of $\rho_\circledast$ (similar to an intrinsic annihilation luminosity except for physical dimensions).}
\ben
\label{eq:defxi}
\xi_{\rm t}(m,c,r) \equiv \int_{x\leqslant x_{\rm t}(m,c,r)} \dd^3\vec{x}\,\left\{\frac{\rho(x,m,c)}{\rho_\circledast}\right\}^2\,,
\een
with $\rho(x,m,c)$ the inner subhalo profile and $x_{\rm t}$ its tidal extension. This tidal annihilation volume can be generalized to the velocity-dependent Sommerfeld enhancement case by writing
\ben
\label{eq:defxiS}
\xi_{{\rm S},{\rm t}}(m,c,r) \equiv \int_{x\leqslant x_{\rm t}(m,c,r)} \dd^3\vec{x}\,\left\{\frac{\rho(x,m,c)}{\rho_\circledast}\right\}^2\,\langle \overline{\cal S} \rangle(x) \,,
\een
where $\langle\rangle$ denotes the velocity average over the 2-particle phase-space volume introduced in Eq.~(\ref{eq:average_O_vrel}). We stress that this average is taken over the subhalo PSDF, which depends on $c$ and $m$ and is very different from the host PSDF. Eventually, the generalized $J$-factor associated with the total subhalo contribution reads
\ben
J_{{\rm S},{\rm sub}} (\theta_{\rm int}) = 2 \pi \int_{0}^{\theta_{\rm int}} \! \dd\psi \, \sin \psi \int \dd s \,  \underline{\rho^2_{{\rm S},{\rm sub}}}(r(s,\psi))\,,
\label{eq:J_S_sub}
\een
with
\ben
\label{eq:rho2sub_generalized}
\underline{\rho^2_{{\rm S},\rm sub}}(r) = n_{\rm sub}(r)\,\rho_\circledast^2\,\langle \xi_{{\rm S},{\rm t}}(m,c,r)\rangle_{(m,c)}(r)
\,.
\een

\paragraph{Total $J$-factor and boost.}

The total generalized $J$-factor is obtained summing up all contributions (host and subhalos):
\ben
J_{{\rm S},{\rm tot}} (\theta_{\rm int}) = 2 \pi \int_{0}^{\theta_{\rm int}} \! \dd\psi \, \sin \psi \int \dd s \,  \underline{\rho^2_{{\rm S},{\rm tot}}}(r(s,\psi))\,,
\label{eq:J_S_tot}
\een
where
\ben
\underline{\rho^2_{{\rm S},{\rm tot}}} = \underline{\rho^2_{{\rm S},{\rm sub}}} + \rho_{{\rm S},{\rm sm}}^2 + 2\, \rho_{{\rm S},{\rm sm}}\,\rho_{\rm sub}\neq \rho_{{\rm S},{\rm host}}^2\,.
\een
All these terms include a Sommerfeld-enhancement correction (subscript $S$). The last term before the inequality is the cross-product between the smooth DM component and subhalos, for which the relevant velocity field is that of the host halo; this term can actually safely be neglected \cite{StrefEtAl2017}. For further technical details on how $J_{\rm S,tot}$ is computed in practice, we refer the reader to \citeapp{app:subhalo_boost}.

Finally, we can formally define the generalized subhalo boost factor as
\ben
{\cal B}_{{\rm S}} \equiv \frac{J_{{\rm S},{\rm tot}}}{J_{{\rm S},{\rm host}}}\approx 1 + \frac{J_{{\rm S},{\rm sub}}}{J_{{\rm S},{\rm host}}}\,,
\een
where $J_{{\rm S},{\rm host}}$ is evaluated from the (squared) smoothed host profile given in \citeeq{eq:rhohost}, also corrected for the Sommerfeld enhancement as above. The approximation on the right-hand-side is valid only when the host is distant enough so that most of the annihilation rate is contained within the angular resolution of the telescope; in that case, $J_{{\rm S},{\rm sm}}\simeq J_{{\rm S},{\rm host}}$. Note that a detailed analytical study of the subhalo boost factor in the context of the Sommerfeld enhancement has been carried out in a companion paper \cite{CompanionPaper}.

\section{Selected targets and mass modelling}
\label{sec:mass_modelling}

We list in this section the astrophysical targets considered for our work (dwarf spheroidal galaxies, dwarf irregular galaxies, and galaxy clusters). For each of these object classes, we motivate our specific selection and discuss the DM density profile used for our analyses.

\subsection{Dwarf spheroidal galaxies}
\label{ssec:dSphs_mass_modelling}

Owing to their close distance (tens of kpc), potentially high DM densities, and negligible astrophysical background, MW dSph satellites are among the most promising targets for indirect DM detection \cite{Lake1990,EvansEtAl2004}. In the absence of a clear signal, the best current limits on WIMP DM candidates in $\gamma$-rays were obtained from the combined analysis of \textit{Fermi}-LAT data on many dSphs \cite{MAGIC2016,AlbertEtAl2017,HoofEtAl2020}. Their DM content is inferred from the velocity dispersion of their stellar population (obtained from spectroscopic measurements), via moments of the Jeans equations \cite{BinneyTremaine2008,Strigari2018}. DSphs are typically separated in two categories: `classical' and `ultra-faint'. The former are brighter, with hundreds to thousands of stars measured, while the latter are fainter, with only tens of known member stars. It ensues that the DM content of ultra-faint dSphs suffers larger uncertainties than that of the classical ones, which translates into less robust constraints on the DM particle properties in the former case.
However, more and more ultra-faint dSphs are discovered thanks to optical surveys \cite{BechtolEtAl2015,KoposovEtAl2015,Drlica-Wagner2015,LaevensEtAl2015a,LaevensEtAl2015b,HommaEtAl2016,HommaEtAl2018,HommaEtAl2019,Torrealba2019,Drlica-Wagner2020}, or will be discovered in the next decade \cite{Mutlu-Pakdil2021}, and those potentially located just tens of kpc away from us could shine even brighter than the classical dSphs in terms of $\gamma$-rays from their annihilating DM halos.

In the last decade, many studies have refined and improved the calculation of $J$-factors, in order to rank the best targets \cite{StrigariEtAl2007,MartinezEtAl2009,CharbonnierEtAl2011,Geringer-SamethEtAl2015,BonnivardEtAl2015b,EvansEtAl2016,SandersEtAl2016,PaceAndStrigari2019,ChiappoEtAl2019,BoddyEtAl2020,AlvarezEtAl2020}. Although these studies are overall in broad agreement, the assumptions made on the underlying ingredients (light and DM profiles, anisotropy distribution, triaxiality), methodology (e.g., using higher moments of the Jeans equation), and statistical analysis framework and priors used (data-driven approach, DM-simulation or mock-data based priors, etc.) can lead to sizeable differences in the expected DM signal and also on the $J$-factor  uncertainties of some dSphs (factor of a few). Discussing the relative merits of each approach to single out the best one goes beyond the scope of this paper, and is in any case a very difficult task: all studies consider slightly different but mostly relevant methodologies (with different limitations) for the reconstruction of DM density profiles. With the improvement on stellar structural parameters \cite{MunozEtAl2018,Simon2019} and new spectroscopic data \cite{SimonEtAl2020,JenkinsEtAl2021}, predictions for the DM halo will hopefully become less uncertain, in particular for ultra-faint dSphs (see, e.g.,~\cite{Simon2019} for a recent review).

For definiteness, we pick here two classical dSphs (Draco and Sculptor, respectively in the Northern and Southern sky) and one prototypical ultra-faint (Reticulum II), which were found to be among the best-ranked targets for DM annihilation in \cite{BonnivardEtAl2015b,BonnivardEtAl2015c}. In the latter studies, the DM profile parameters were reconstructed from a Markov Chain Monte Carlo (MCMC) engine coupled to a Jeans analysis with the CLUMPY code\footnote{\url{https://clumpy.gitlab.io/CLUMPY/}} \cite{BonnivardEtAl2016,HuettenEtAl2019}; see \cite{BonnivardEtAl2015a} for more details on the methodology. We use these chains to calculate the median profile, that we adopt as a reference for our analyses here\footnote{We do not use the best-fit profile parameters, because the scarcity of data in ultra-faint dSphs make them display an unphysical behaviour (e.g., a very flat and extended profile). Using `effective' structural parameters matching the median profile cures this issue. The generalised $J$-factors calculated from these effective parameters are also found to be very close to the median generalised $J$-factors calculated over the MCMC values.}.
We gather in Table~\ref{tab:dsphs_mass_models} the position and DM profile parameters for the three selected dSphs, modelled following an Einasto profile:
\begin{equation}
\rho_{\rm Ein}(r)=\rho_{-2}\, \exp\left\{-\frac{2}{\alpha}\left[\left(\frac{r}{r_{-2}}\right)^\alpha -1\right]\right\}\;.
\label{eq:Einasto}
\end{equation}
The parameters $r_{-2}$, $\rho_{-2}$, and $\alpha$ are the radius for which the slope is $-2$, the DM density at this radius, and the slope of the Einasto profile, respectively. 
\begin{table}[t!]
\begin{center}
\begin{tabular}{|l|cc|ccc|}\hline
  dSph &  (l, b)  &  D & $\rho_{-2}$ &  $r_{-2}$ & $\alpha$ \\
       &  [deg]  & [kpc] &[10$^7$M$_\odot$\,kpc$^{-3}$]& [kpc] & - \\
\hline 
Reticulum II (Ret2)    & (266.3, -49.7) & 30 & 2.53 & 0.92 & 0.46  \\
Sculptor             (scl)  & (287.5, -83.2) & 79 & 2.87 & 0.50 & 0.31 \\
Draco                (dra)  & (86.4, +34.7)  & 82 & 1.06 & 2.09 & 0.46 \\
\hline
\end{tabular}
\caption{\label{tab:dsphs_mass_models} Relevant parameters of the three MW dSphs selected for this analysis. The first columns report the dSph position (Galactic longitude and latitude) and distance from the observer: for consistency with the analysis of Ref.~\cite{BonnivardEtAl2015b}, that we follow here, we take the distance from Ref.~\cite{Mateo1998}, although more recent estimates can slightly differ \cite{McConnachie2012}. The last three columns list the Einasto profile parameters (normalisation $\rho_{-2}$, scale radius $r_{-2}$, and slope $\alpha$) corresponding to the median profile calculated over our MCMC sample.}
\end{center}
\end{table}

The generalised $J$-factors for velocity-dependent cross-sections rely on the DM phase space distribution (see \citesec{ssec:phase-space}). To properly and fully propagate the DM profile uncertainties to the generalised $J$-factor, we start from 1000 profile parameter samples taken from the analysis of \cite{BonnivardEtAl2015b} (\cite{BonnivardEtAl2015c} for Reticulum II), apply the Eddington calculation to obtain the phase-space associated to each profile (see also \citeapp{app:phase_space_modelling}), and then calculate the associated generalised $J$-factor. From the distribution of the 1000 calculated $J$-factors, we can calculate any quantile to derive the mean of the distribution and its uncertainties.

\subsection{Dwarf irregular galaxies}
\label{ssec:dIrrs_mass_modelling}

Dwarf irregular (dIrrs) galaxies have recently entered in the list of prime targets for indirect $\gamma$-ray DM searches. Indeed, the existence of these isolated galaxies within the Local Group, at $\mathcal{O}(1~\rm{Mpc})$ distances, makes them interesting targets given both their proximity and typical masses $M_{\rm 200}\approx 10^7-10^{10} \rm M_{\odot}$. DIrrs are rotationally-supported objects, allowing to reconstruct the underlying DM density profiles from their measured rotation curves (RCs). Such RC studies show that dIrrs are DM-dominated objects at all radii \cite{Oh:2015xoa, oh1, Gentile:2006hv}. Unlike dSphs, dIrrs are star-forming galaxies, yet the $\gamma$-ray emission associated to astrophysical processes has been estimated to be negligible compared to that expected from DM annihilation \cite{Winter:2016wmy, Gammaldi:2017mio}.  
One more reason that makes dIrrs promising targets for DM searches is the fact that, given their typical host halo masses, the so-called subhalo boost is expected to be significant in their case, reaching values up to $\sim$5 \citep{2021PhRvD.104h3026G}, depending on the definition of the boost factor. This is in contrast to the case of dSphs, which are not only less massive than dIrrs but also tidally stripped objects, thus with expected subhalo boosts of the order of only a few percent \citep{MolineEtAl2017}. 
Despite the above considerations, dIrrs have not been used for $\gamma$-ray DM searches up to just recently \citep{Gammaldi:2017mio, 2021PhRvD.104h3026G}. 

With the current available observational data, the study of dIrrs RCs is not conclusive and, indeed, there is still a debate in the literature about the precise inner shape of the DM density profile in these objects. Fits to the RCs favor core-like profile~\cite{Salucci:2007tm}, yet this conclusion is in contrast with that expected from N-body cosmological simulations, that point to a universal cuspy profile like NFW \citep{Navarro:1995iw, NavarroEtAl1997} or Einasto \citep{Einasto1965}. Multiple studies have investigated the source of this apparent disagreement --- not unique to this type of objects --- between data and $\Lambda$CDM expectations, providing different solutions mainly based on the impact of baryonic feedback on the DM distribution and its ability to shallow the initial cusps in the innermost regions of the DM density profiles, especially at some particular mass scales~\cite{Gomez-Vargas:2013bea, Schaller:2014uwa, 2017MNRAS.472.2153P,Bose:2018oaj, 2019MNRAS.488.2387B}. As this issue is far from being solved, the authors in \cite{2021PhRvD.104h3026G} adopted an agnostic path and decided to perform a DM modelling for dIrrs using the two different types of profiles, i.e. (i) a Burkert, core-like profile \cite{Burkert:1995yz}
\begin{equation}
    \rho_\mathrm{Bur}(r) = \frac{\rho_c\,r_c^3}{(r+r_c)\,(r^2+r_c^2)}\,,
\label{eqn:Burkert}    
\end{equation}
where $r_{c}$ and $\rho_{c}$ are, respectively, a core radius and DM density, and (ii) an NFW cusp-like profile
\begin{equation}\label{eqn:NFW}
 \rho_\mathrm{NFW}(r)=\frac{\rho_{\rm 0}}{\left(\frac{r}{r_{\rm s}}\right)\left(1+\frac{r}{r_{\rm s}}\right)^{2}}\,,
\end{equation}
where $r_{\rm s}$ and $\rho_{\rm 0}$ are, respectively, a scale radius and a characteristic DM density.

In this work, for each of these profiles, we simply use the best-fit parameters obtained in~\cite{2021PhRvD.104h3026G}, where authors analyze the RCs of 7 dIrrs and obtain a prediction of the J-factors for the two different models of the DM density profile under consideration here. According to the observed RCs, NGC6822, IC10 and WLM are the ones with more available data, thus in these cases the fits are more robust and stable than for the rest of objects in their sample. The mentioned three objects also yield the highest J-factor values independently of the selected DM profile or substructure boost values. Taking these findings in~\cite{2021PhRvD.104h3026G} into account, we thus decided to include NGC6822, IC10 and WLM in our sample, whose parameters are gathered in Table~\ref{tab:dIrr}.
%
\begin{table}[t!]
\begin{center}
\begin{tabular}{|c|cc|c|cccc|}
\hline
\multirow{2}{*}{dIrr} & ($l$, $b$) & $D$ & $M_{\rm 200}$ & Profile & $\rho$ & $r$ & $R_{\rm 200}$ \\
 & [deg] & [kpc] & [$10^{10}\; \rm M_\odot$] &  & [$10^{7}\;\rm M_\odot$\,kpc$^{-3}$] & [kpc] & [kpc]\\
\hline 
\multirow{2}{*}{NGC6822} & \multirow{2}{*}{(25.34, -18.40)} & \multirow{2}{*}{480} & \multirow{2}{*}{3.16} & Burkert* & 3.16 & 3.3 & 62.9  \\ 
 &  &  &  & NFW & 0.79 & 5.9 & 62.6 \\
\hline
\multirow{2}{*}{IC10} & \multirow{2}{*}{(118.96, -3.33)} & \multirow{2}{*}{790} & \multirow{2}{*}{3.98} & Burkert* & 15.85 & 2.0 & 71.3  \\ 
 &   &  &  & NFW & 0.63 & 6.8 & 70.3 \\
\hline
\multirow{2}{*}{WLM} & \multirow{2}{*}{(75.87, -73.86)} & \multirow{2}{*}{970} & \multirow{2}{*}{0.40} & Burkert* & 6.31 & 1.3 & 33.3  \\ 
 &   &  &  & NFW & 1.00 & 2.8 & 33.6 \\
\hline
\end{tabular}
\caption{\label{tab:dIrr}Parameters of the three dIrrs in our sample. The first column reports the dIrrs position (galactic longitude and latitude) and distance from the observer (see \cite{2021PhRvD.104h3026G} and references therein). We then list the best-fit profile parameters for Burkert and NFW (mass, normalisation and scale radius), as well as the virial radius assuming an overdensity of 200 times the critical density of the Universe. For columns 6 and 7, $\rho$ and $r$ stand for $\rho_c$ and $r_c$ in case of the Burkert profile, and $\rho_{\rm 0}$ and $r_{\rm s}$ for NFW. Profiles marked with $^{*}$ are used as reference for all calculations and figures unless indicated otherwise.}
\end{center}
\end{table}
As for the dSphs, the calculation of the generalised $J$-factors relies on the DM phase space distribution described in \citesec{ssec:phase-space} (based on the inversion of the DM profile). However, at variance with the dSphs, the uncertainties for dIrrs are estimated from the comparison of the results obtained from the Burkert and NFW profiles. The modelling of subhalos in the context of Sommerfeld enhancement relies on the formalism described in \citesec{ssec:subhalo_boost}.

\subsection{Galaxy clusters}
\label{ssec:galaxy-clusters-smooth}

Galaxy clusters are the largest gravitationally-bound objects in the Universe. Their masses are between $M_{\rm 200}\approx 10^{14}-10^{15}\, \rm M_{\odot}$ and up to 80\% of this mass is expected to be DM \cite{RevModPhys.77.207}. The rest is baryonic matter, in the form of galaxies, hot gas and dust in the intra-cluster medium (ICM). Even though clusters are supposedly stable and virialized objects at present, the presence of hot gas, galaxies, and even Active Galactic Nuclei (AGNs), produces turbulence phenomena and complex baryonic feedback reactions in the ICM (where also significant high magnetic fields are involved). All these astrophysical processes end up acting as acceleration mechanisms, leading to the presence of cosmic rays (CRs), that have been confirmed through the observation of diffuse synchrotron emission produced by the leptonic CRs at different wavelengths \cite{vanWeeren:2019vxy}. Galaxy clusters have avoided detection in $\gamma$-rays so far \cite{Ackermann:2013iaq, Colavincenzo:2019jtj}\footnote{There is a growing evidence, though, for a potential detection in the vicinity of the Coma cluster \cite{AckermannEtAl2016,XiEtAl2018,AdamEtAl2021,BaghmanyanEtAl2021}.}, but this high-energy emission is indeed expected from hadronic CRs \cite{Blasi:2007pm,2010MNRAS.409..449P,AdamEtAl2021}.

Despite their expected CR-induced $\gamma$-ray emission, galaxy clusters are still considered excellent targets for $\gamma$-ray DM searches in the WIMP scenario (from DM annihilation or decay). The DM science case of galaxy clusters soon resulted in studies aimed at determining which galaxy clusters meet the most appropriate conditions to be searched in $\gamma$-rays \cite{JeltemaEtAl2009,2010JCAP...05..025A,PinzkeEtAl2011,CombetEtAl2012,NezriEtAl2012,2012JCAP...07..017A,Ackermann:2015fdi,AckermannEtAl2016,Acciari:2018sjn} and at disentangling both the CR- and DM-induced $\gamma$-ray emissions from each other \cite{JeltemaEtAl2009,MaurinEtAl2012}. First, there exists a significant number of local galaxy clusters ($z<0.1$) for which substantial DM-induced fluxes are expected. Second, DM searches should focus on those with the lowest expected CR backgrounds \cite{JeltemaEtAl2009}. In \cite{2011JCAP...12..011S,NezriEtAl2012}, the authors studied the annihilation flux of the most promising galaxy clusters, once DM halo substructures --- particularly relevant for clusters --- were taken into account. 
It was found that the brightest galaxy clusters can yield total annihilation fluxes as large as some of the dSphs. Furthermore, for clusters, the annihilation flux profiles become comparatively more spatially extended, as most subhalos are located in the outer halo regions. Overall, this subhalo boost to the annihilation signal is expected to play a key role for clusters as compared to other targets, such as dSphs and dIrrs, for which the boost is negligible or much smaller, respectively \cite{CharbonnierEtAl2011,2011JCAP...12..011S, MolineEtAl2017, 2021PhRvD.104h3026G}.
We note, however, that the inclusion of halo substructure, in the case of expanding the annihilation cross-section to $p$-waves and in the framework of Sommerfeld enhancement, becomes more complex and requires a specific approach that is addressed in Sec. \ref{ssec:subhalo_boost}.

\paragraph{Halo mass modelling.}
For this work, we follow Ref.~\cite{2011JCAP...12..011S} as a starting point to build our sample of most promising galaxy clusters for DM searches. Their sample was constituted by Virgo, Coma, Fornax, Ophiuchus and Perseus.\footnote{A comprehensive and systematic ranking of galaxy clusters in terms of their expected annihilation signals can be found in \cite{NezriEtAl2012}, where other targets were also found at the level of those selected for this study.} Yet, some of these clusters present major observation inconveniences. While Virgo exhibits the highest $J$-factor, it is currently going through a major merger event with the neighbouring M49 galaxy cluster \cite{Ackermann:2015fdi}. Also, its proximity to Earth results in an angular extension of several degrees. The observation of an object of this size is extremely challenging given the field of view of existing IACTs. On the other hand, the galactic diffuse emission should be ideally avoided as to simplify any potential DM analysis. This can be easily addressed by removing from our sample those objects located close to the Galactic plane and centre, where this emission is most extreme. This requirement leaves out Ophiuchus, less than 10 degrees far from the Galactic centre. Thus, in the following we will obtain predictions for Coma, Fornax and Perseus and will remove both Virgo and Ophiuchus from our list of clusters. Note that this number of targets is also similar to the numbers in our sample of dSphs and dIrrs.
We build the DM density profile of galaxy clusters starting from their measured mass. For nearby galaxy clusters as the ones in our sample, $M_{\rm 200}$ can be obtained from X-ray observations of the surface brightness profiles. Indeed, these observations have been used to create catalogues containing the most relevant cluster parameters \cite{2002ApJ...567..716R, 2011A&A...534A.109P, Snowden:2007jg, Vikhlinin:2008cd}. In our work, we adopt the mass estimates in \cite{Schellenberger:2017wdw} for Coma and Fornax, while for Perseus we use data from \cite{2002ApJ...567..716R} (rescaled to our cosmology). 
First, we assume the NFW DM density profile given in Eq.~(\ref{eqn:NFW}). 
Assuming a spherical collapse model with an overdensity $\Delta=200$ times the critical density of the Universe, we can obtain the corresponding virial radius $R_{\rm 200}$ than contains the mass $M_{\rm 200}$.
Now, in order to obtain the two NFW profile parameters we need to assume a concentration-mass ($c-M$) relation. We adopt the parametrization proposed in \cite{Sanchez-CondeEtAl2014} for main halos. 
From the value of the concentration and the already obtained $R_{\rm 200}$, we can then compute the NFW scale radius $r_{\rm s}$ as well as the scale density $\rho_{\rm 0}$.

\paragraph{Mass modelling uncertainties.}
The main uncertainties in our DM modelling come from (i) the estimate of the mass as derived from X-rays data  and (ii) the intrinsic scatter of the concentration-mass relation. 
Indeed, it is well known that different observational methods can yield different mass estimates for galaxy clusters. Deviation of mass estimates from surface brightness X-ray measurements with respect to the masses obtained by other observation methods is typically referred to as the hydrostatic bias. Yet, at present there is an on-going debate in the community about how to precisely quantify and treat its value \citep{Pratt:2019cnf}. A complementary approach is to compare X-ray masses, usually labelled as $M_{\rm hydro}$, with the masses provided by other methods, whenever available. For example, the authors in~\cite{Schellenberger:2017wdw} concluded that cluster masses in their catalogue showed a good agreement with the ones obtained from velocity dispersion measurements \cite{2017A&A...599A.138Z}, while this was not the case for the objects in their sample for which SZ measurements \cite{2016A&A...594A..27P} were also available. More precisely, for clusters with masses $M_{\rm 200}<5\times 10^{14} h^{-1}\rm M_{\odot}$, they narrowed down the discrepancy to $M_{\rm hydro}/M_{SZ} = 0.86 \pm 0.01$, and for clusters with larger masses to $M_{\rm hydro}/M_{SZ} = 1.46 \pm 0.08$. From these results, we can conclude that the X-ray mass can be underestimated by $\sim 20\%$ in the case of less massive clusters, while $M_{\rm 200}$ can be overestimated by $\sim 50\%$ for the most massive ones.

Following these results, in this work we adopt two mass estimates for each galaxy cluster, that will translate into a bracketing of the $J$-factor uncertainties (due to the cluster mass uncertainty). 
Our default model is built starting from $M_{\rm hydro}$ and, in addition, we assign each cluster a second mass depending on the above bias.
For Fornax, a light cluster, we use $1.2\times M_{\rm hydro}$ as a second, upper bound mass estimate. In contrast, we adopt $0.5\times M_{\rm hydro}$ as a lower bound for Perseus and Coma, both massive clusters according to the classification scheme in \cite{Schellenberger:2017wdw}. 
As for the uncertainty associated to the scatter of concentrations values for a given mass, we adopt a value of 0.14 dex as suggested by the authors of Ref.~\cite{Sanchez-CondeEtAl2014}. 
In order to keep a limited number of models, we take advantage of the fact that the $J$-factor $\propto (M_{\rm 200}^2~c_{\rm 200}^{3})/D^{2}$ to further increase the previous uncertainties by considering extreme values of the concentration scatter. To do so, we consider, for both the upper and lower mass bounds previously derived, $M_{\rm 200}^{(\rm min)}$ and $M_{\rm 200}^{(\rm max)}$, the concentrations $c_{\rm 200}(M_{\rm 200}^{(\rm min)}) \times 10^{-\sigma_{\rm c}}$ and $c_{\rm 200}(M_{\rm 200}^{(\rm max)}) \times 10^{+\sigma_{\rm c}}$. The obtained DM density profile parameters for our sample of galaxy clusters are given in Table~\ref{tab:clusters_mass_models}.

\begin{table}[t!]
\begin{center}
\scalebox{0.87}{
\begin{tabular}{|c|cc|ccccc|}
\hline
\multirow{2}{*}{Cluster} & ($l$, $b$) & $D$  & Mass  & $M_{\rm 200}$  & $R_{\rm 200}$ & $\rho_{\rm 0}$ & $r_{\rm s}$  \\
& [deg] & [Mpc] &   estimate   & $[10^{14}\; \mathrm{M}_{\odot}]$ & [$10^2\;$kpc]     & $[10^6\;\rm M_{\odot}\,kpc^{-3}]$    & [$10^2\;$kpc]    \\
\hline
\multirow{2}{*}{Coma} & \multirow{2}{*}{(58.09, 87.96)}  & \multirow{2}{*}{102.18}    & Hydrostatic        &       $13.16$      &        $23.19$    &      $2.29$           &  $3.38$  \\ 
&  & & Lower*       &       $8.77$        &     20.26      &    $5.37$        &      $5.58$  \\
\hline
\multirow{2}{*}{Fornax} & \multirow{2}{*}{(236.72, -53.64)} & \multirow{2}{*}{20.35}    & Hydrostatic*        &       $0.51$         &   $7.83$     &       $7.42$         &  $1.86$   \\ 
&  & & Upper      &       $0.61$          &   $8.32$      &       $3.20$         & $1.05$  \\ 
\hline
\multirow{2}{*}{Perseus} & \multirow{2}{*}{(150.57, -13.26)} & \multirow{2}{*}{80.69}   & Hydrostatic    &    $7.71$     &   $19.41$     &       $2.35$       & $2.80$  \\ 
&  & & Lower*       &       $5.14$         &     $16.96$    &   $5.57$  &  $4.59$ \\ 
\hline
\end{tabular}}
\caption{\label{tab:clusters_mass_models} DM density profile parameters for the three galaxy clusters in our sample. For each target, we show two mass models in order to bracket uncertainties in the corresponding $J$-factors. See text for details. Profiles marked with $^{*}$ are used as reference for all calculations and figures unless indicated otherwise.}
\end{center}
\end{table}

\paragraph{Impact of baryons.}

As introduced before in this section, most of galaxy clusters' mass is in the form of non-visible DM, and the rest is accounted for baryonic matter. This baryonic content is mostly encoded in the form of super-heated ionized plasma, the so-called ICM, that accounts for $\sim$15\% of the cluster mass, while the remaining $\sim$5\% is in the form of galaxies.
Because of this, the effect of these baryonic components on the DM modelling of the galaxy cluster's main halo can be neglected (as done above), as their contribution to the total mass of the system is even smaller than the size of the uncertainty in the mass estimates themselves. 
However, the inclusion of the baryonic content in the mean gravitational potential may play a relevant role in the modelling of substructures and the computation of the boost factor since it directly impacts the tidal field experienced by these objects. Indeed, given the typical mass range of the substructures, this second-order effect could lead to different distributions and properties of the subhalo population, meaning that, ideally, we would need to obtain a density model for the baryonic matter.

We thus wanted to quantify this effect for our work, neglecting in a first approximation the galaxies and focusing on the ICM alone, e.g.,~\cite{Chen:2007sz}. Starting from standard X-ray gas density profiles, we built baryon density profiles that included not only electrons, but also protons and Helium following the methodology in \cite{Adam:2020atc}. The cluster X-ray parameters were taken from \cite{Chen:2007sz}. We found that including baryons in the modelling of the mean gravitational impacts the final boost factors at the level of one percent at most. Thus, in the following, we implicitly neglect the baryonic content in clusters and only show results related to their DM content.

\section{Generalised $J$-factors for host halos without substructures}
\label{sec:uncertainties_phase_space_smooth}

In this section, we describe, for the host DM halos of our selected targets, the salient features of $J_{\mathrm{S}}$, the generalised $J$-factor, as a function of $\epsilon_{\phi}$  (\citesec{ssec:general_features_J_factors}). We then show how systematic errors on parameters of the smooth DM profiles translate into systematic uncertainties on $J_{\mathrm{S}}$ (\citesec{ssec:host_uncertainties}). Next, we discuss the ranking of our targets in the various regimes of the Sommerfeld enhancement (\citesec{ssec:host_ranking}).
We stress that this section only deals with $J_{\mathrm{S}}$ from the smooth DM distribution in our targets; the full calculation of the generalised $J$-factors including the contribution of DM substructures is postponed to the next section (\citesec{sec:subhalo_boost_and_classification}).
 
We emphasise that all our results, based on full numerical calculations, have been cross-validated (for all regimes) thanks to the analytical calculations presented in the companion paper \cite{CompanionPaper}. This gives us a strong confidence in these results and the conclusions we draw.

\subsection{General features for $s$-wave and $p$-wave annihilations}
\label{ssec:general_features_J_factors}

The generalised $J$-factors for host halos (i.e., smooth DM distribution) are shown in \citefig{fig:JS_smooth_error_bars} for dSphs (top), dIrrs (middle) and galaxy clusters (bottom), for $s$-wave annihilations (left panels) and $p$-wave annihilations (right panels). 
The behavior of $J_{\mathrm{S}}$ as a function of $\epsilon_{\phi}$ results from the convolution of the Sommerfeld enhancement factor, ${\cal S}(v)$, with the velocity distribution in each target. The results are directly associated with the various regimes of ${\cal S}$ discussed in \citesec{ssec:sommerfeld}. 

\begin{table}[t!]
\begin{center}
\begin{tabular}{|c|c|c|}
\hline
Target class & Target  & $\epsilon_{\phi}^{\star}$\\   
\hline
\multirow{3}{*}{dSphs} & Draco & $1.4 \times 10^{-2}$ \\
 &  Sculptor & $5.7 \times 10^{-3}$ \\
 & Reticulum II & $9.5 \times 10^{-3}$ \\
\hline
\multirow{3}{*}{dIrrs} & IC10 & $2.2 \times 10^{-2}$ \\
 & NGC6822 & $1.6 \times 10^{-2}$\\
 & WLM & $9 \times 10^{-3}$\\
 \hline
 \multirow{3}{*}{Clusters} & Fornax & $1.8 \times 10^{-1}$ \\
 & Coma & $5.2 \times 10^{-1}$\\
 & Perseus & $4.4 \times 10^{-1}$\\
 \hline
\end{tabular}
\caption{\label{tab:targets}Summary table of the characteristic values $\epsilon_{\phi}^{\star}$ of the $\epsilon_{\phi}$ parameter, corresponding to the transition between the Coulomb and saturation regimes of Sommerfeld enhancement, associated with the typical velocity in the object of interest, for $\alpha_{\rm D} = 10^{-2}$.}
\end{center}
\end{table}

The main scale of the problem is the characteristic value $\epsilon_{\phi}^{\star}$ at which the Sommerfeld enhancement saturates for a given object. In practice, this transition between the Coulomb and (resonant) saturation regimes can be well reproduced by
\begin{equation}
    \epsilon_{\phi}^{\star} \sim \dfrac{\bar{v}}{\alpha_{\rm D} c}\,,
    \label{eq:def_epsphi_star}
\end{equation}
where $\bar{v}$ is the characteristic velocity of the object, for which a good order-of-magnitude estimate\footnote{Roughly speaking, the characteristic velocity is of order $\sqrt{4\pi G_{\rm N} \rho_{0} r_{\rm s}^{2}}$, where $\rho_{0}$ and $r_{\rm s}$ refer generically to the characteristic density and scale radius of the DM profile considered for each class of object (be it NFW, Einasto or Burkert).} is given by the circular velocity at the scale radius of the DM profile, 
\begin{equation}
    \bar{v} \sim \sqrt{\dfrac{G_{\rm N} m(r_{\rm s})}{r_{\rm s}}}\,.
    \label{eq:circ_velocity}
\end{equation}
The corresponding values of $\epsilon_{\phi}^{\star}$ are given in the last column of Table~\ref{tab:targets}; these values are also relevant for the boost from DM substructure (see \citeapp{app:subhalo_boost}).
Depending on the ordering of $\epsilon_{\phi}$ and $\epsilon_{\phi}^{\star}$, three different regimes for $J_{\mathrm{S}}(\epsilon_{\phi})$ can be identified.

\begin{figure}[th!]
\centering
\includegraphics[width=0.485\linewidth]{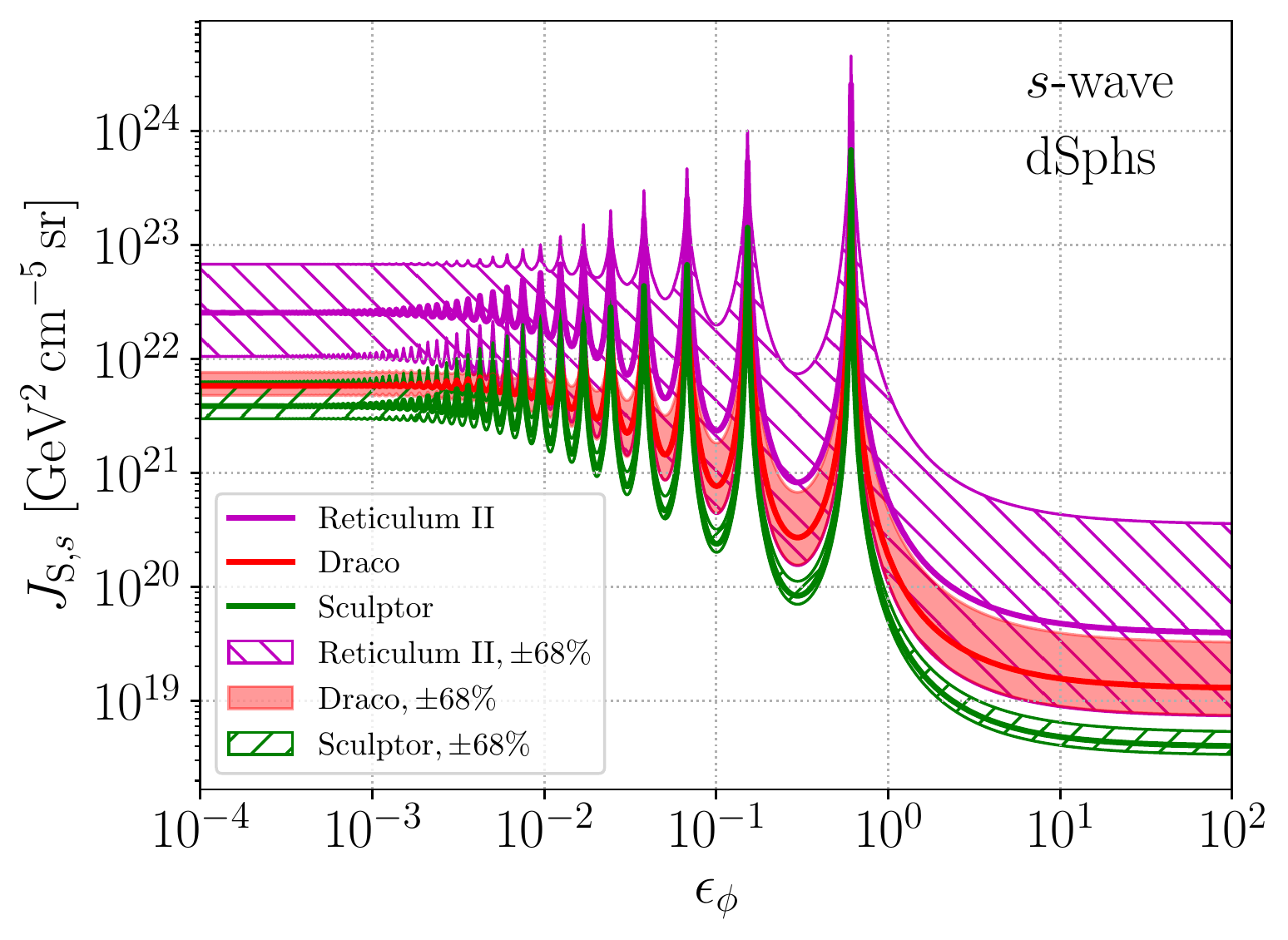} \hfill
\includegraphics[width=0.485\linewidth]{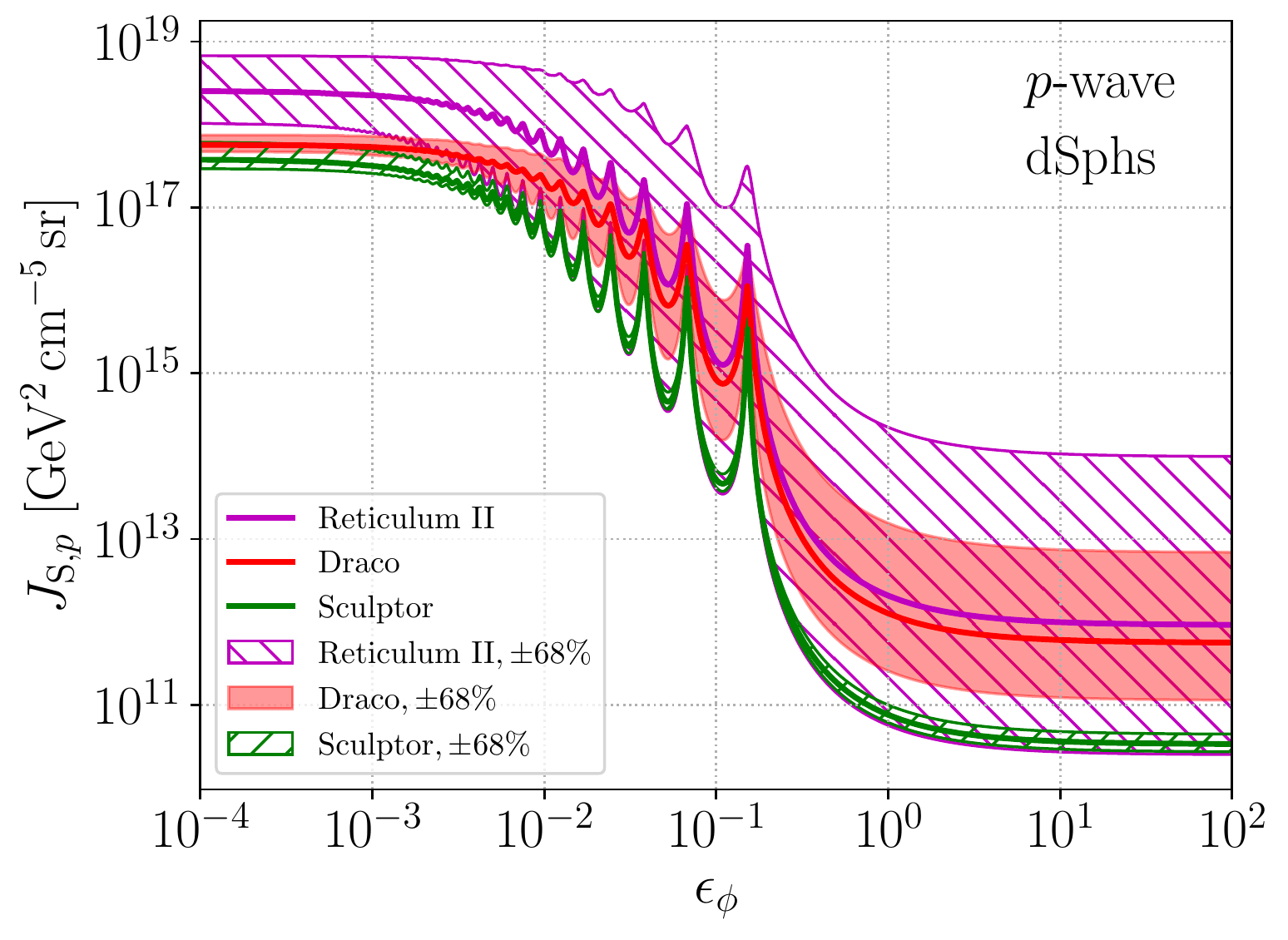}
\includegraphics[width=0.485\linewidth]{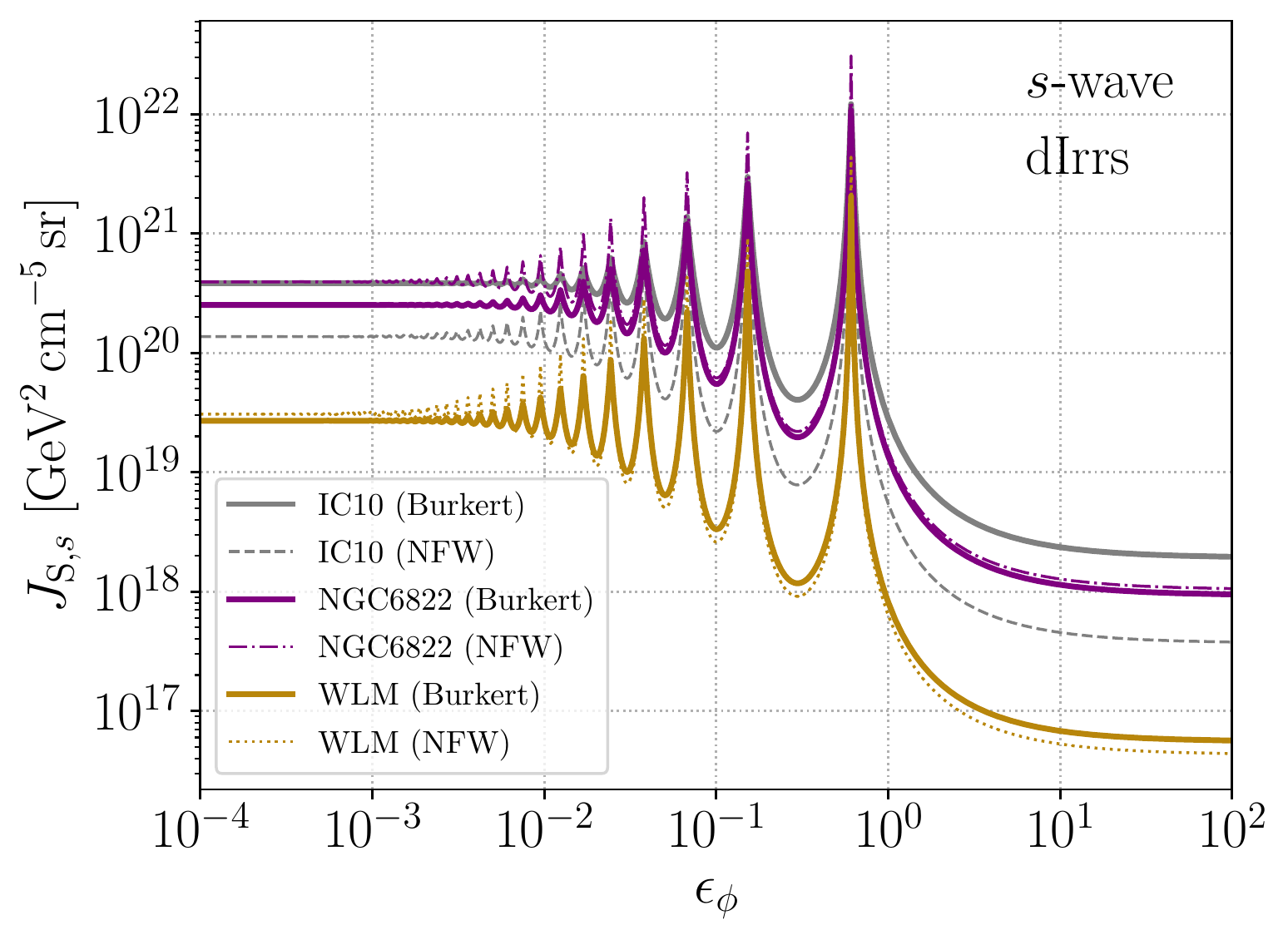} \hfill \includegraphics[width=0.485\linewidth]{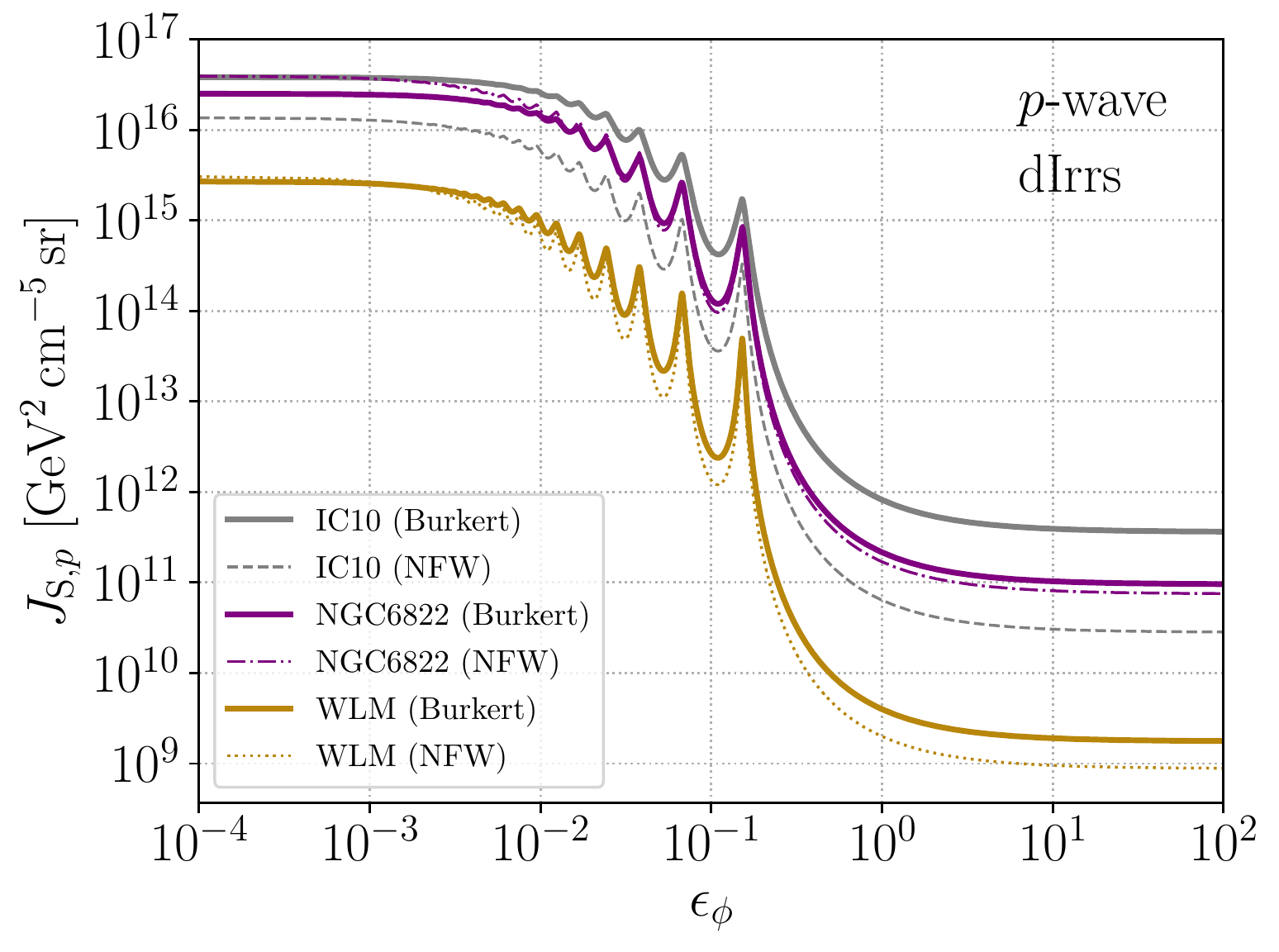}
\includegraphics[width=0.485\linewidth]{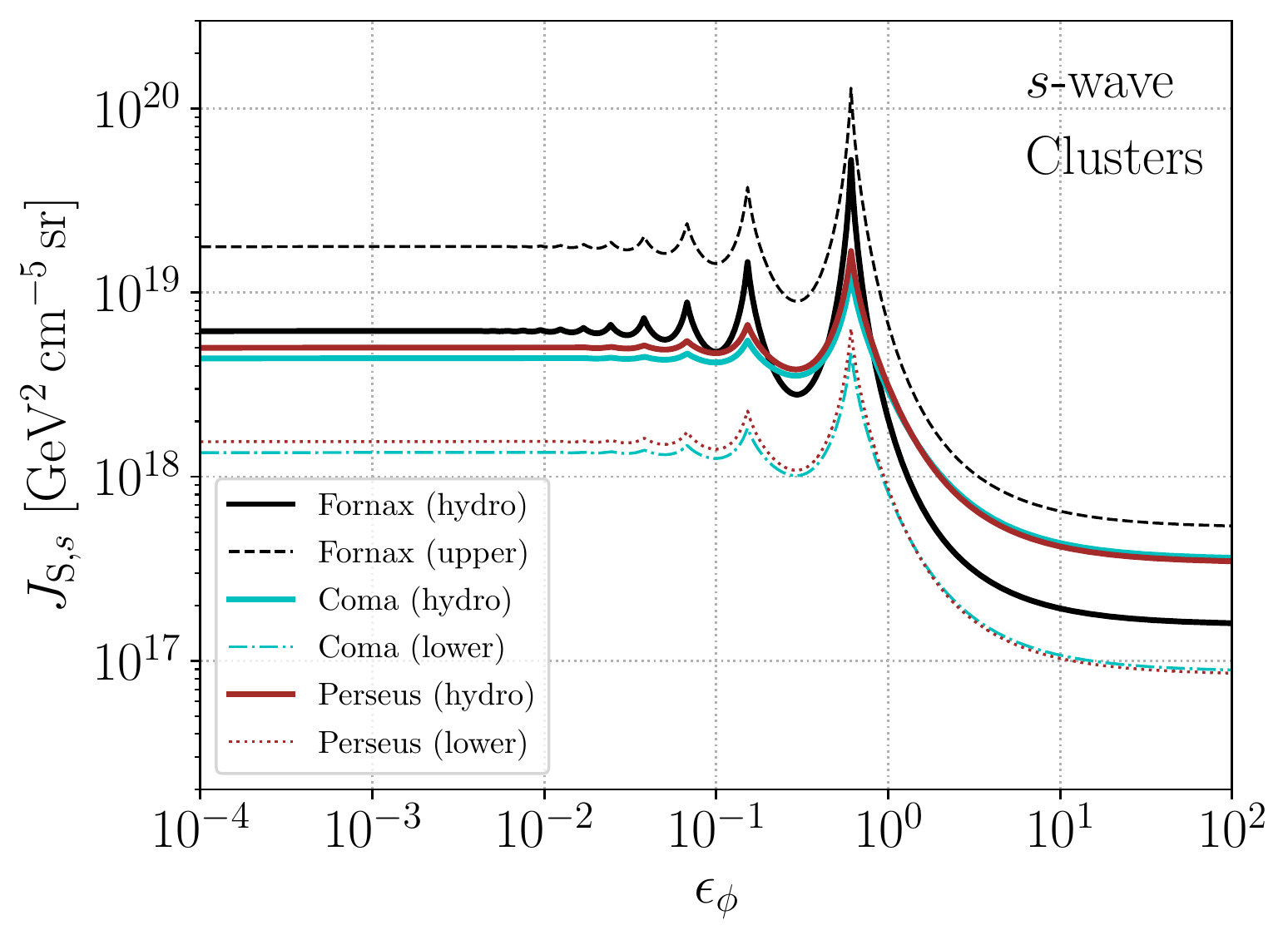} \hfill
\includegraphics[width=0.485\linewidth]{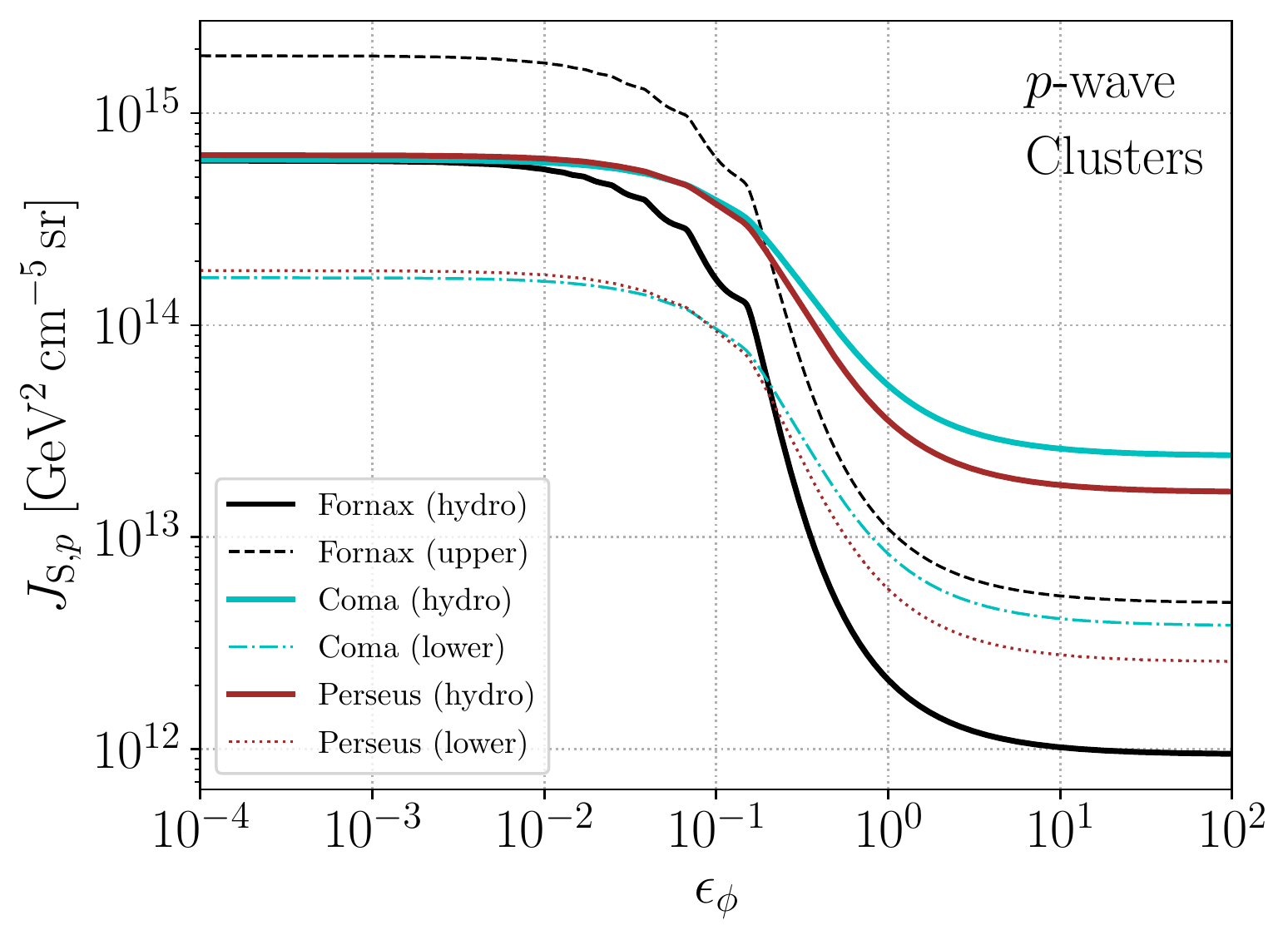}
\caption{Generalised $J$-factors $J_{\rm S}$ for the host halos of the selected targets as a function of $\epsilon_{\phi}$, for $s$-wave (left panels) and $p$-wave (right panels) annihilations; all calculations rely on the Eddington (isotropic) PSDF. We recall that $\theta_{\rm int}=0.5^\circ$ for dSphs and dIrrs, and $\theta_{\rm int}=R_{\rm 200}/D$ for clusters; {\bf Top panels:} Draco (red), Sculptor (green) and Reticulum II (violet) dSphs, where solid colored lines and shaded/hatched bands represent the median and 68\% confidence regions computed from 1000 samples of the DM profile parameters (see \citesec{ssec:dSphs_mass_modelling}).
{\bf Middle panels:} IC10 (gray), NGC6822 (purple), and WLM (golden) dIrrs, for Burkert (solid lines) and NFW (dashed) DM density profiles (see \citesec{ssec:dIrrs_mass_modelling}).
{\bf Bottom panels:} Coma (cyan), Fornax (black) and Perseus (brown) galaxy clusters, for `hydro' (solid lines) and `upper' (dashed) DM density profiles (see \citesec{ssec:galaxy-clusters-smooth}).
}
\label{fig:JS_smooth_error_bars}
\end{figure}

\paragraph{$s$-wave annihilation (left panels of \citefig{fig:JS_smooth_error_bars}).} 

\begin{itemize}
    \item for $\epsilon_{\phi} \lesssim \epsilon_{\phi}^{\star}\, \ll 1$, the Sommerfeld enhancement is in the Coulomb regime, i.e., $\overline{{\cal S}}\propto 1/\epsilon_{v}$: as a result, $J_{\mathrm{S}}$ does not depend on $\epsilon_{\phi}$ ---but is roughly proportional to $\bar{v}^{-1}$--- and displays a plateau below $\epsilon_{\phi}^{\star}$ (left-hand side of the curves);
    
    \item for $\epsilon_{\phi} \gg 1$, there is no enhancement, and $J_{\mathrm{S}}$ boils down to the standard $J$-factor (right-hand side of the curves);
    
    \item for $\epsilon_{\phi}^{\star} \lesssim \epsilon_{\phi} \lesssim 1$, this is the resonant (saturation) regime, where the behaviour depends whether $\epsilon_{\phi}$ falls at, or between, resonances (between the two plateaus in the curves): at resonance, $J_{\mathrm{S}}$ is roughly proportional to $\bar{v}^{-2}$ and follows a $1/n^2$ power law (where $n$ is the integer defining each resonance, see \citesec{ssec:sommerfeld}). Between resonances, $J_{\mathrm{S}} \propto 1/\epsilon_{\phi}$ and does not depend on the velocity.
\end{itemize}

\paragraph{$p$-wave annihilation (right panels in \citefig{fig:JS_smooth_error_bars}).} 

\begin{itemize}
    \item for $\epsilon_{\phi} \lesssim \epsilon_{\phi}^{\star} (\ll 1)$, $\overline{{\cal S}}\propto 1/\epsilon_{v}$ similar to the $s$-wave case and is also independent of $\epsilon_{\phi}$, so $J_{\mathrm{S},p}$ also features a plateau in this regime. The $p$-wave plateau is lower by a factor $\alpha_{\rm D}^{2}$ compared to the $s$-wave case (left-hand side of the curves);
    
    \item for $\epsilon_{\phi} \gg 1$, as for the $s$-wave, there is no enhancement and $J_{\mathrm{S}}$ boils down to the standard $p$-wave $J$ (right-hand side of the curves);

    \item for $\epsilon_{\phi}^{\star} \lesssim \epsilon_{\phi} \lesssim 1$, this is also the resonant (saturation) regime: at resonances, $\overline{\cal S}$ is independent of $\epsilon_{v}$ and $J_{\mathrm{S}} \propto 1/\epsilon_{\phi}^{3}$; however, between resonances, the Sommerfeld factor is independent of $\epsilon_{v}$ so $J_{\mathrm{S}}$ is shaped by the $p$-wave velocity-dependence $\left\langle v_{\rm rel}^{2} \right\rangle \sim \bar{v}^{2}$. Because of this dependence,  $p$-wave annihilation resonances are more clear-cut for objects with a low characteristic velocity, like dSphs (top right panel), compared to dIrrs and galaxy clusters (middle and bottom right panels).
    
\end{itemize}

\subsection{Uncertainties from the mass modelling}
\label{ssec:host_uncertainties}

Systematic errors on the parameters describing the smooth DM density profiles (see \citesec{sec:mass_modelling}) translate into systematic uncertainties on the generalised $J$-factors, which we briefly discuss quantitatively in the following.

\paragraph{DSphs (top panels of \citefig{fig:JS_smooth_error_bars}).} As described in \citesec{ssec:dSphs_mass_modelling}, for dSphs we compute the uncertainty on the generalised $J$-factors from the posterior distribution on $J_{\rm S}$ obtained from the kinematic analysis of~\cite{BonnivardEtAl2015b}. For $s$-wave annihilation and in the absence of any Sommerfeld enhancement, the uncertainty on the reconstruction of the DM profile parameters leads to factors of a few for classical dSphs (Draco and Sculptor), and a factor $\sim 20$ for the Reticulum II ultra-faint dSph (top-left panel of \citefig{fig:JS_smooth_error_bars}). This is the same in the saturation regime, off resonance, where the Sommerfeld factor does not depend on the velocity. Yet, in the Coulomb regime, and at resonance peaks in the saturation regime, the additional velocity dependence goes in the opposite direction with respect to $\rho^{2}$. For instance, for a given value of the scale radius $r_{-2}$, a larger value of $\rho_{-2}$ gives larger $\rho^{2}(r)$ but at the same time a larger typical velocity which enters the Sommerfeld factor through $\bar{v}^{-1}$ or $\bar{v}^{-2}$, leading to a reduction of the generalised $J$-factors. This leads to 68\% uncertainty bands that are typically smaller in the Coulomb regime --- \eg,~less than an order of magnitude for Reticulum II --- than in the no-Sommerfeld case. 

For $p$-wave annihilation (top-right panel of \citefig{fig:JS_smooth_error_bars}), the uncertainty band spans about two orders of magnitude for Draco in the no-Sommerfeld regime, but `only' one for Sculptor. For Reticulum II, the uncertainty reaches almost four orders of magnitude, owing to the loose kinematic constraints that affect both $\rho^{2}$ and $\left\langle v_{\rm rel}^{2} \right\rangle$. In the Coulomb regime, the uncertainty on $\rho^{2}$ is again balanced by the $\bar{v}^{-1}$ dependence, leading to small 68\% bands for all dSphs in our sample.

\paragraph{DIrrs (middle panels of \citefig{fig:JS_smooth_error_bars}).} For these objects, we bracket the systematic error on the generalised $J$-factors by considering the Burkert and NFW mass models obtained from fitting rotation curve data (see \citesec{ssec:dIrrs_mass_modelling}). This error is encoded in the ratio ${\cal R}_{J_{\rm S}}(\epsilon_{\phi})=J_{{\rm S}}^{\rm (Burkert)}/J_{{\rm S}}^{\rm (NFW)}$, which differs for the $s$-wave and $p$-wave cases. First, for NGC6822 and WLM, the two DM profiles are almost degenerate for the rotation curve fits, so that all associated ratios are close to one (for all $\epsilon_{\phi}$ values). However, for IC10, the best fit using the NFW profile differs more appreciably from the Burkert one: (i) in the Coulomb regime (small $\epsilon_{\phi}$ values), ${\cal R}_{J_{\rm S}}\sim 3$ for both the $s$- and $p$-wave cases; (ii) in the saturation regime off-resonance (intermediate $\epsilon_{\phi}$), and in the standard regime with no enhancement (large $\epsilon_{\phi}$), ${\cal R}_{J_{\rm S}}\sim 5$ for $s$-wave while ${\cal R}_{J_{\rm S}}\sim 10$ for $p$-wave; (iii) on resonance, ${\cal R}_{J_{\rm S}}\sim 1$ for $s$-wave while ${\cal R}_{J_{\rm S}}\sim 5$ for $p$-wave. It should be noted that for $p$-wave on resonances, $J_{{\rm S}}$ does not depend on the velocity, thus the difference between Burkert and NFW is the same as for $s$-wave without Sommerfeld enhancement.

\paragraph{Galaxy clusters (bottom panels of \citefig{fig:JS_smooth_error_bars}).} In this case, the uncertainties on the generalised $J$-factors are related to uncertainties on the derived $X$-ray masses and the scatter on the mass-concentration relation. This allows to define a lower and upper bound on the modelling of the DM density profiles (see \citesec{ssec:galaxy-clusters-smooth}). The ratio ${\cal R}_{J_{\rm S}}(\epsilon_{\phi})$ of these two bounds is $J_{{\rm S}}^{\rm (hydro)}/J_{{\rm S}}^{\rm (lower)}$ for Coma and Perseus, and $J_{{\rm S}}^{\rm (upper)}/J_{{\rm S}}^{\rm (hydro)}$ for Fornax. We find, in the $s$-wave case, that ${\cal R}_{J_{\rm S}}\sim$~3-4 in all regimes for all the clusters in our sample. We have otherwise in the $p$-wave case ${\cal R}_{J_{\rm S}}\sim$ 3 in the Coulomb regime (small $\epsilon_{\phi}$ values), and 5-6 in both the inter-resonance saturation regime (intermediate $\epsilon_{\phi}$ values) and the pure $p$-wave case with no enhancement (large $\epsilon_{\phi}$ values).

\subsection{Impact of uncertainties on the ranking of targets}
\label{ssec:host_ranking}

From the above discussion, we conclude that systematic errors --- that stem from the data-driven modelling of the smooth DM density profile --- have a strong impact on the generalised $J$-factors. As such, they can affect the hierarchy of targets according to their potential for $\gamma$-ray DM searches. In contrast, as discussed in \citeapp{app:phase_space_modelling}, the ${\cal O}(10\%)$ uncertainty on the PSDF itself --- in particular how the anisotropy of the velocity distribution is accounted for --- has little impact on the generalised $J$-factors, and does not affect the ranking of targets.

\begin{figure}[t!]
\centering
\includegraphics[width=0.49\linewidth]{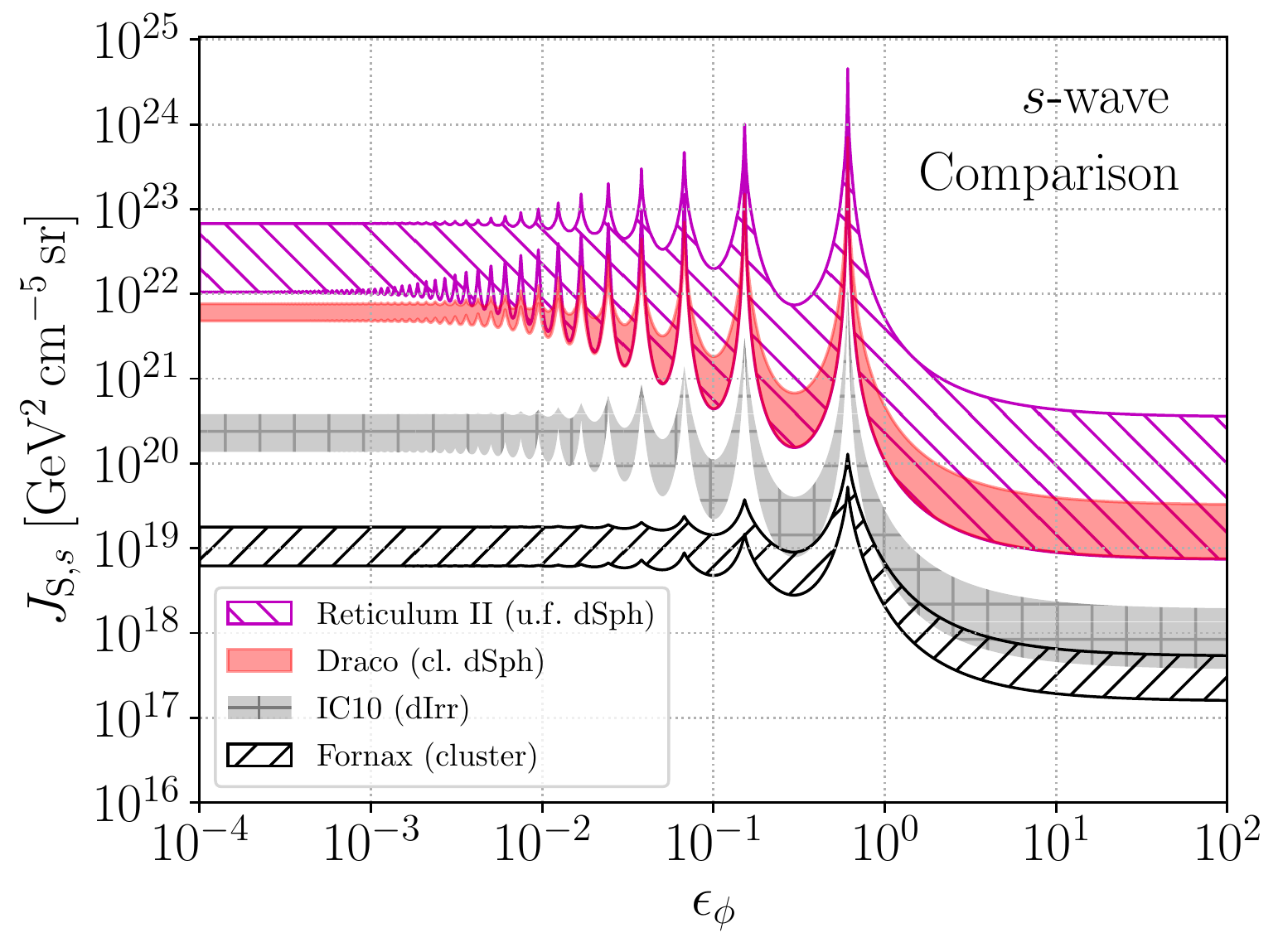} \hfill
\includegraphics[width=0.49\linewidth]{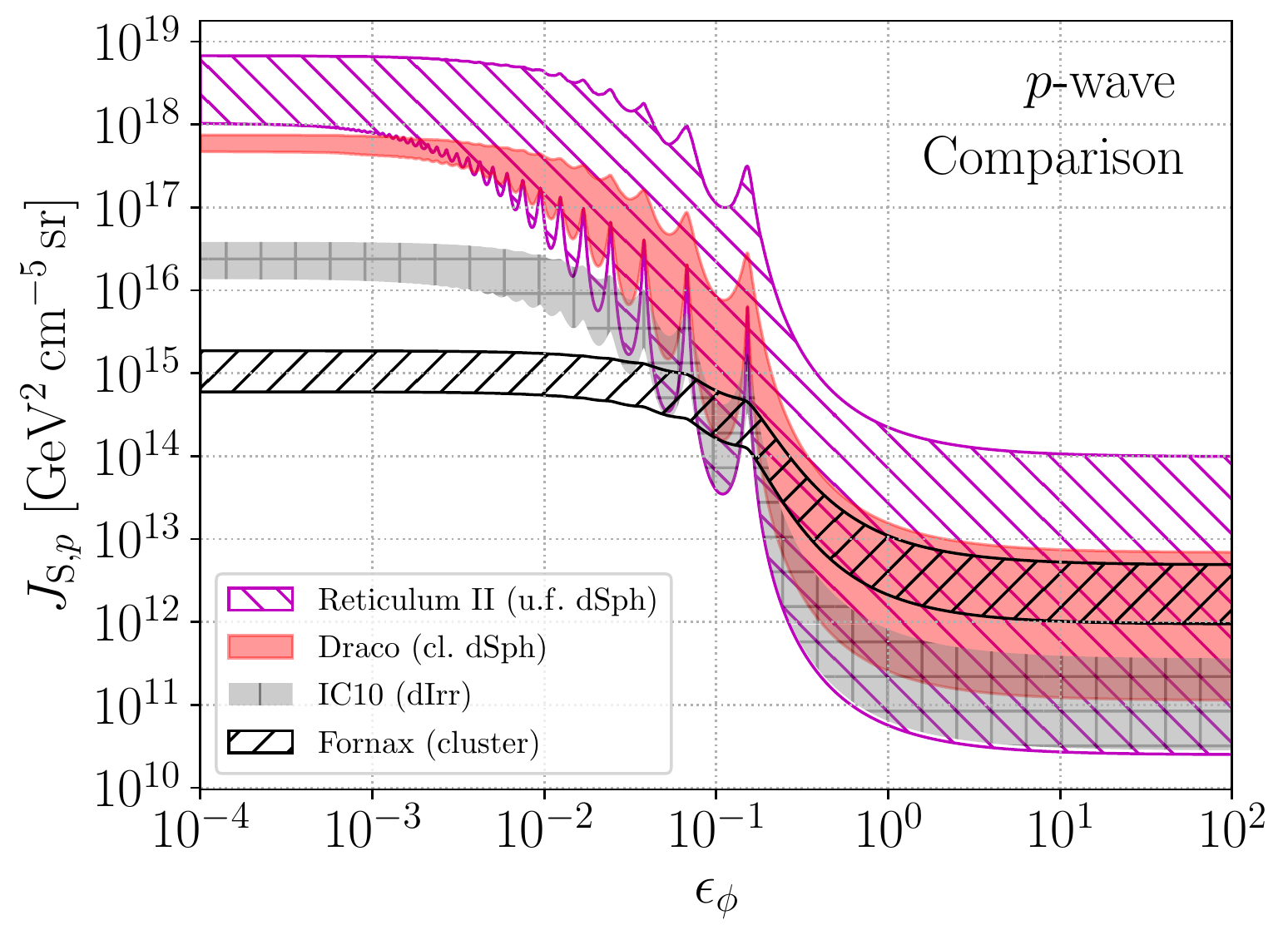} 
\caption{
Comparison of generalised $J$-factors for the smooth (host) halo of selected targets as a function of $\epsilon_{\phi}$ for $s$-wave (left panel) and $p$-wave (right panel) annihilations. We reproduce here some of the curves shown in \citefig{fig:JS_smooth_error_bars}, highlighting as shaded/hatched areas the uncertainties estimated from mass modelling uncertainties. The selected objects are an ultra-faint dSph (Reticulum~II), a classical dSph (Draco), a dIrr (IC10), and a galaxy cluster (Fornax). We recall that $\theta_{\rm int}=0.5^\circ$ for dSphs and dIrrs, and $\theta_{\rm int}=R_{\rm 200}/D$ for clusters.
}
\label{fig:comparison_host_halo}
\end{figure}

In \citefig{fig:comparison_host_halo}, we compare the $J$-factors (as a function of $\epsilon_{\phi}$) for a selection of representative objects among the target classes considered in this work, namely Reticulum II (ultra-faint dSph), Draco (classical dSph), IC10 (dIrr), and Fornax (galaxy cluster). To ease the comparison, we highlight our estimated uncertainties as shaded/hatched areas, although we remind that the different bands neither have the same origin nor the same statistical meaning (see discussion in \citesec{ssec:host_uncertainties}).
For $s$-wave annihilation (left panel), the uncertainty bands overlap in the regime with no Sommerfeld enhancement (large $\epsilon_{\phi}$ values) for Fornax and IC10, while this overlap disappears and is replaced by a gap for decreasing $\epsilon_{\phi}$; there is for instance almost a factor 10 difference between the  Fornax lower edge (black-hatched band) and IC10 upper edge (gray-hatched band) in the Coulomb regime (small $\epsilon_{\phi}$ values). The situation is qualitatively similar for Draco and Reticulum II. 
For $p$-wave annihilation, the uncertainty bands overlap for the four representative targets in the regime of no enhancement (large $\epsilon_{\phi}$ values), whereas a clear hierarchy also appears when Sommerfeld enhancement becomes important, especially in the Coulomb regime (small $\epsilon_{\phi}$ values) due to the $1/v$ dependence. 

As a conclusion from this section, which focused only on the signal from the host DM halo, we see that for $s$-wave annihilation, dSphs represent (in all regimes) the most promising targets for $\gamma$-ray searches in terms of generalised $J$-factors, even accounting for modelling uncertainties. Yet, the situation is less clear-cut for $p$-wave annihilation in the regime of no enhancement (large $\epsilon_{\phi}$ values). Indeed, in the latter case, within the uncertainties, some galaxy clusters can become the best targets. Nevertheless, as already highlighted in the literature (mostly for the standard $J$-factor calculations), accounting for DM substructures in all these different targets may change these conclusions. We discuss and detail in the next section how DM substructures are expected to boost the annihilation signal and impact the computation of the generalised $J$-factors.

\section{Generalised $J$-factors with substructure boost}
\label{sec:subhalo_boost_and_classification}

A fraction of the DM in halos is in the form of subhalos, which can boost the annihilation signal (compared to the case in which all the DM mass is smoothly distributed within the main halo). While the impact of these substructure boosts has been discussed extensively in the literature for the `classical' $J$-factors (see, e.g.~\cite{PieriEtAl2011,NezriEtAl2012,Sanchez-CondeEtAl2014,BonnivardEtAl2016,MolineEtAl2017,StrefEtAl2017}), they have been discussed with lesser details in the context of generalised $J$-factors \cite{Arkani-HamedEtAl2009,LattanziAndSilk2009,Bovy2009,KuhlenMadauSilk2009,KamionkowskiEtAl2010}.

The results derived in this section rely on the general formalism and methodology presented in \citesec{ssec:subhalo_boost}, and our calculations are based on up-to-date models for both the properties of the subhalo population  and the velocity distribution in each subhalo (determined by a phase space); the numerical calculations in this section (for subhalos) have also been cross-checked and validated with analytical approximations (see the companion paper, \cite{CompanionPaper}). First, we discuss the boost factors obtained for our representative targets and highlight the differences observed between the $s$- and $p$-wave cases (\citesec{ssec:JS_boost_factors}). We then show the full calculation of the generalised $J$-factors for all our targets, and rank them according to their expected signals, also depending on the regime considered for $s$- or $p$- wave annihilations (\citesec{ssec:JSwboost_target_comparison}). We finally  show how these boosted signals compare to the `foreground' DM annihilation signal coming from the smooth DM distribution in the MW, and briefly discuss the prospects for $\gamma$-ray searches (\citesec{ssec:JSwboost_prospects}).

\subsection{Impact of subhalos: boost factors for generalised $J$-factors}
\label{ssec:JS_boost_factors}

The generalised boost factor ${\cal B}_\textrm{S}$, calculated for the generalised $J$-factor $J_\textrm{S}$, is given by \begin{equation}
{\cal B}_{\rm S} = \frac{J_{\rm S,tot}}{J_{\rm S,host}}\,.
\label{eq:def_boost_factor}
\end{equation}
In this definition, the denominator $J_{\rm S,host}$ is the generalised $J$-factor for the host halo without substructures (i.e., assuming all the DM to be smoothly distributed), already calculated and presented in \citesec{ssec:general_features_J_factors}. The numerator $J_{\rm S,tot}$ is the sum of the signals from the smooth halo of the host (which is now all the DM not in substructures), the population of subhalos and the cross-annihilation between host and subhalos (the latter is in general negligible compared to the sum of the other two contributions). In this definition, when the fraction of DM into substructures goes to zero, the boost goes to one, i.e., the overall signal is not boosted.

\begin{figure}[t!]
\centering
\includegraphics[width=0.49\linewidth]{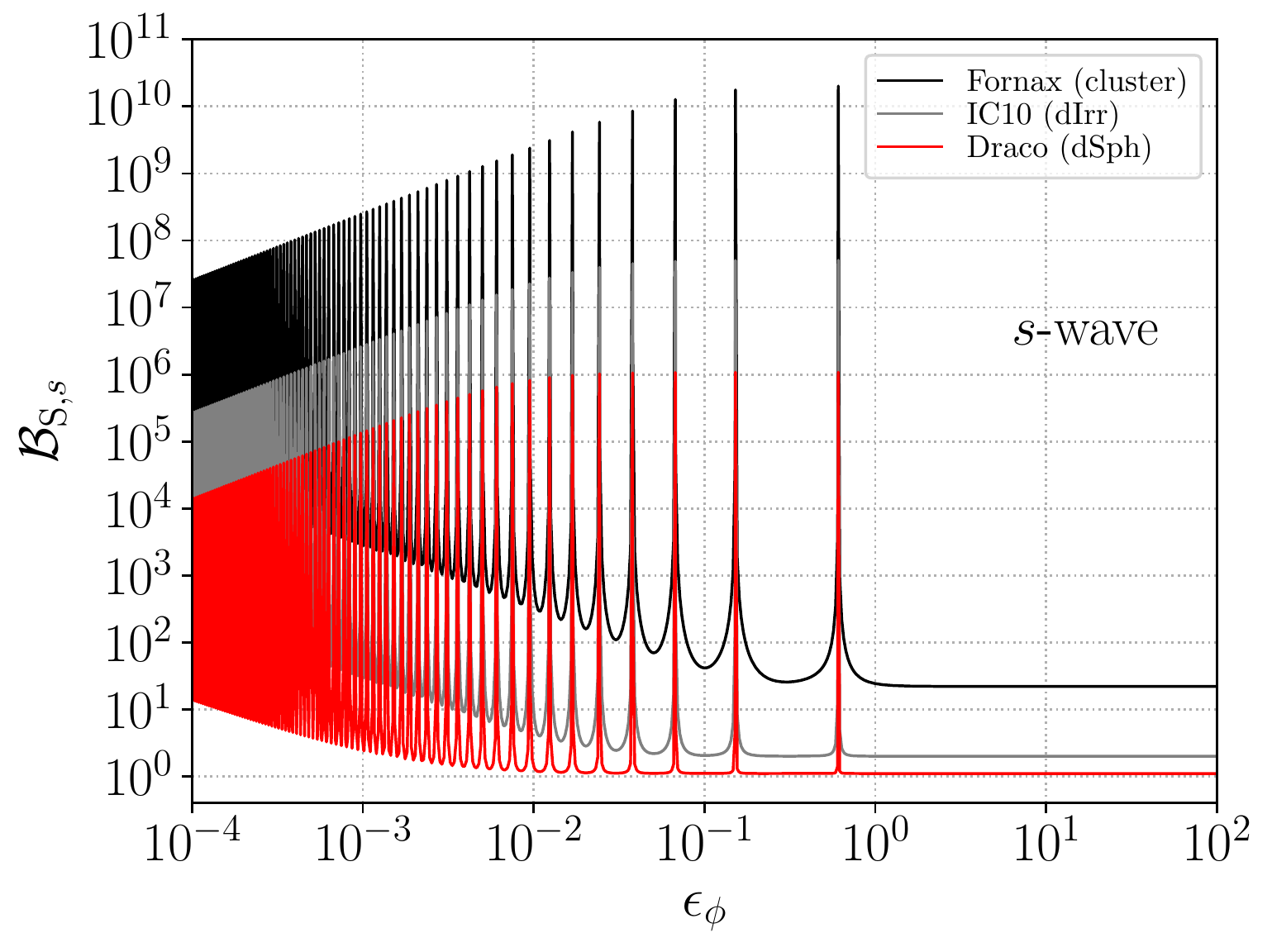} \hfill \includegraphics[width=0.49\linewidth]{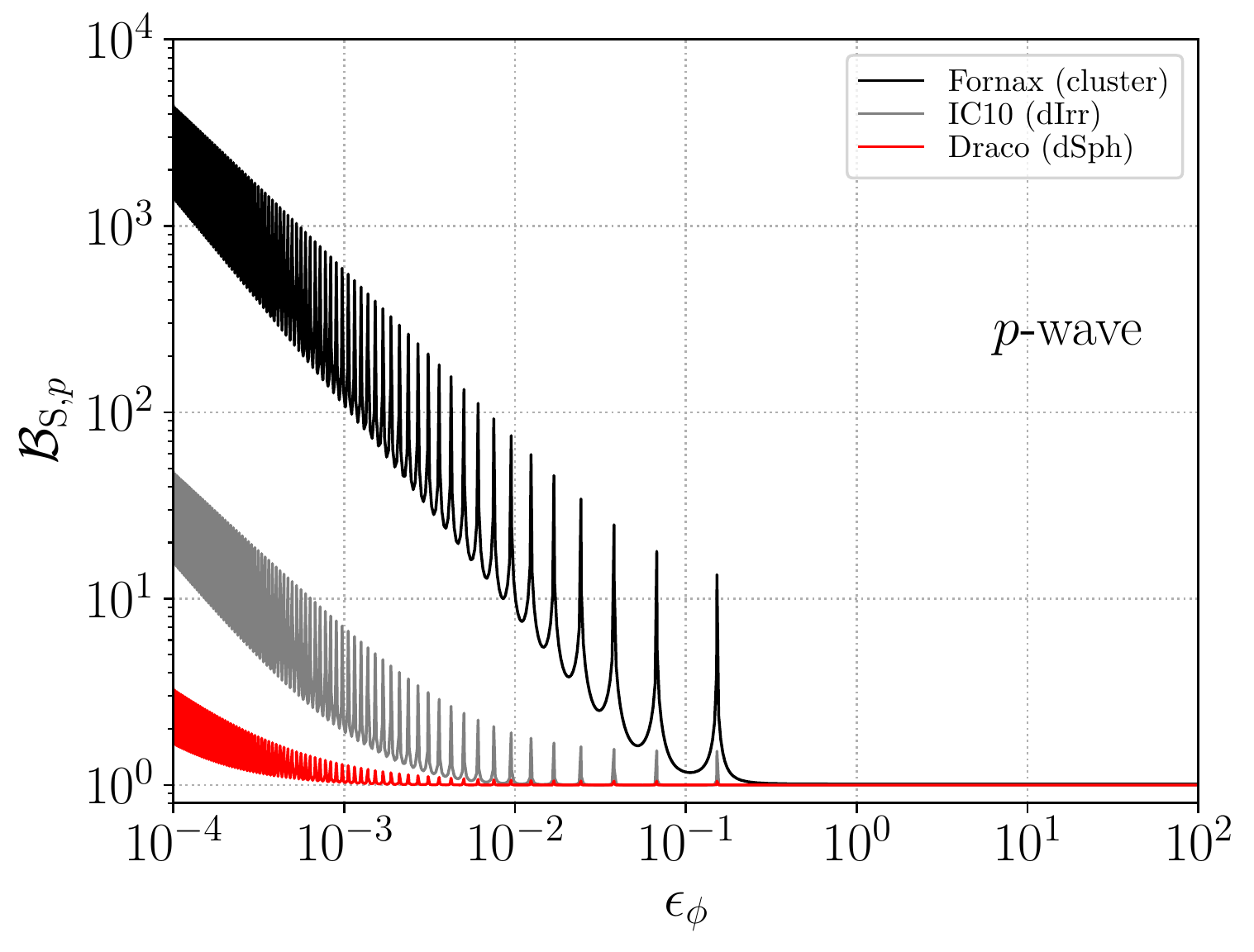}
\caption{
 Generalised boost factors ${\cal B}_{\rm S}$ (see Eq.~\ref{eq:def_boost_factor}) as a function of $\epsilon_\phi$ for $s$-wave (left panel) and $p$-wave (right panel) annihilation. We illustrate the boosts obtained for each family of targets considered, i.e., galaxy clusters (Fornax, black lines), dIrrs (IC10, gray lines), and dSphs (Draco, red lines). The more massive the object, the more boosted the signal (see text for discussion). We recall that $\theta_{\rm int}=0.5^\circ$ for dSphs and dIrrs, and $\theta_{\rm int}=R_{\rm 200}/D$ for clusters.}
\label{fig:boost_swave_comparison}
\end{figure}

We show in \citefig{fig:boost_swave_comparison} the generalised boost factors ${\cal B}_{\rm S}$ as a function of $\epsilon_\phi$, for three representative objects from our three families of targets. 
Several striking features are observed. 
First, contrarily to the smooth host halo case where resonances disappear below $\epsilon_\phi\lesssim 10^{-2}$ (see, e.g., \citefig{fig:comparison_host_halo}), resonances are present down to much smaller $\epsilon_\phi$ values here. This is because subhalos are less massive with smaller velocity dispersion, hence a smaller $\epsilon_\phi^\star$ (see Eq.~\ref{eq:def_epsphi_star}) below which the Coulomb regime is reached (compared to their host halo counterpart). 
Second, we see that a larger host is more boosted. This is a well-known feature of $s$-wave annihilation without Sommerfeld enhancement due to each decade in subhalo mass contributing to the annihilation at roughly the same level. When the Sommerfeld effect is included, this dependence on the host halo mass is preserved although the scaling is slightly modified. Moreover, it now extends to the $p$-wave case as well, and the scaling with the host mass is identical for both $s$-wave and $p$-wave.
Third, a different scaling with $\epsilon_\phi$ is observed in the $s$-wave and $p$-wave case. This difference can be explained by considering which subhalos contribute most to the annihilation. We find that the $s$-wave signal is dominated by subhalos near the free-streaming cutoff, while the $p$-wave signal is dominated by subhalos near the mass scale which sets the transition between the Coulomb and saturation/resonant regime, which depends on $\epsilon_\phi$. Details and scaling relations are provided in the companion paper \cite{CompanionPaper}.
We stress that the resonances do not appear at arbitrarily low $\epsilon_\phi$, because subhalos cannot form below the free-streaming scale. We fixed this scale to $m_{\rm min}=10^{-6}\,\rm M_\odot$, which translates into a value $\epsilon_\phi^\star\sim 10^{-7}$ below which all subhalos, and therefore all the DM in the object, are in the Coulomb regime. We chose to limit the $x$-axis to $\epsilon_\phi=10^{-4}$ however, because lower values have little motivations from the model-building point of view. 
The sensitivity of the $s$-wave annihilation to this minimum low-mass scale also explains why the $s$-wave boost is generically much larger than the $p$-wave boost. Indeed, for $s$-wave processes, the baseline and the resonant peaks have their amplitudes fixed by $\epsilon_\phi^{-/+1}\,\mmin^{-\widetilde{\alpha}_s}$, respectively,\footnote{Formally, as explained in the companion paper \cite{CompanionPaper}, the amplitudes of resonant peaks saturate at vanishingly small DM velocity, which translates into a universal unitarity cutoff mass $m_{\rm unit}\neq\mmin$, extremely sensitive to the DM fine structure constant, as it scales like $\alpha_{\rm D}^{12}$. For $\alpha_{\rm D}\sim 0.01$, we have $m_{\rm unit}\sim 10^{-3}\Msun>\mmin$, which means that it is actually $m_{\rm unit}$, still a universal parameter related to particle physics, that sets the peaks amplitudes in our calculations. Had we taken $\alpha_{\rm D}\sim 10^{-3}$ instead, then $\mmin$ would have been the peaks maker.} where $\widetilde{\alpha}_s$ is some effective mass index that depends on the subhalo mass index, found positive here ($\sim 0.1$ on the baseline and $\sim 0.8$ on peaks---see \cite{CompanionPaper} for details). This explains why the relative peak amplitude decreases as $\epsilon_\phi^2$ as $\epsilon_\phi$ decreases. In contrast, for $p$-wave processes, only the peaks have their amplitudes that scale like $\epsilon_\phi^{-1}\,m_{\rm min}^{-\widetilde{\alpha}_p}$, with $\widetilde{\alpha}_p\sim 0.1$, while the baseline scales only $\propto \epsilon_\phi^{-1.3}$. In both the $s$- and $p$-wave cases, the overall amplitude of the boost, once the host smooth halo lies in the Coulomb regime, is further modulated by the host halo mass to some positive power, which explains the hierarchy between the different curves. All this allows to understand how changing $m_{\rm min}$ may affect the final results.

To be more quantitative, the boost factors in the different regimes of the Sommerfeld enhancement are as follows: for large $\epsilon_\phi$, we have ${\cal B}_{\rm S,s}\sim 1-20$ going from dSphs to galaxy clusters,\footnote{For $s$-wave, this is the regime where standard $J$-factor calculations are recovered, and the boost values obtained are in line with standard boost factors found in the literature, e.g. \cite{BonnivardEtAl2016,MolineEtAl2017}.} while ${\cal B}_{\rm S,p}=1$ (no boost) for all targets. Moving down towards the saturation regime, and for the dSphs, dIrrs and galaxy clusters, respectively, we have for the first resonance ${\cal B}_{\rm S,s}\sim10^6,~10^8,~10^{10}$ and ${\cal B}_{\rm S,p}\sim1,~2,~20$.

\subsection{Ranking of target classes}
\label{ssec:JSwboost_target_comparison}

Now that we have detailed the behaviour of the generalised boost factors, we can go back to the generalised $J$-factors. 

\begin{figure}[t!]
\centering
\includegraphics[width=0.49\linewidth]{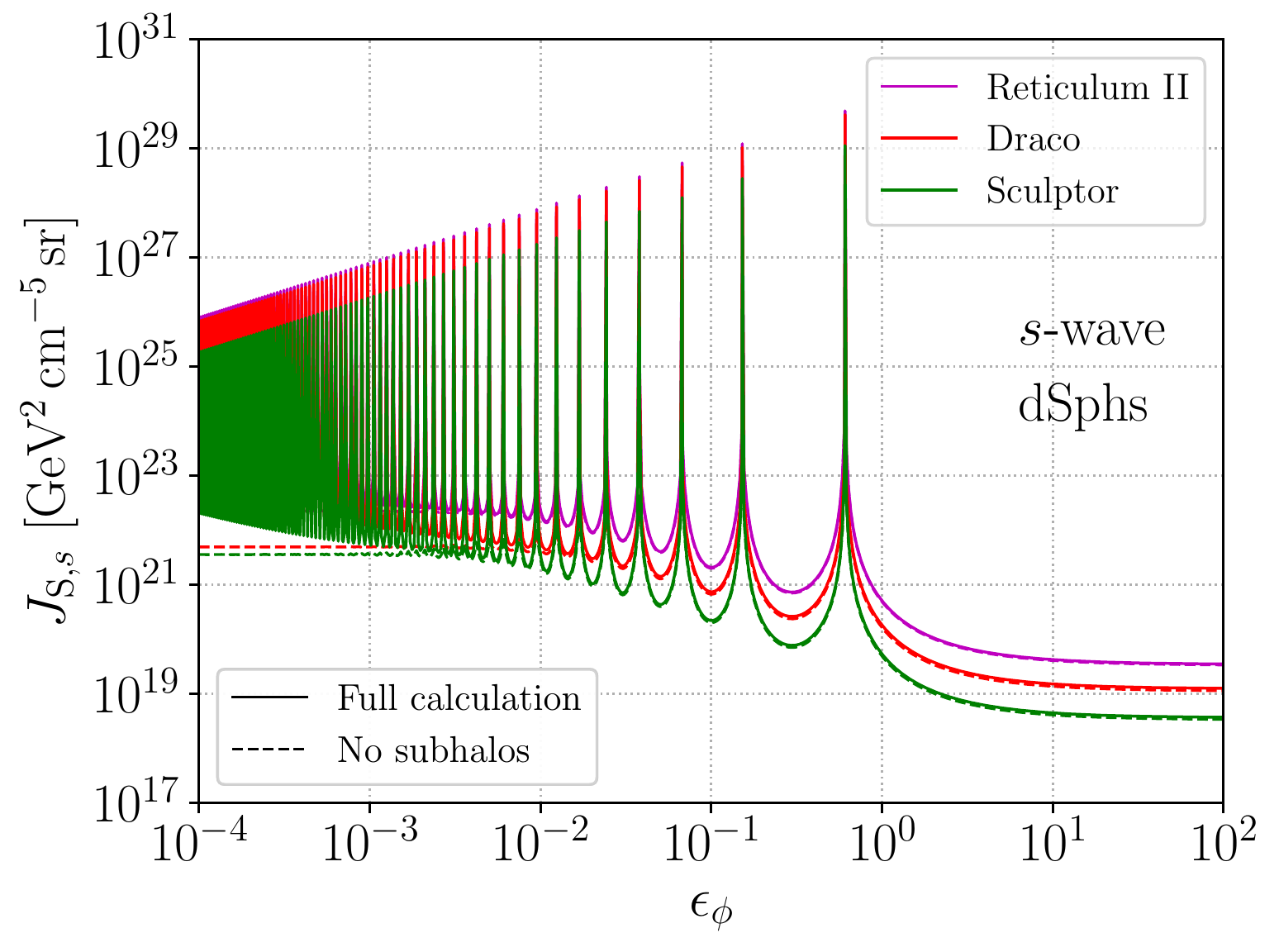} \includegraphics[width=0.49\linewidth]{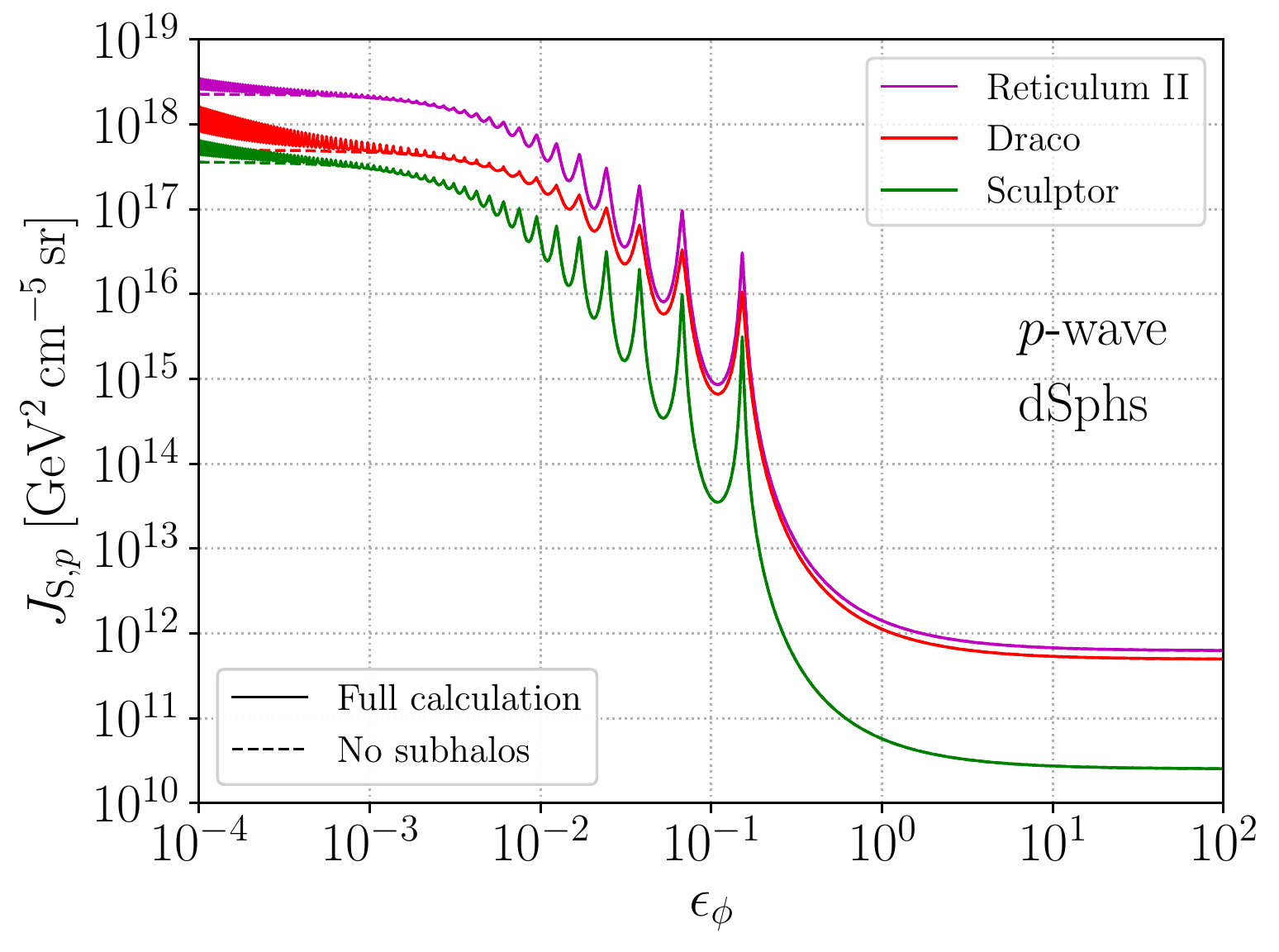} \hfill 
\includegraphics[width=0.49\linewidth]{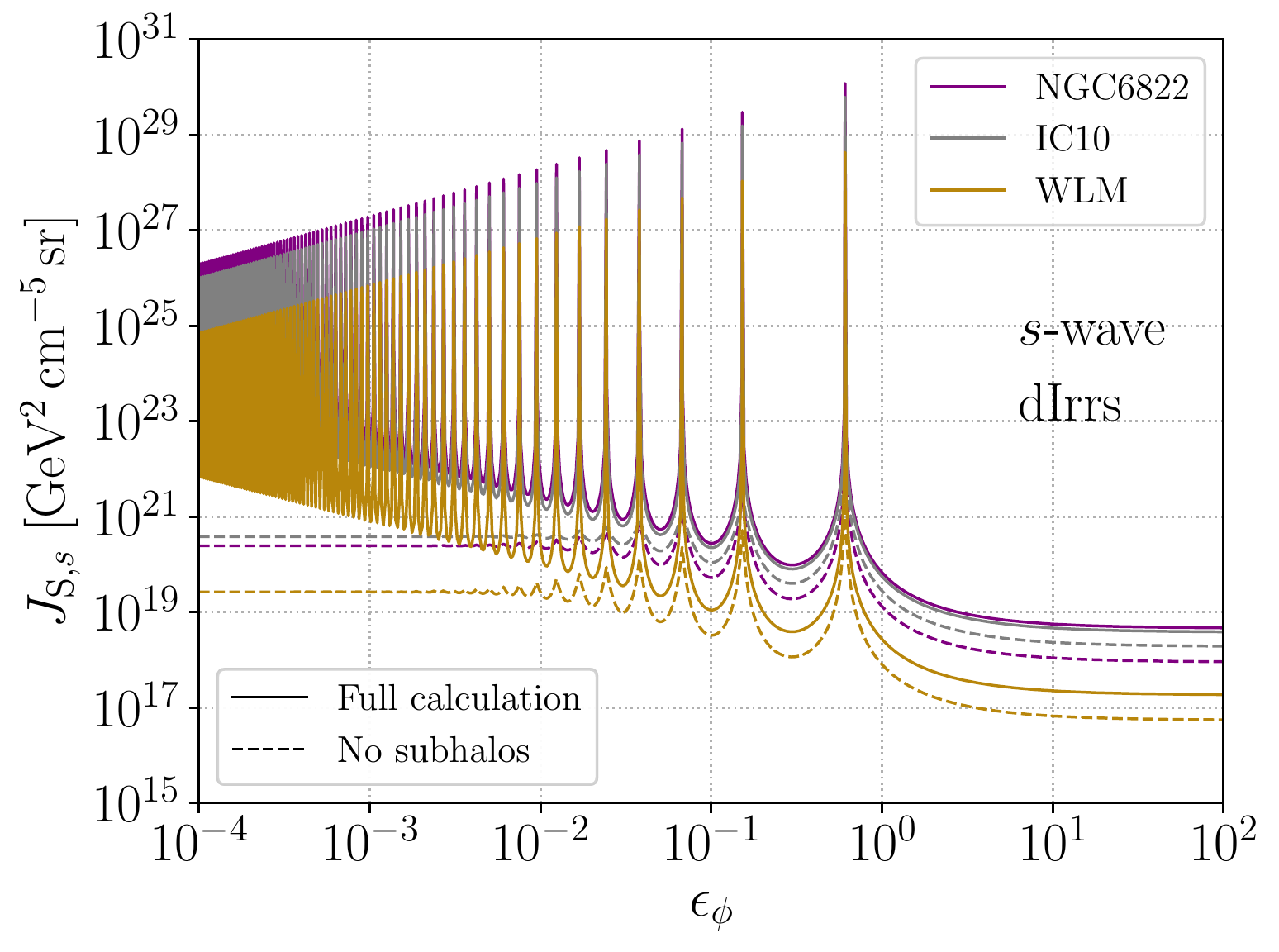} \includegraphics[width=0.49\linewidth]{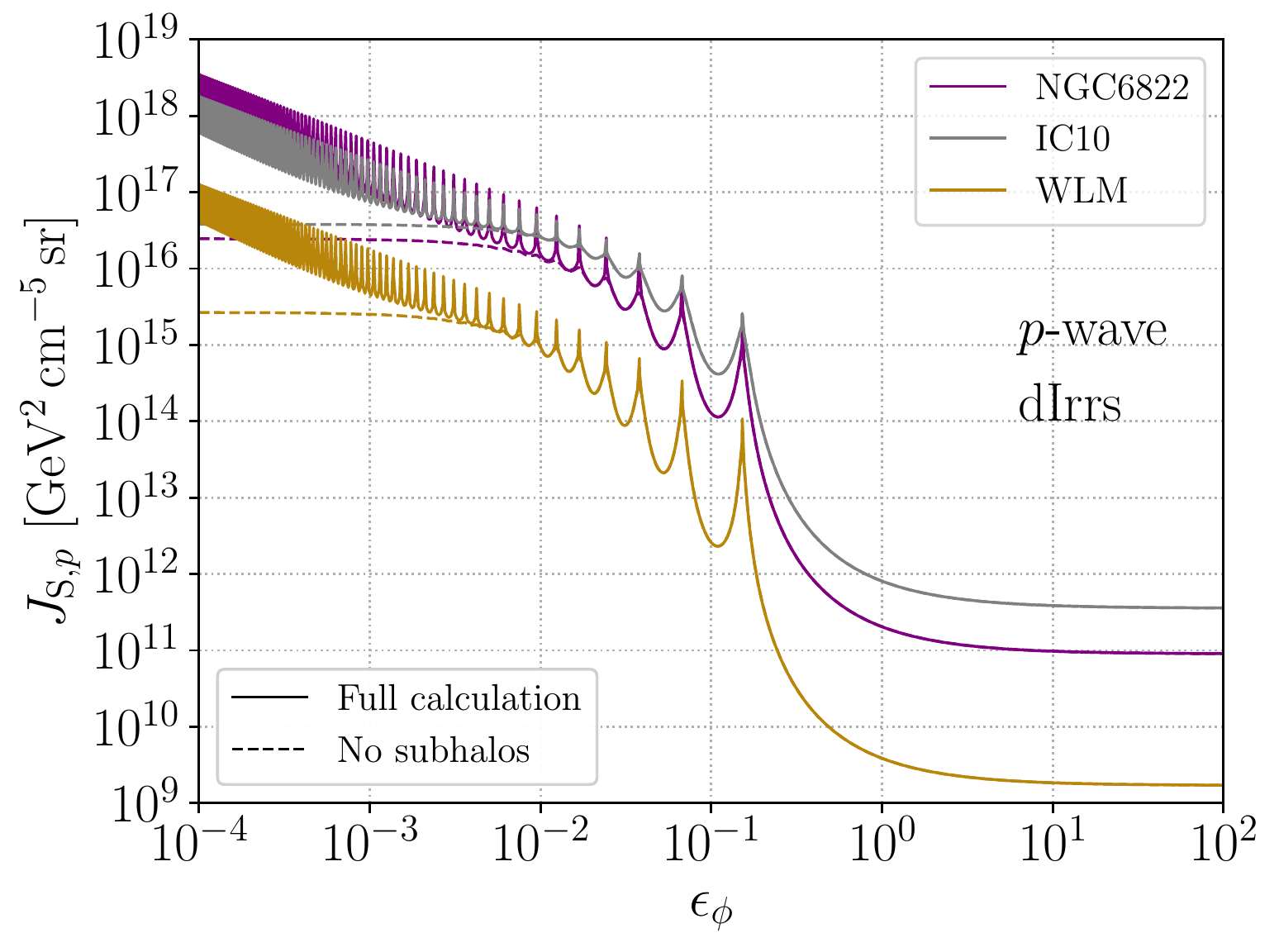} \hfill 
\includegraphics[width=0.49\linewidth]{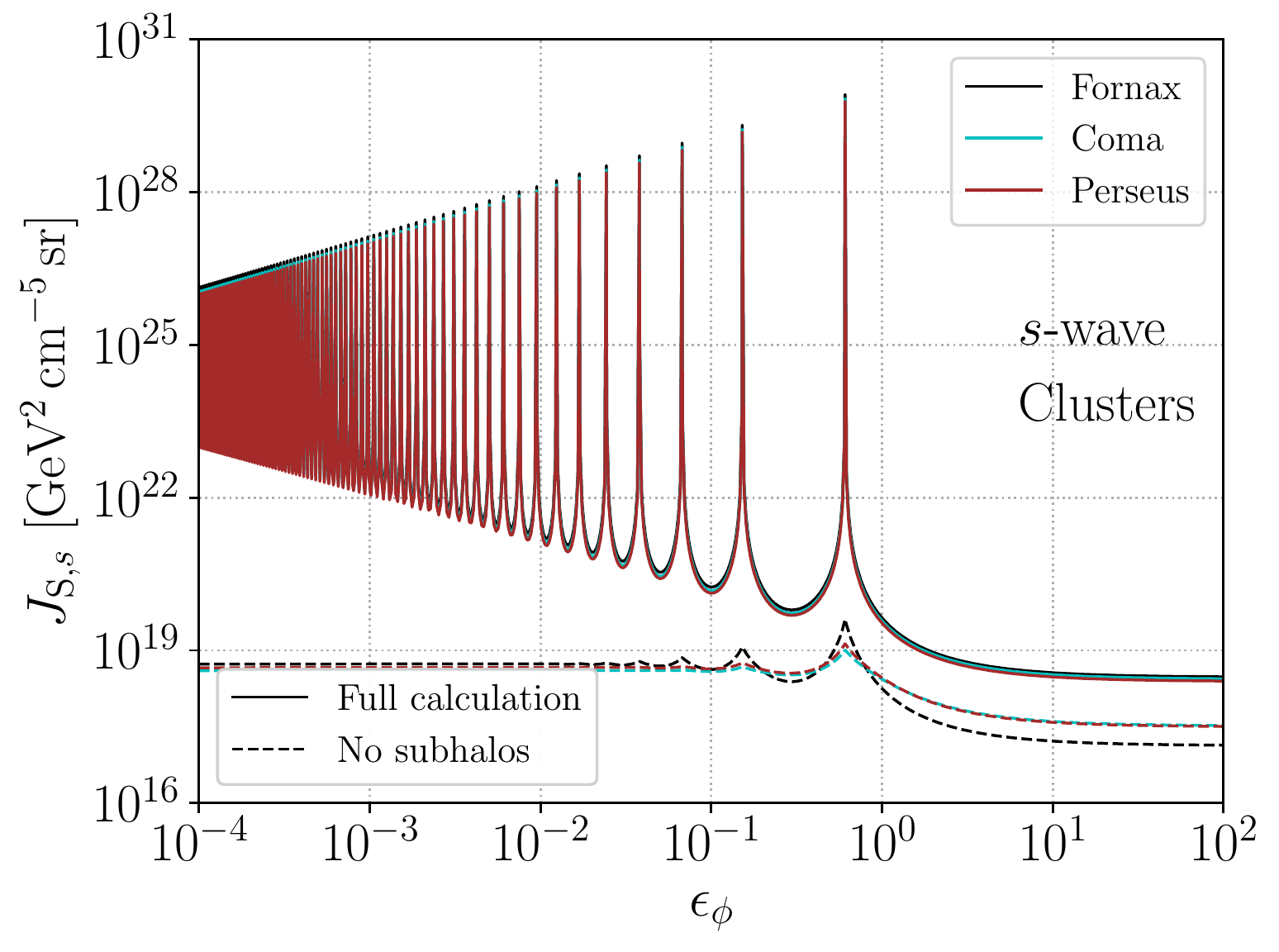}
\includegraphics[width=0.49\linewidth]{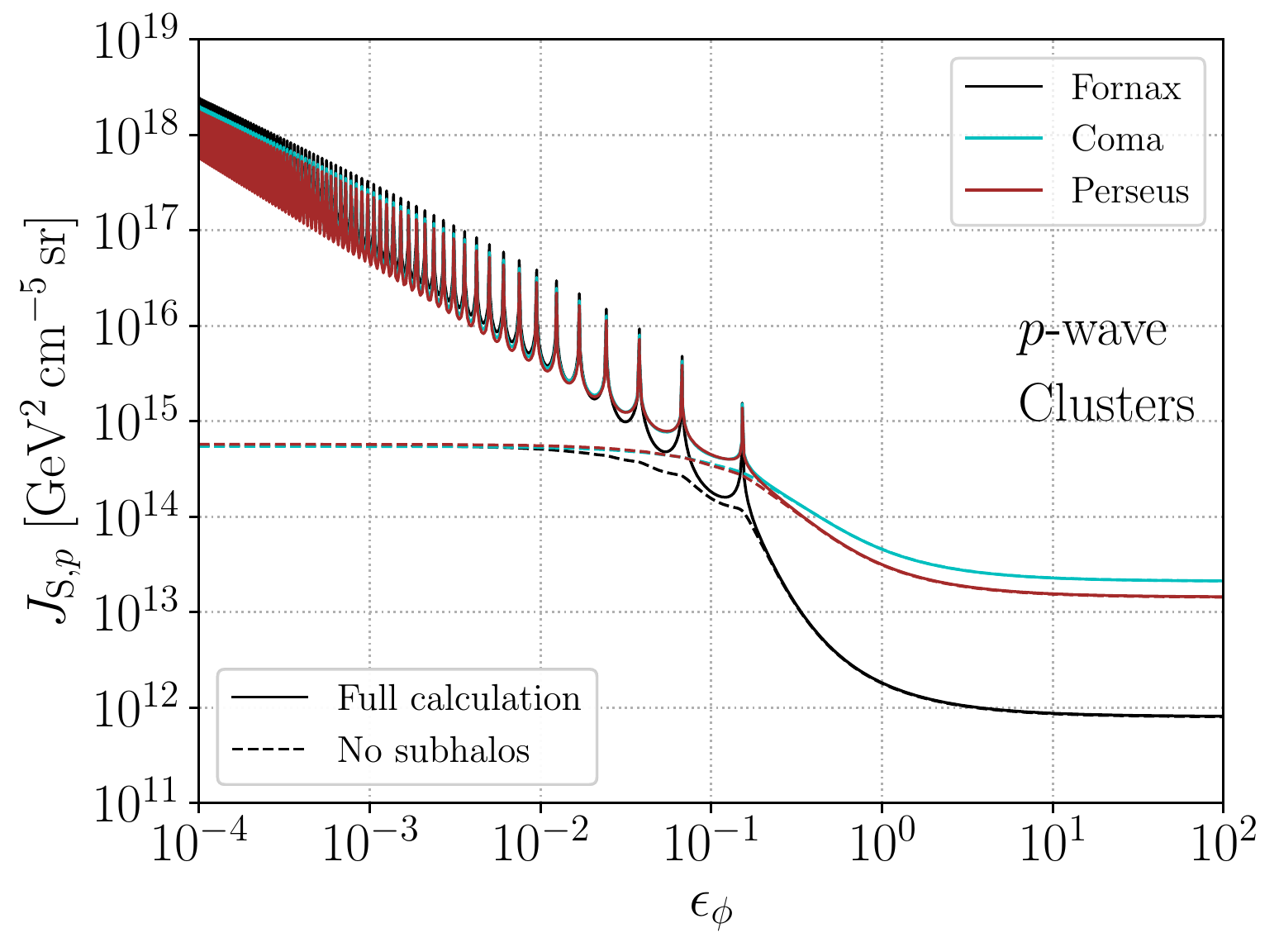}
\caption{Same as \citefig{fig:JS_smooth_error_bars}, i.e., generalised $J$-factors vs $\epsilon_{\phi}$ for our sample of dSphs (top panels), dIrrs (middle panels), and galaxy clusters (bottom panels), for both $s$-wave (left panels) and $p$-wave (right panels) annihilation. Two different calculations are shown: solid lines correspond to the full calculation accounting for the contribution of subhalos (see Sec.~\ref{ssec:JS_boost_factors}), whereas dashed lines (for comparison purpose) correspond to the `no-subhalos' case already shown in \citefig{fig:comparison_host_halo} (we only show our `best' mass modelling here).}
\label{fig:comparison_s+p_wave_subhalos_various_targets}
\end{figure}

We show in \citefig{fig:comparison_s+p_wave_subhalos_various_targets} the $J_{\rm S}$ values as a function of $\epsilon_\phi$ for the full calculation including the boost from substructures (solid lines). To our knowledge, these $J_{\rm S}$ are the most complete and up-to-date estimates for such a variety of targets. For comparison purpose, we also reproduce some of the values shown in \citefig{fig:comparison_host_halo} for the case with no substructures (dashed lines). We note that the ratios between the pairs of solid and dashed lines in each panel are directly the boosts discussed in the previous section; we refer the reader to the details therein rather than repeating the discussion here. For brevity, it is enough to summarise the most salient features of the full calculation (solid lines): (i) $J_{\rm S}$ values in the no-enhancement regime (large $\epsilon_\phi$) reach a plateau for both $s$- and $p$-waves, and these plateaus actually correspond to the minimum value of $J_{\rm S}$ over $\epsilon_\phi$; (ii) the saturation regime at resonances gives the most favourable (and tremendous) signal for $s$-wave annihilation, but this is comparatively only mildly significant for $p$-wave annihilation; (iii) off-resonance and moving down towards the Coulomb regime (small $\epsilon_\phi$ values), the $J_{\rm S}$ factor is increasing for both the $s$- and $p$- waves, but it increases less and converges faster towards a plateau in the former case.

\begin{figure}[t!]
\centering
\includegraphics[width=0.49\linewidth]{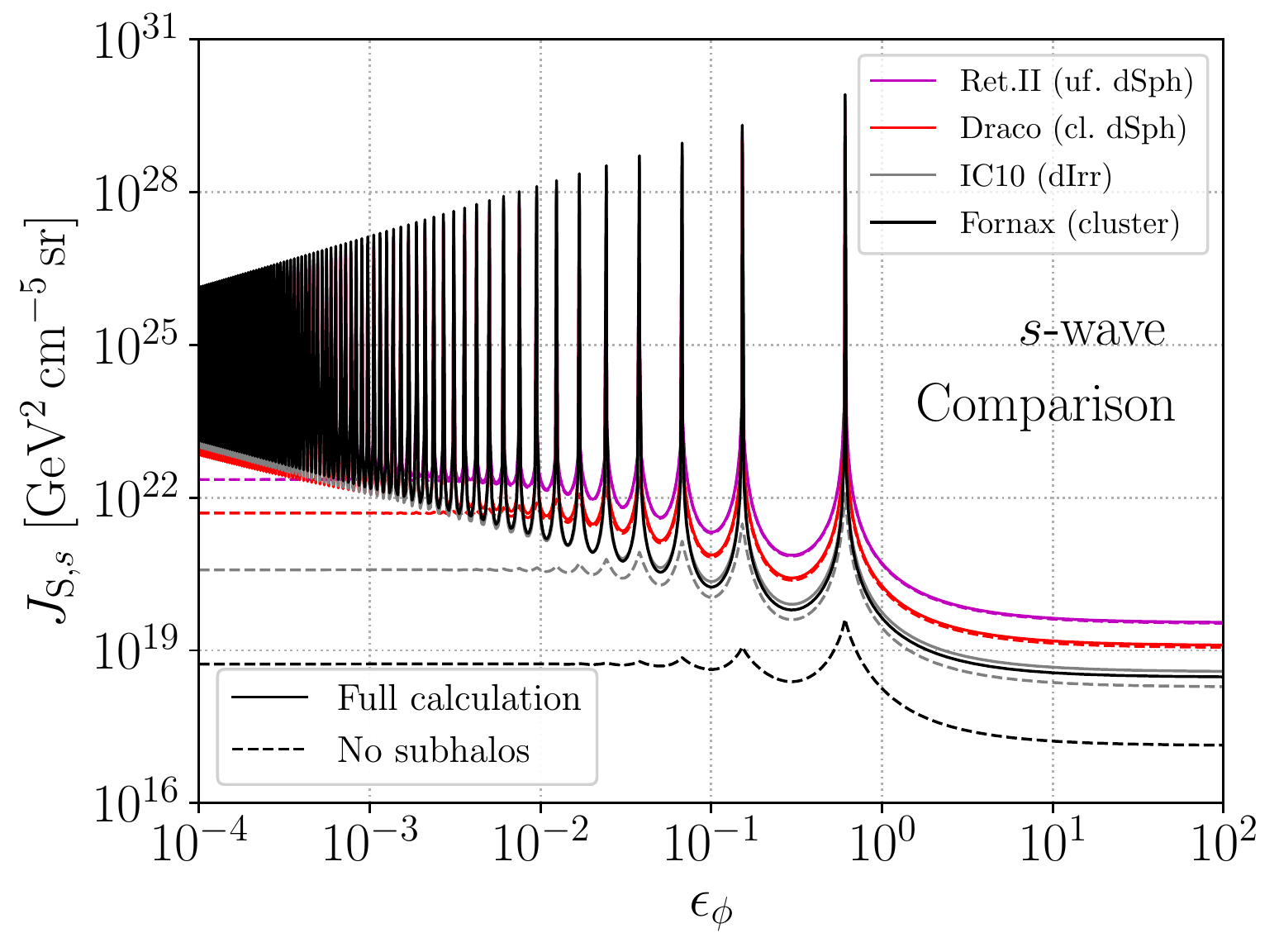} \hfill \includegraphics[width=0.49\linewidth]{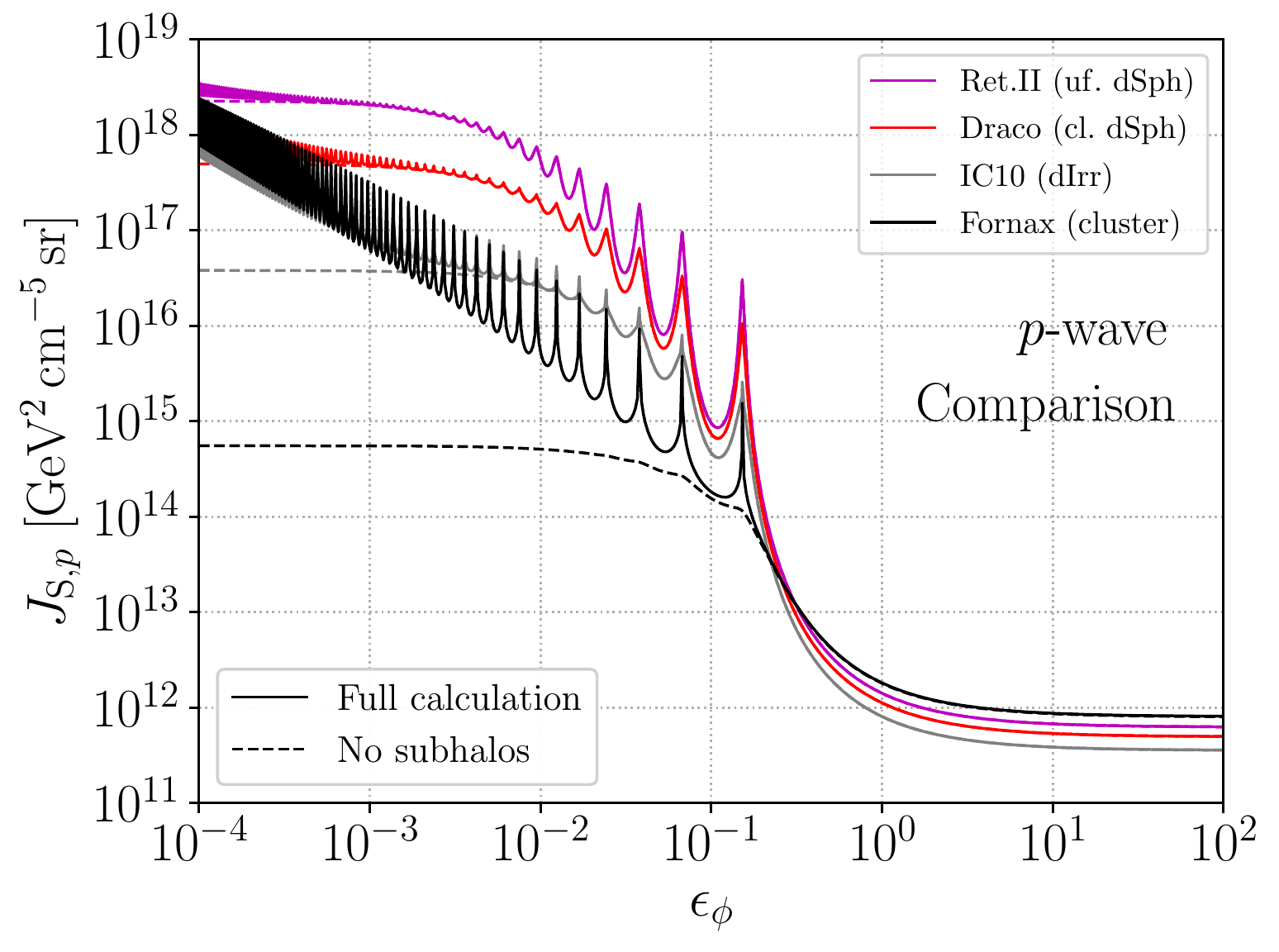}
\caption{Same as in \citefig{fig:comparison_host_halo}, i.e., comparison of $J$-factors for $s$-wave (left panel) and $p$-wave (right panel) annihilation, for one object in each considered target class but dSphs, for which we include both a ``classical'' and a ``ultra-faint'' dwarf. Two different calculations are shown: solid lines correspond to the full calculation accounting for the contribution of subhalos (see Sec.~\ref{ssec:JS_boost_factors}), whereas dashed lines (for comparison purpose) correspond to the `no-subhalos' case already shown in \citefig{fig:comparison_host_halo} (without uncertainties here).}
\label{fig:comparison_subhalos_all}
\end{figure}

With these results, we can now revisit our discussion on the ranking of the best targets (to either detect them or to set stringent constraints on DM particle candidates). 
We show in \citefig{fig:comparison_subhalos_all} a comparison between targets picked among each of the families considered in this study, namely dSphs  (Draco and Reticulum II, with violet and red lines, respectively, and as representative examples of both classical and ultra-faint dSphs), dIrrs (IC10; gray lines), and  galaxy clusters (Fornax; black lines). As before, solid lines correspond to our final results (with substructure boost), and dashed lines show their `no-substructures' counterparts just for comparison purpose (taken from \citefig{fig:comparison_host_halo}). From this \citefig{fig:comparison_subhalos_all}, we notice that we can have a complete inversion of the standard ranking for both the $s$- and $p$-wave cases. Indeed, instead of dSphs being the best targets (for standard $J$-factor calculation), galaxy clusters can now outrank dIrrs, which themselves outrank dSphs. This inversion can happen, e.g., at Sommerfeld resonances in the $s$-wave, and in both the Coulomb ($\epsilon_\phi\ll 1$) and no-Sommerfeld enhancement ($\epsilon_\phi\gg1$) regimes for $p$-wave.

We recall that this inversion in the ranking of targets arises because of the role of substructures, that boost differently the different target classes. While the exact value of these boosts may vary by a factor of a few (due to uncertainties in the subhalo distribution, abundance and structural parameters), the trend of these boosts is not expected to change significantly. 
We also conclude that, for the generalised $J$-factors, the mass modelling uncertainties of the host halos play a subdominant part in almost all regimes: such uncertainties only impact the ranking in the no-enhancement regime ($\epsilon_\phi\gtrsim1$) in $p$-wave annihilation (see discussion in  \citesec{ssec:host_uncertainties} and also \citefig{fig:comparison_host_halo}).

\subsection{Comparison to previous works and prospects for $\gamma$-ray DM searches}
\label{ssec:JSwboost_prospects}

With the results in previous sections, we can now draw some conclusions regarding the selection of the best targets for $\gamma$-ray DM searches, depending on the Sommerfeld regime considered (in $s$- or $p$-wave). Alternatively, given some $\gamma$-ray observations, we can also highlight the regimes where DM candidates are expected to be constrained the most.

\paragraph{Comparison with previous studies.}
It is interesting to compare our findings to what was previously obtained in other works. The largest body of results in the literature is for the standard $J$-factors --- corresponding to $\epsilon_\phi\gg1$ in the $s$-wave case. In this regime, which boils down to the calculation of the boost factors, our results agree with previous determinations; this is not a surprise since we recall that we rely, for the most part, on input ingredients taken from some of our previous works (e.g., \cite{BonnivardEtAl2015b} for dSphs, \cite{2021PhRvD.104h3026G} for dIrrs, and \cite{2011JCAP...12..011S} for galaxy clusters). 
As for the calculation of generalised $J$-factors, there is no study to compare to for dIrrs. For galaxy clusters, to our knowledge, the only previous study is that of \cite{AbramowskiEtAl2012}, where the authors do not directly calculate the generalised $J$-factors, but show limits on DM candidates from the observation of Fornax; hence it is difficult to make comparisons. 
There are several generalised $J$-factor calculations in the literature for dSphs, mostly ignoring substructures. Comparing the results obtained in the `no-substructure' case (\citefig{fig:comparison_host_halo}, top panels), we find our results to be comparable to those of \cite{BoddyEtAl2017,BergstromEtAl2018,PetacEtAl2018,BoddyEtAl2020,AndoAndIshiwata2021} for $s$-wave in all regimes.\footnote{Note that, in some cases, a rescaling of $\alpha_{\rm D}$ is needed to perform these comparisons.}

The interplay between subhalos and velocity-dependence has been investigated in several studies \cite{KuhlenEtAl2009,Bovy2009,LattanziAndSilk2009,PieriEtAl2009,SlatyerEtAl2012,ZavalaEtAl2014a,BoddyEtAl2019,RunburgEtAl2021,PiccirilloEtAl2022}. Comparison with our results is difficult in most cases as alternatives targets and different regions of the parameter space are considered.
The subhalo models are also quite different from the one we have used.
In \cite{Bovy2009} the author performed a calculation of the subhalo boost factor in the presence of Sommerfeld enhancement, focusing on the $s$-wave case and dSph-sized hosts. 
The author found results that are qualitatively similar to ours: the resonant regime extends to very low values of $\epsilon_\phi$ when subhalos are considered, and the boost factor can reach extremely high values. Quantitatively, the boost factors in \cite{Bovy2009} seem to be higher than ours by one or two orders of magnitude. A possible reason for this discrepancy is the subhalo mass-concentration relation used in \cite{Bovy2009} which leads to subhalos that are much denser, and therefore over-annihilate, compared to what has been found in more recent numerical simulations (see \cite{Sanchez-CondeEtAl2014} for the mass-concentration we have used instead).

\paragraph{Angular extension of the signal.}
The morphology or, for our targets, the radial dependence of the $\gamma$-ray signal is directly linked to the underlying emission processes and source spatial distribution (here, annihilations in the smooth halo and substructures).
As advocated in past studies on standard $J$ factors, the angular extension of the signal could be used to identify decaying from annihilating DM in dSphs \citep{Palomares-RuizEtAl2010} (objects in which boost factors are mostly irrelevant), or to disentangle CR-induced from DM-induced $\gamma$-ray signals in galaxy clusters \cite{JeltemaEtAl2009,MaurinEtAl2012} (objects in which substructures both boost and enlarge the size of the object on the sky for annihilating DM; see also \cite{2011JCAP...12..011S}).   
This reasoning has been further developed in the context of velocity-dependent annihilations in \cite{BoddyEtAl2019,2021arXiv211009653B}, where the authors discuss how the radial dependence of the signal could help identifying the underlying particle physics model. However, these studies mostly focus on a single halo (although \cite{BoddyEtAl2019} briefly comments on the consequences for a distribution of subhalos), whereas we have shown that substructures may be important even for dSphs in some specific regimes. Another difference is that we perform a full numerical calculation while \cite{BoddyEtAl2019,2021arXiv211009653B} rely on analytical approximations.

\begin{figure}[t!]
    \centering
    \includegraphics[width=0.49\linewidth]{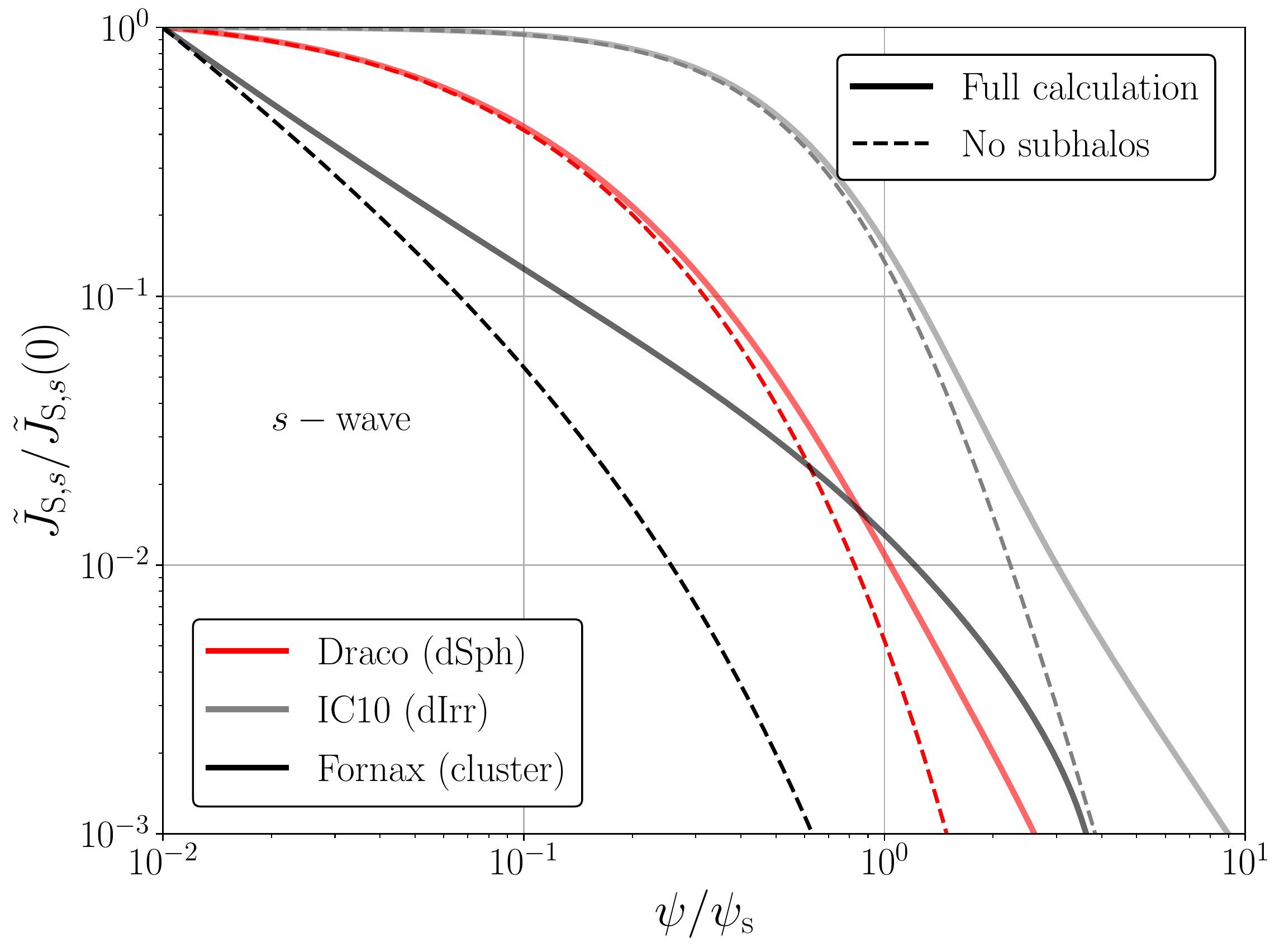}
    \includegraphics[width=0.49\linewidth]{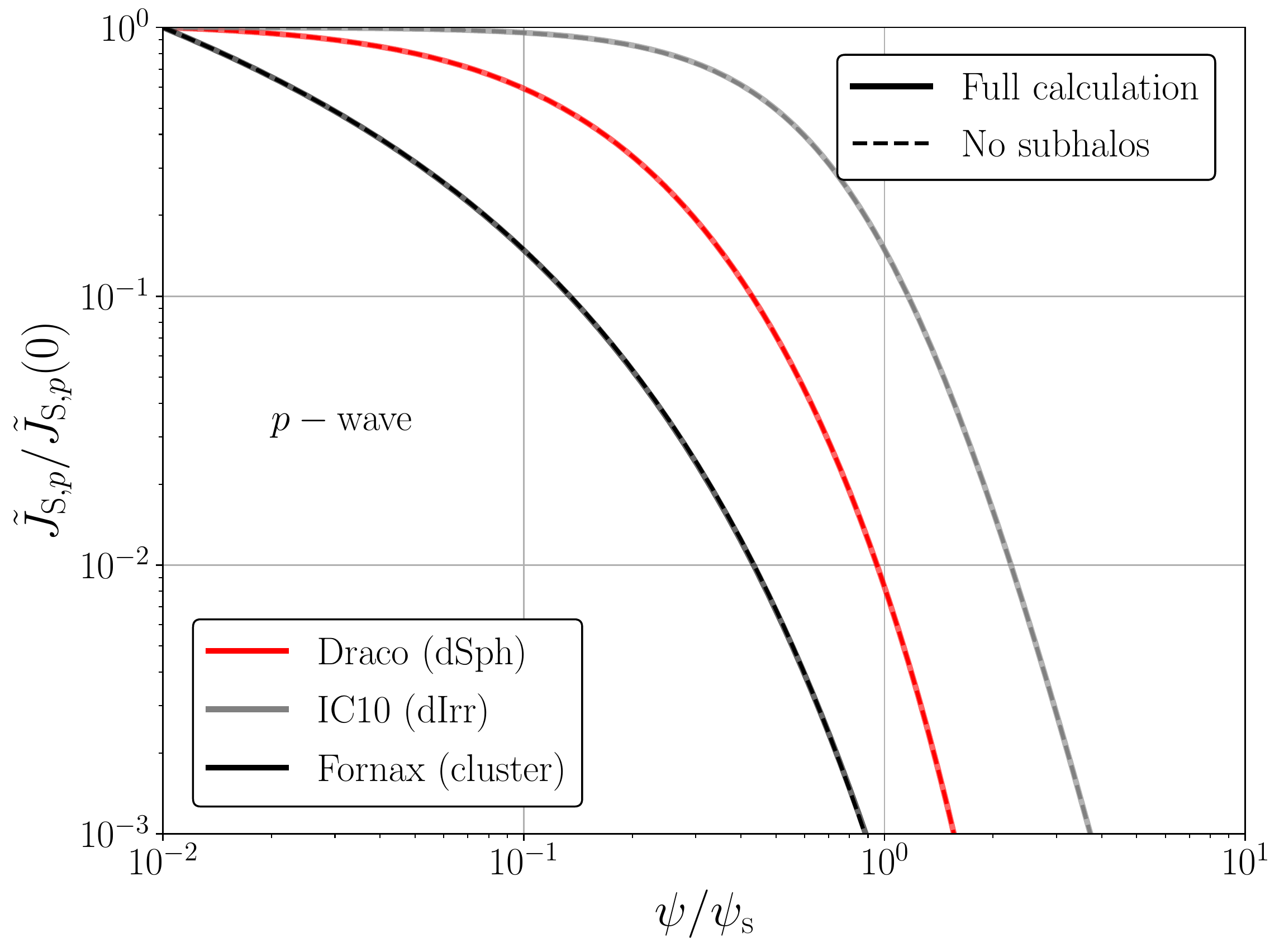}
    \includegraphics[width=0.49\linewidth]{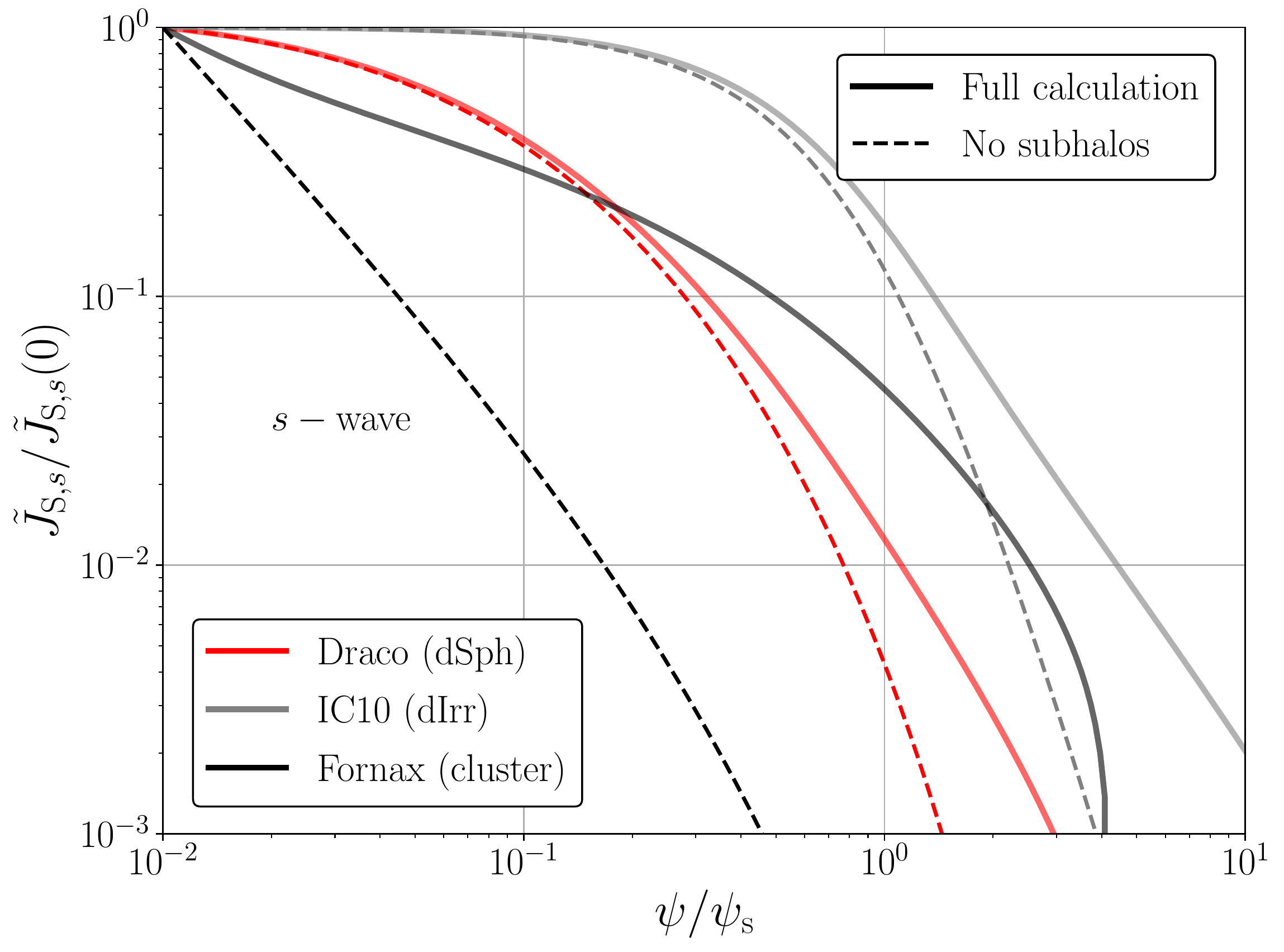}
    \includegraphics[width=0.49\linewidth]{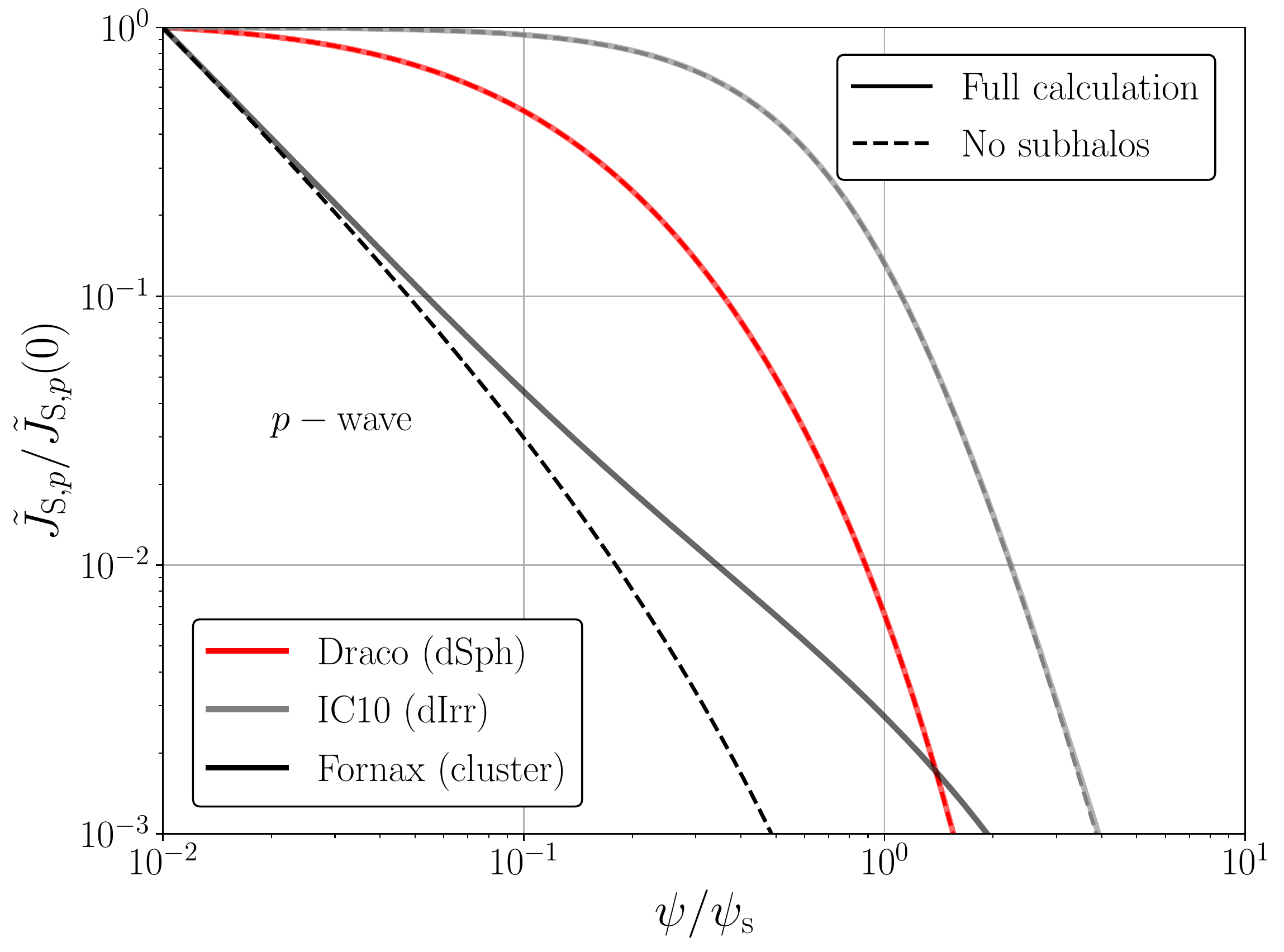}
    \caption{Differential $J_{\rm S}$-factor as a function of the scaled angle $\psi/\psi_{\rm s}$ from the centre of the object, for three targets belonging to our three classes of interest. To ease the comparison, all profiles have been normalised to unity at their centre. The top row displays the results without Sommerfeld enhancement ($\epsilon_\phi\gg 1$), while the bottom row shows the results for $\epsilon_\phi\simeq0.011$ (off resonance). The results for $s$-wave are shown in the left panels and those for $p$-wave are in the right panels. The reference angular size is $\psi_{\rm s}=1.4^\circ/0.15^\circ/0.5^\circ$ for Draco/IC10/Fornax.
    }
    \label{fig:angular_extension}
\end{figure}

Figure~\ref{fig:angular_extension} shows the differential $J$-factor $\tilde{J}_{\rm S}(\psi)$ defined as
\begin{eqnarray}
J_{\rm S}=\int_0^{\theta_{\rm int}}\mathrm{d}(\cos\psi)\,\tilde{J}_{\rm S}(\psi)
\end{eqnarray}
without (dashed lines) or accounting for (solid lines) substructures, as a function of the angle $\psi$ from the target centre. 
We recall that the relevant angular size of the $\gamma$-ray signal will be a combination of the physical size of the target, the distance to the observer, the steepness of the DM distribution, and, also for generalised $J$-factors, the velocity distribution profile.
The latter is illustrated in Fig.~\ref{fig:angular_extension} with the calculation of the signal without subhalos (dashed lines). Indeed, for standard $J$-factors, i.e., $\epsilon_\phi\gg1$ in the $s$-wave case (top left panel), the typical radial extension of a given target boils down to $\psi_{\rm s}=\arctan(r_{\rm s}/D)$ (most of the emission is within $r_{\rm s}$), i.e., $1.4^\circ$ for Draco, $0.15^\circ$ for IC10, and $0.5^\circ$ for Fornax (using numbers taken from Tables~\ref{tab:dsphs_mass_models}, \ref{tab:dIrr}, and \ref{tab:clusters_mass_models} respectively). These numbers compare well to the radial extensions seen in the top left panel of Figure~\ref{fig:angular_extension}, but we already see the impact of the velocity distribution function comparing the top left and bottom left dashed curves (no subhalos, no Sommerfeld enhancement) for $s$- and $p$-wave respectively. These radial extensions are also slightly different in the Sommerfeld regime (bottom panels).

The relatively small differences observed between different particle physics models for a smooth halo (without subhalos) are qualitatively similar to those highlighted in \cite{2021arXiv211009653B}. Yet, the presence of subhalos (solid lines) on $\tilde{J}_{\rm S}$ is strongly model dependent. Subhalos have almost no impact for $\epsilon_\phi\gg1$ in the $p$-wave case (right panels in Fig.~\ref{fig:angular_extension}) and a maximal impact in the $s$-wave resonant regime (bottom left panel), where a factor of ten increase of the typical angular size is observed for Fornax (compare the dashed and solid black lines). All in all, these results reinforce the case for the use of the angular dependence (or radial extension) of the $\gamma$-ray signal as a tool to discriminate between different particle models (if a DM signal is seen). Further study is necessary to decide/optimise which combinations of different targets are best to discriminate among particle physics models. Also, it is not clear whether degeneracies (and uncertainties) between both the DM and particle physics modellings would prevent the use of this strategy in some cases (not to mention the observational challenge to characterise the spatial morphology of a putative $\gamma$-ray signal: current $\gamma$-ray telescopes possess typical angular resolutions of about one to few tenths of degree).

\paragraph{Galactic DM foreground.} 

So far, we have considered $\gamma$-ray signals from isolated targets. However, the diffuse $\gamma$-ray emission originated from DM annihilations happening in the MW halo itself can be a sizable foreground when searching for $\gamma$-rays from various targets (the DM extragalactic diffuse signal also adds up to this foreground but will not be considered here; see, e.g., \cite{2015JCAP...09..008F} for a comparison between both DM diffuse components in terms of their intensity flux). To optimise the detectability of such targets in $\gamma$-rays, their DM-induced signal should ideally lie above this Galactic DM foreground (typically, for a field of view corresponding to the angular resolution of the detector). 

For completeness, we compute the astrophysical factor of the MW halo for a solid angle defined by $\Delta \Omega = 2 \pi (1 - \cos \theta_{\rm int})$ around the line of sight (l.o.s.) at an angle $\psi_{\rm GC}$ from the Galactic centre, $J_{\rm S}(\psi_{\rm GC})$. We use Eq.~(\ref{eq:J_S_fov}) for the calculation of the annihilation. For the MW mass model, we use an NFW profile with $\rho_{0} = 8.5\times10^6\,\rm M_\odot/kpc^3$ and $r_{\rm s} = 19.6\,\rm kpc$ \cite{McMillan2017}. For the subhalos, we use again the SL17 model. 
An important difference when studying subhalos in the MW compared to the other targets is the role of baryons in shaping their distribution. In SL17, both the smooth tidal stripping induced by the baryonic potential and the gravitational shocking induced by the stellar disk are taken into account.\footnote{We have checked that baryonic tidal effects are completely negligible in the other targets, even in galaxy clusters which have a sizeable baryonic content.} 
The calculation for the contributions of the MW smooth halo and subhalos (see \citesec{ssec:subhalo_boost}) is otherwise similar to that of the other targets, the main differences being that (i) the observer is now sitting inside the host halo, and (i) the baryonic potential can no longer be neglected (see \citeapp{app:SL17} on how it is accounted for in the modelling).

\begin{figure}[t!]
    \centering
    \includegraphics[width=0.49\linewidth]{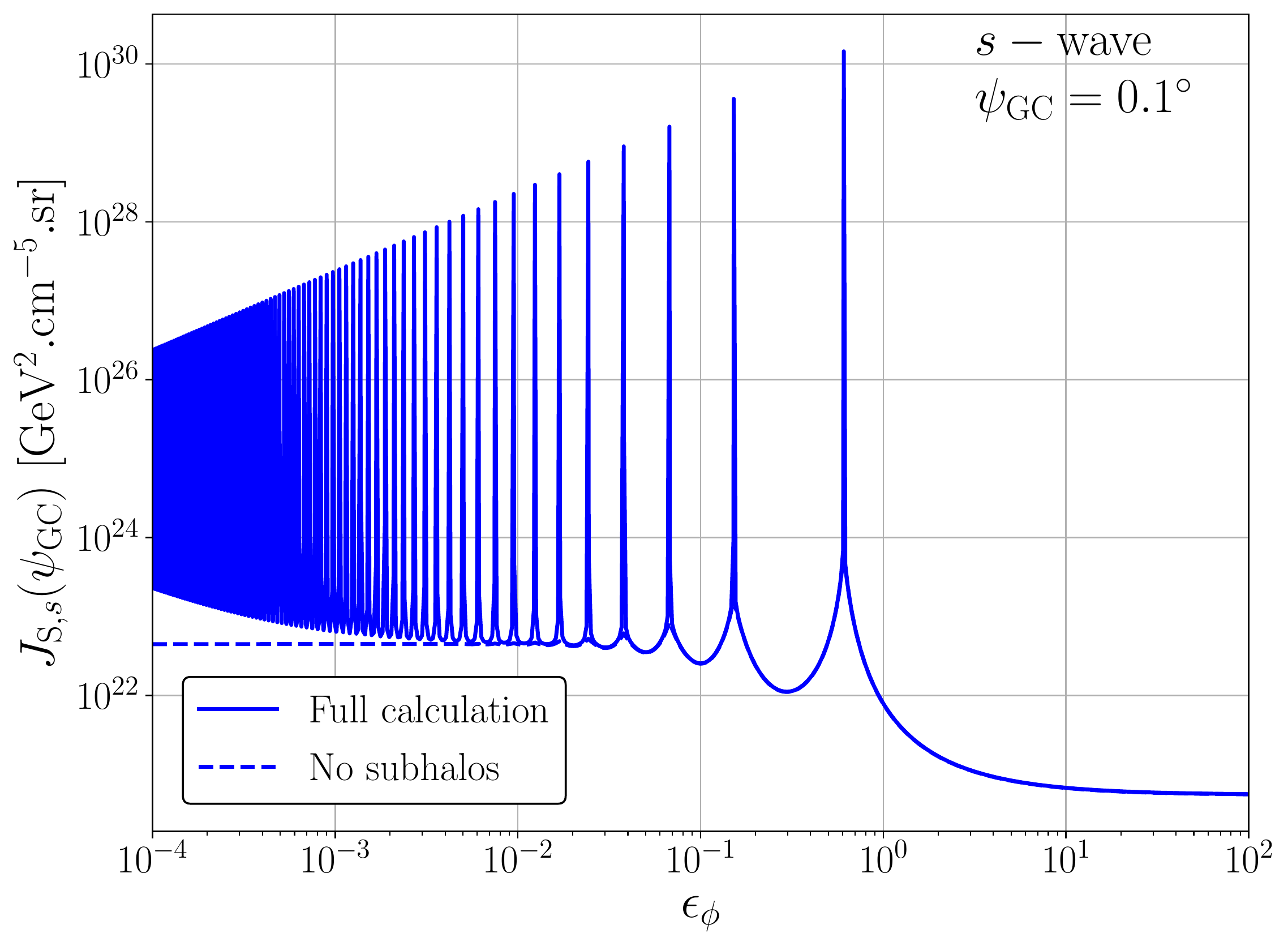}
    \includegraphics[width=0.49\linewidth]{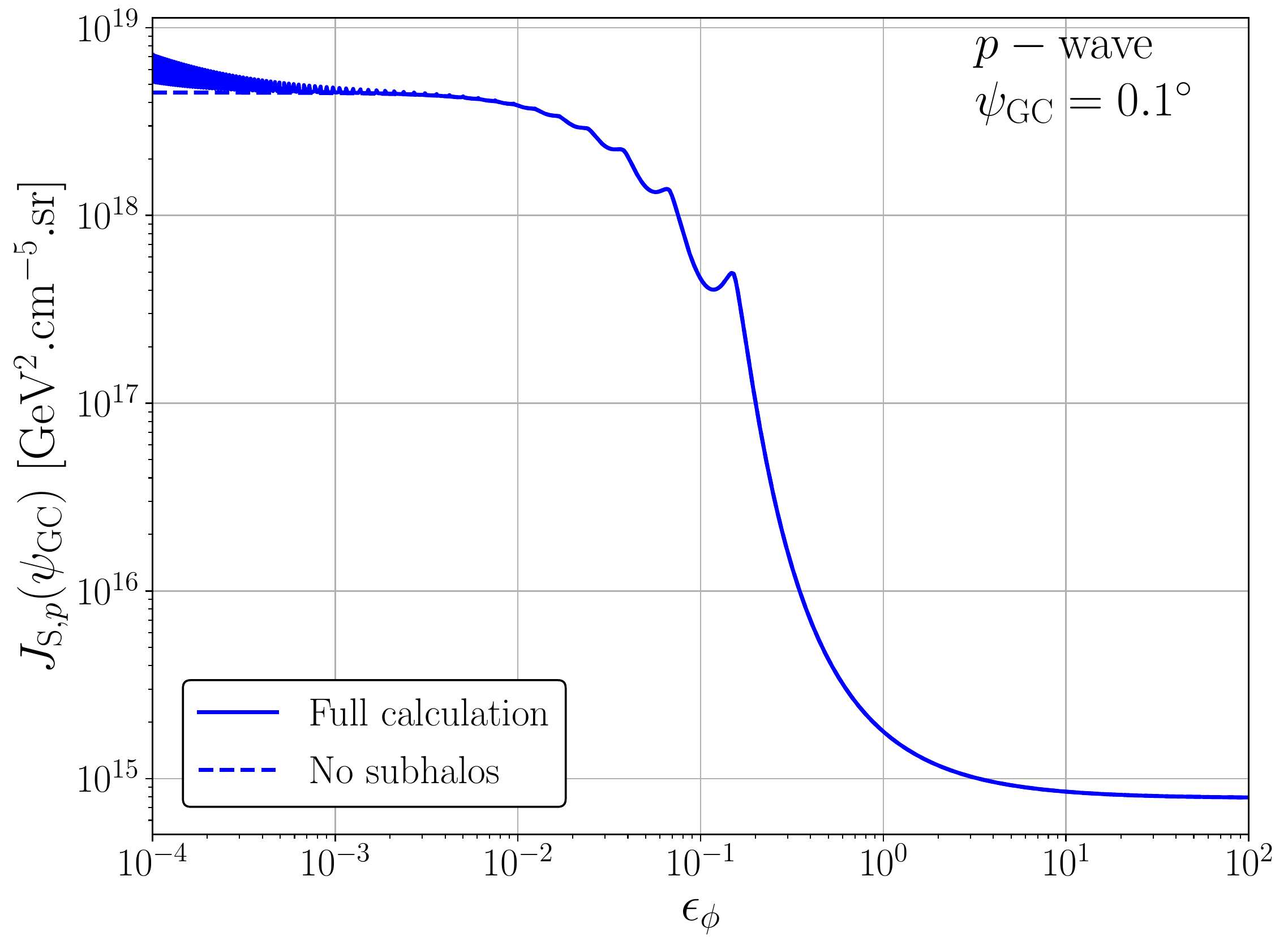}
    \includegraphics[width=0.49\linewidth]{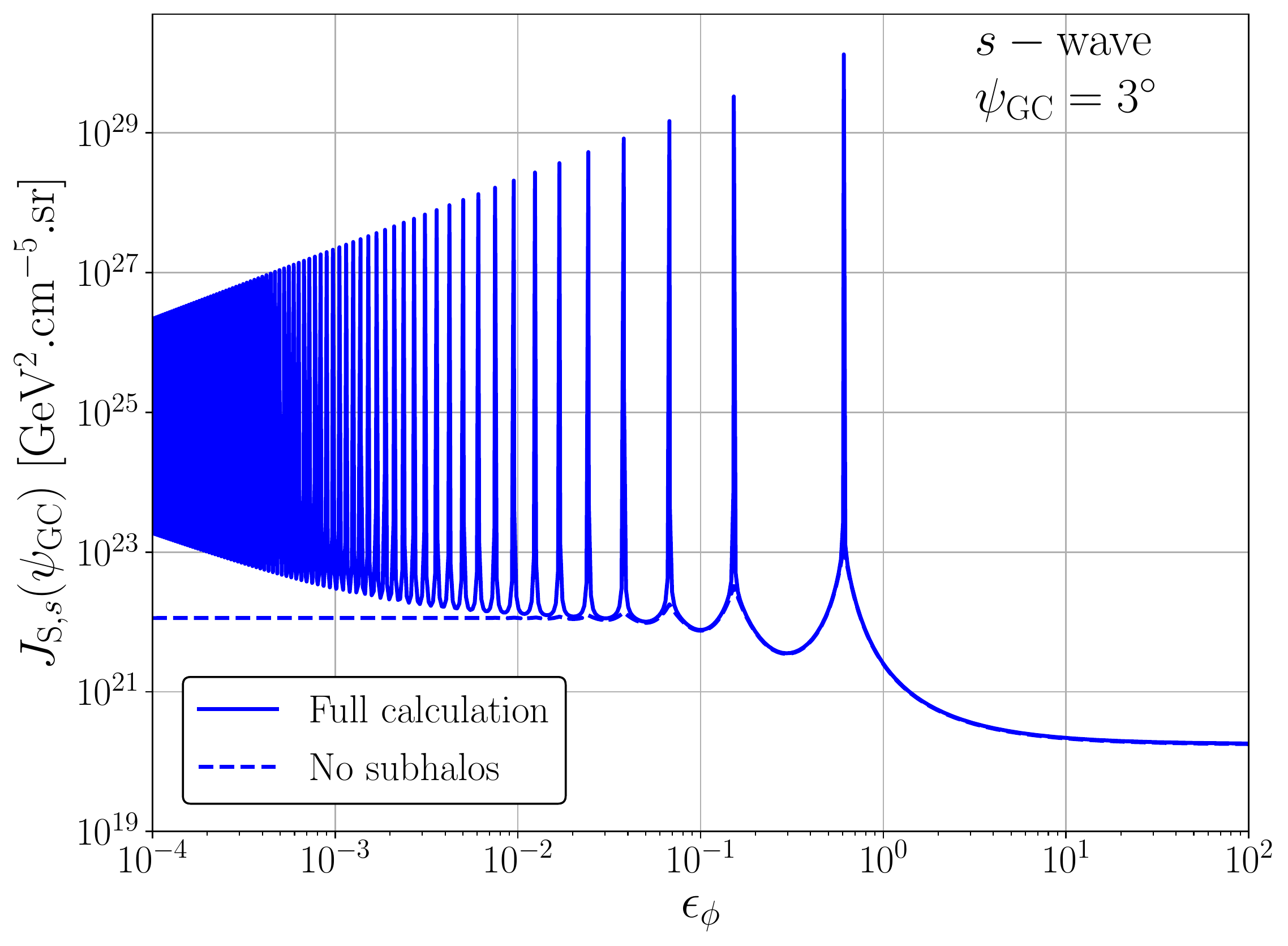}
    \includegraphics[width=0.49\linewidth]{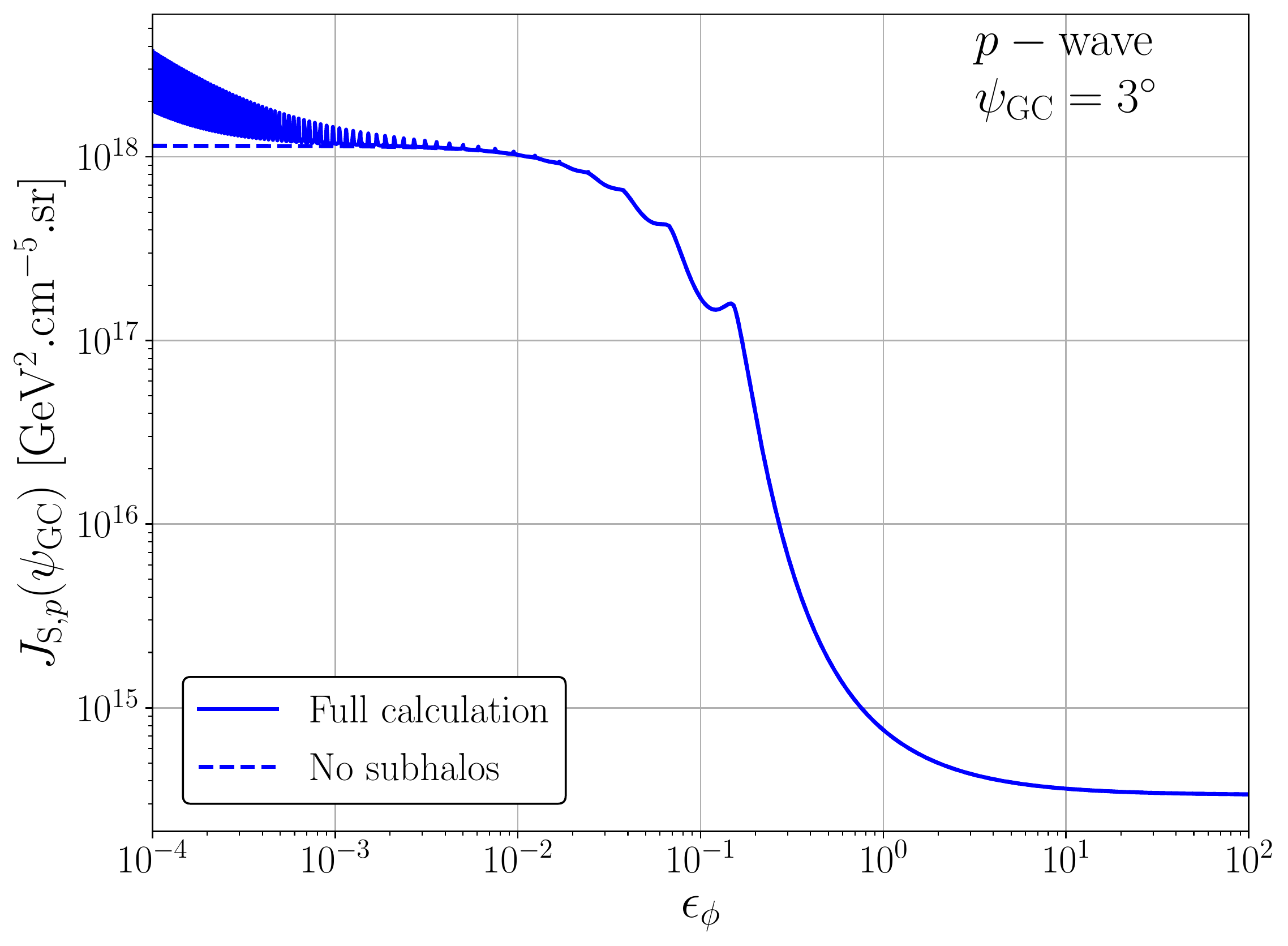}
    \includegraphics[width=0.49\linewidth]{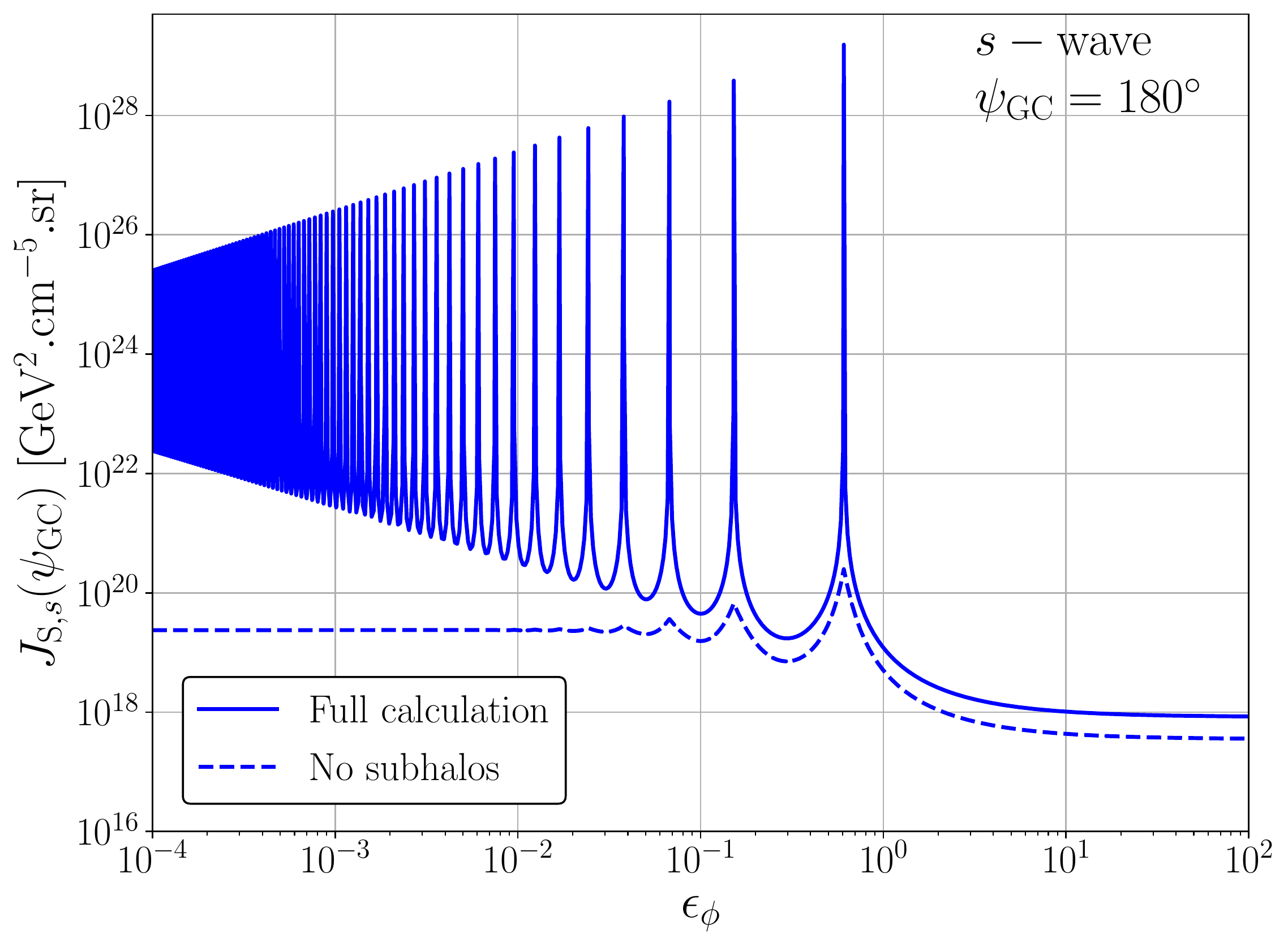}
    \includegraphics[width=0.49\linewidth]{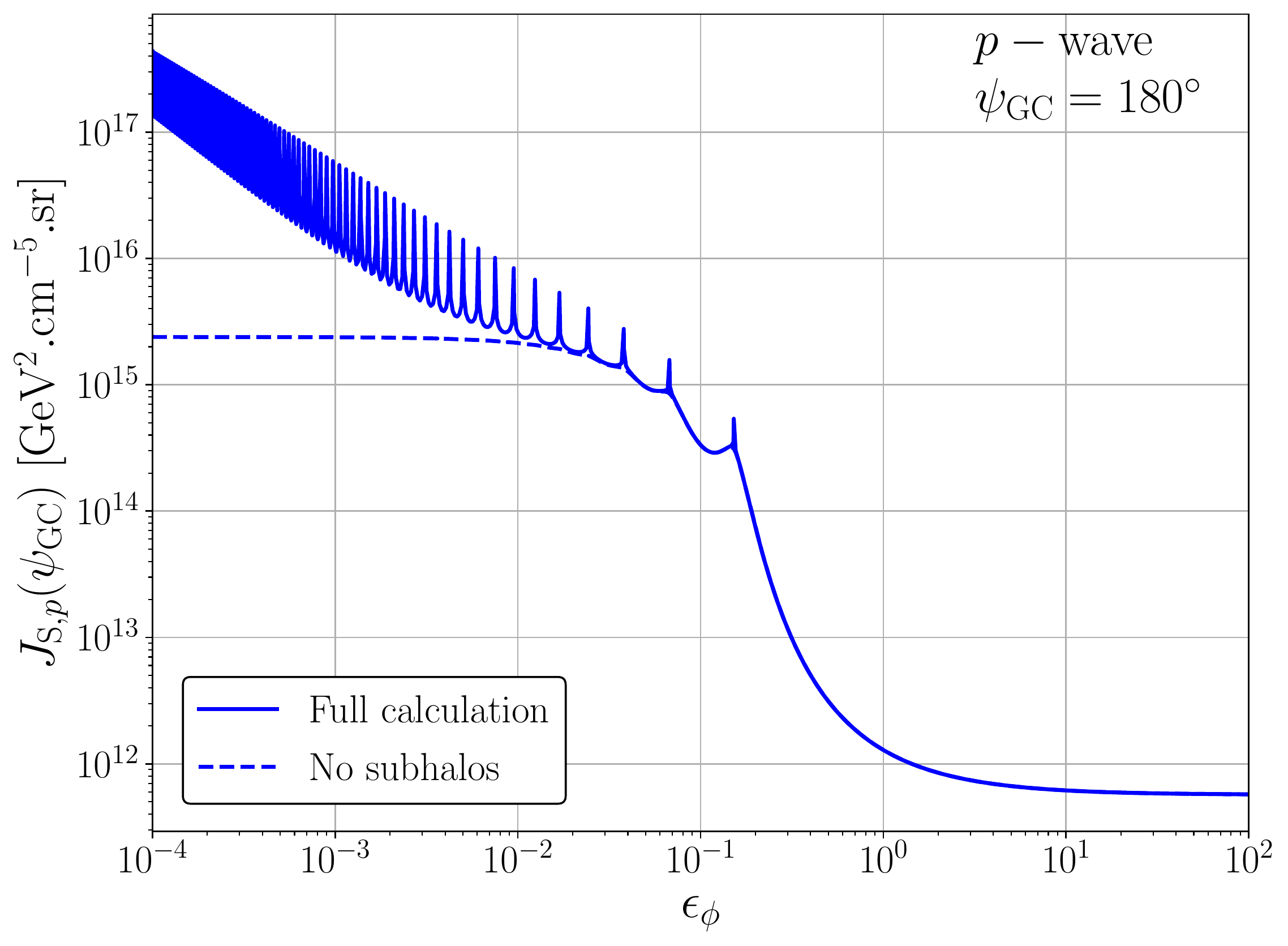}
    \caption{Generalised MW $J$-factors (integrated over $\theta_{\rm int}=0.5^\circ$) vs. $\epsilon_{\phi}$ for $s$-wave (left panels) and $p$-wave (right panels) annihilation, in various directions $\psi_{\rm GC}$ from the Galactic centre: from top to bottom, $\psi_{\rm GC}=0.1^\circ$, $3^\circ$, and $180^\circ$ (anti-centre). Solid lines correspond to the full calculation accounting for the contribution of subhalos (see Sec.~\ref{ssec:subhalo_boost}), whereas dashed lines correspond to the `no-subhalos' case. 
    }
    \label{fig:JS_MW}
\end{figure}

Figure~\ref{fig:JS_MW} shows the resulting MW $J_{\rm S}$ factors for $s$- and $p$-wave (left and right panels, respectively), without (dashed lines) or with (solid lines) subhalos. From top to bottom, going from a l.o.s. slightly offset from the Galactic centre ($\psi_{\rm GC}=0.1^\circ$ and $3^\circ$, top and middle panels) and moving towards the anticentre ($\psi_{\rm GC}=180^\circ$, bottom panel), we observe in all regimes, as expected, that the $J_{\rm S}$ factors decrease. Also, as expected in the velocity-independent case ($\epsilon_\phi\gg1$ in $s$-wave), subhalos only boost the signal towards the anticentre (the signal is dominated by the smooth halo towards the halo centre). Note that there is no boost from subhalos in the no-Sommerfeld enhancement regime ($\epsilon_\phi\gg1$) for $p$-wave (right panels). This larger impact of subhalos, away from the Galactic centre, is also recovered in the  $s$- (left panels) and $p$- (right panels) wave at intermediate and small $\epsilon_\phi$. Actually, the pattern is very similar to the one seen in \citefig{fig:comparison_subhalos_all}, where the presence of subhalos, down to the cutoff mass, leads to larger peaks at resonances and growing boost with decreasing $\epsilon_\phi$ (see discussion in \citesec{ssec:JS_boost_factors} for more details). 

Based on numerical simulations, \cite{BoardEtAl2021} recently showed that the predicted $J_{\rm S}$ factor from the smooth halo of the MW is very sensitive to the DM velocity distribution function, comparing predictions from DM-only or hydrodynamical simulations. These authors also concluded that the impact of subhalos is subdominant in their work, given that their simulation only resolves the largest subhalos. Our results show that the distribution of subhalos down to the smallest masses is actually critical to correctly predict the MW signal on resonances and in the regime $\epsilon_\phi\ll1$.

\paragraph{Summary view of all targets against the Galactic DM foreground.} 

Thanks to the above calculation, we can now assess the contrast between the DM signal from all our targets and that from the MW DM foreground. 
This contrast is shown as a function of $\psi_{\rm GC}$ (angle between the l.o.s. and the Galactic centre; the Galactic halo is spherically-symmetric) in \citefig{fig:comparison_mw_foregrounds}. The panels in this figure are for two regimes of the $s$-wave (left panels) and $p$-wave (right panels) annihilations, namely the no-Sommerfeld enhancement regime at $\epsilon_\phi\gg1$ (top panels), the saturation regime off-resonance at $\epsilon_\phi\simeq 10^{-2}$ (middle panels) and the saturation regime on-resonance also at $\epsilon_\phi\simeq 10^{-2}$ (bottom panels). 
We recall that $\epsilon_\phi\gg1$ corresponds to a mediator mass comparable or larger than the DM mass, while $\epsilon_\phi=10^{-2}$ corresponds to a light mediator (\eg, $m_\phi=1\,\rm GeV$ for $m_\chi=10\,\rm TeV$ and our choice of $\alpha_{\rm D}=10^{-2}$).
On each plot, the solid line corresponds to the MW $J_{\rm S}$ values (see previous paragraph), and the symbols represent our various target values,\footnote{Each target position $\psi_{\rm GC}$ is computed from its Galactic longitude $l$ and latitude $b$ as $\psi_{0,\rm target} = \arccos{(\cos{l} \cos{b})}$.} $J_{{\rm S}_{\psi_{\rm GC}}}$. Both the MW and the target values are calculated for an integration region $\theta_{\rm int} = 0.5^{\circ}$.\footnote{This value is motivated by the typical angular extensions of our targets. We note though that this integration angle may not enclose their total DM signals in some cases, especially for galaxy clusters. Yet, most of the signal will be still originated from this inner $0.5^{\circ}$. Also, from the data analysis point of view, more extended objects are more difficult to deal with, thus our compromise in terms of the chosen $\theta_{\rm int}$.}

\begin{figure}[t!]
\centering
\includegraphics[width=0.49\linewidth]{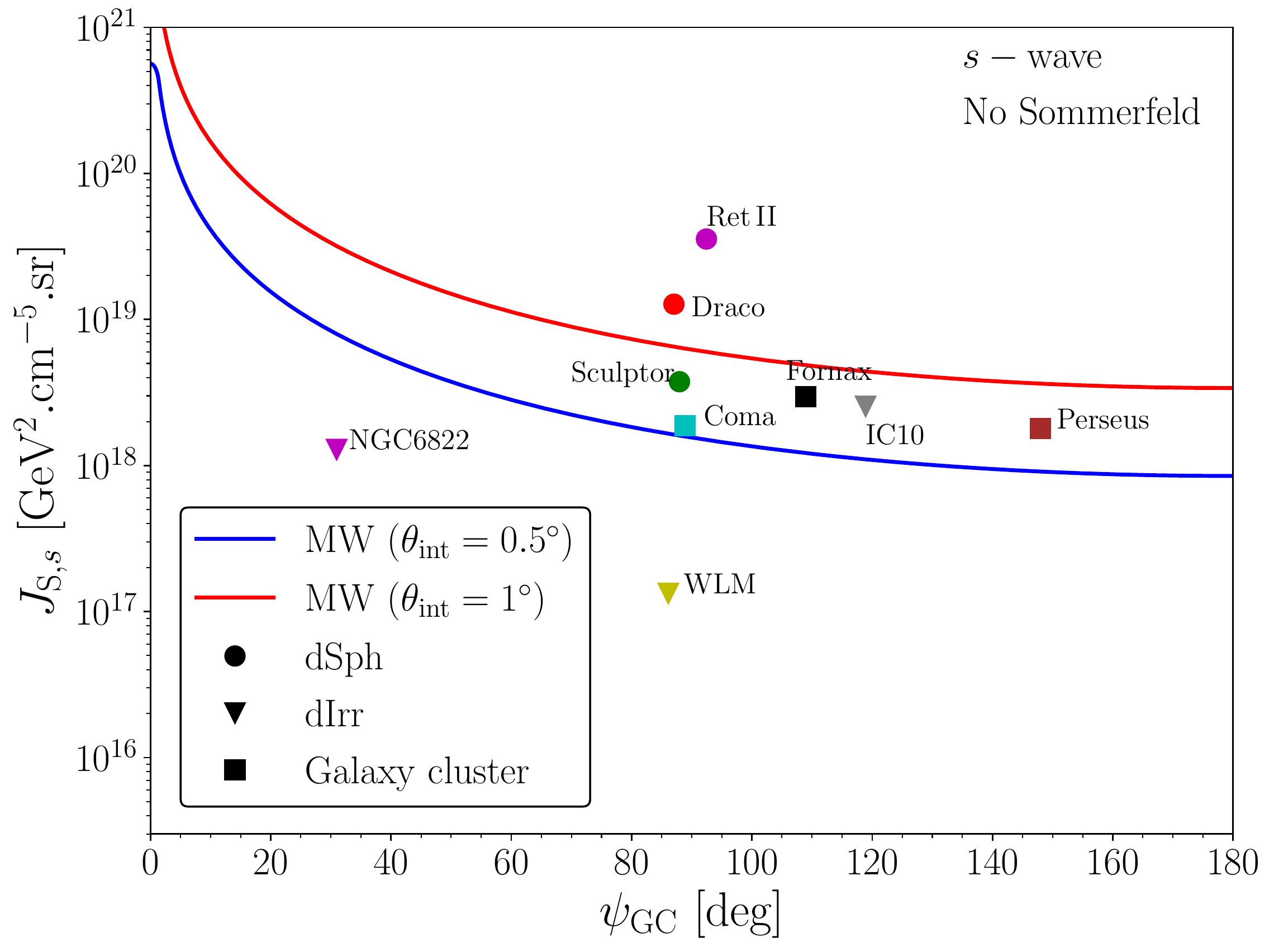} \hfill
\includegraphics[width=0.49\linewidth]{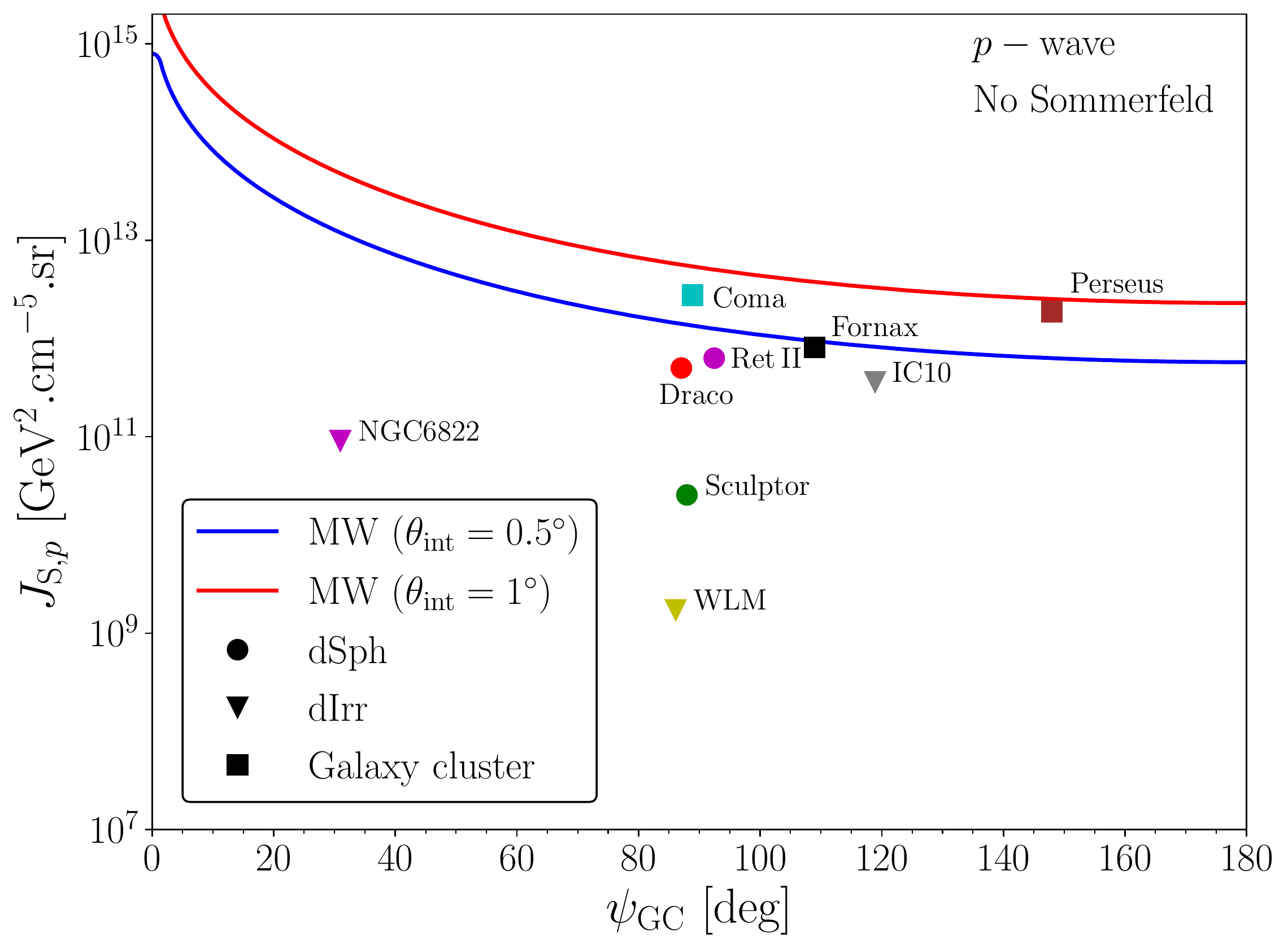} \hfill
\includegraphics[width=0.49\linewidth]{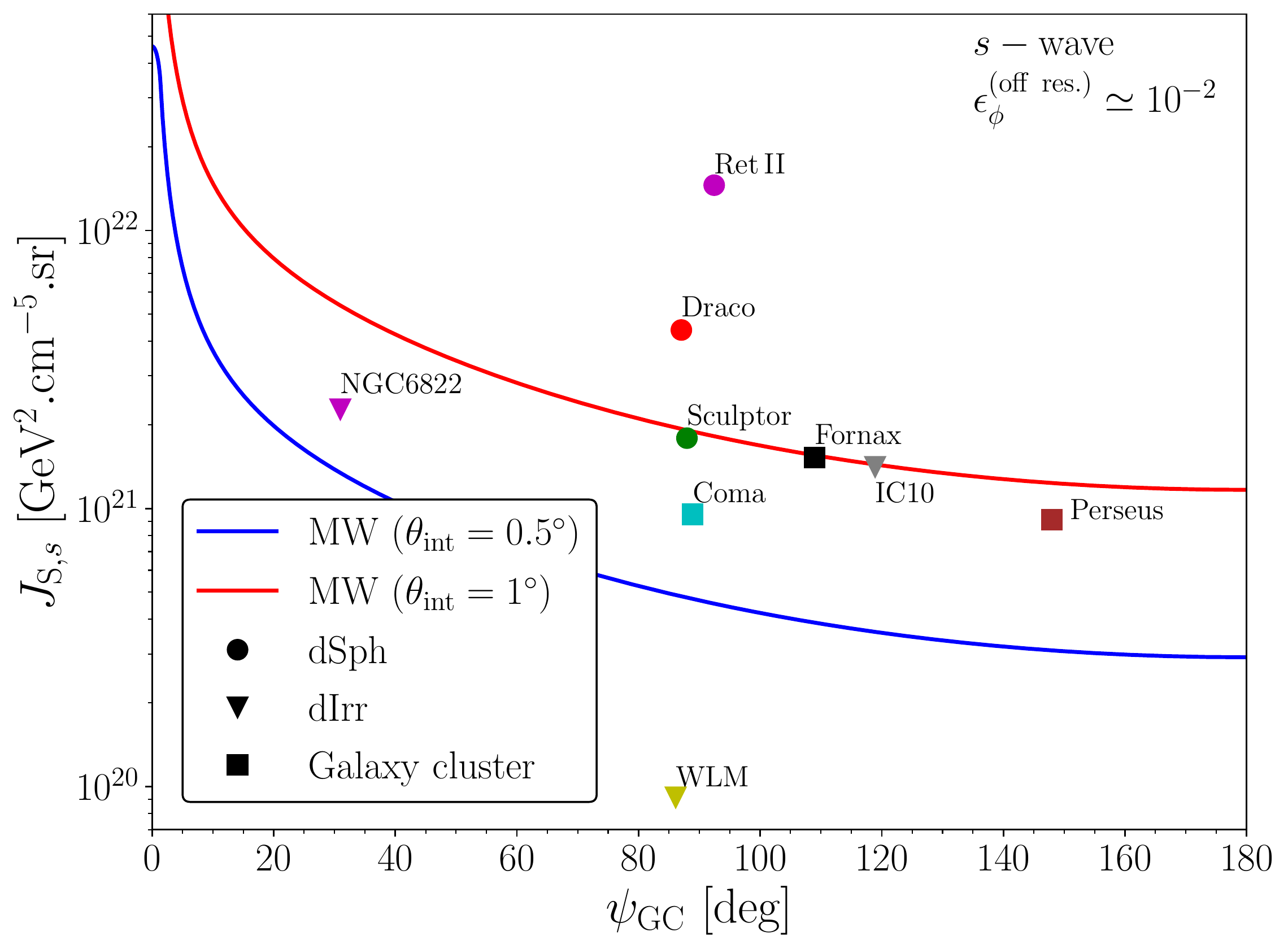} \hfill
\includegraphics[width=0.49\linewidth]{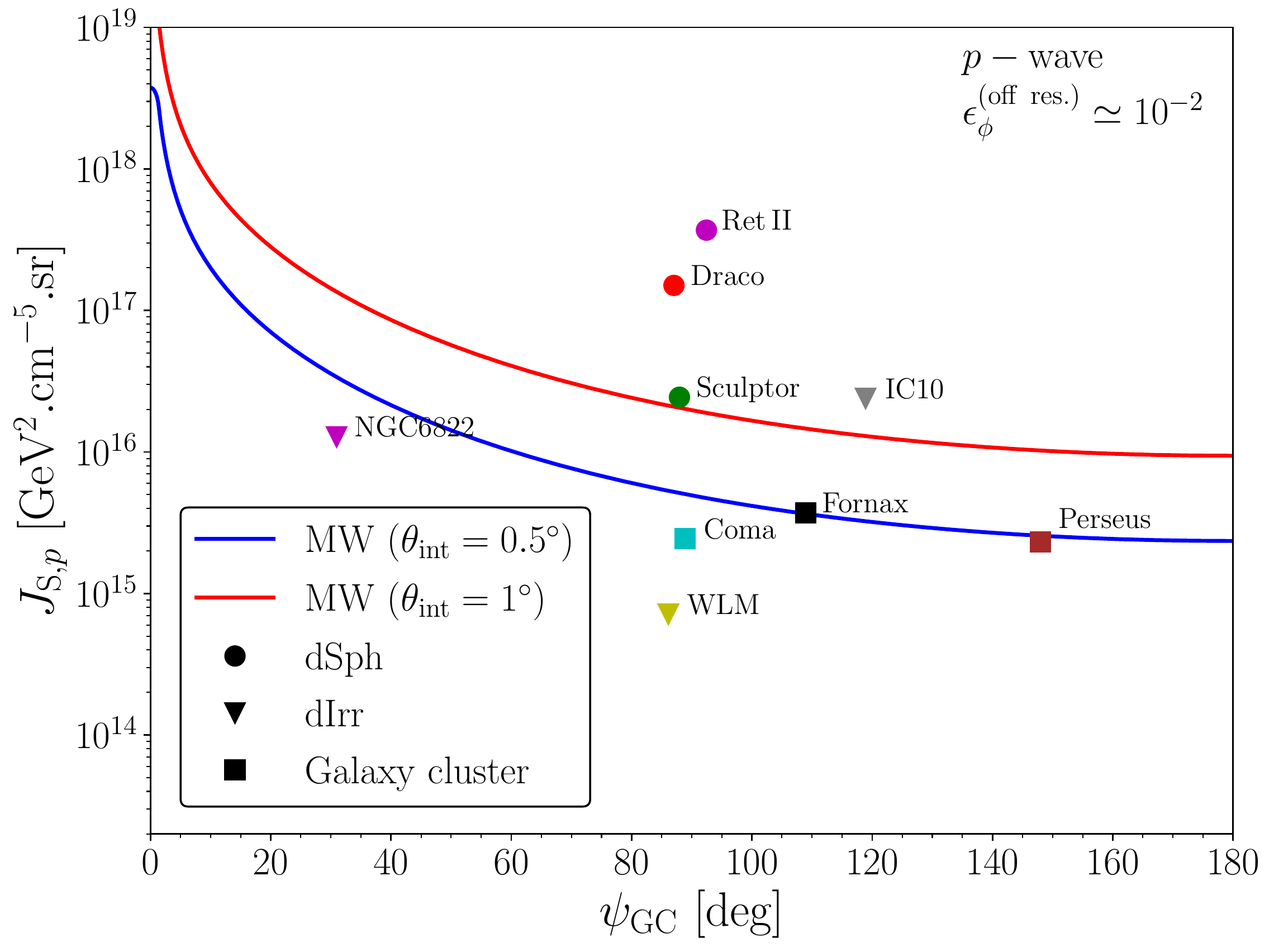} 
\includegraphics[width=0.49\linewidth]{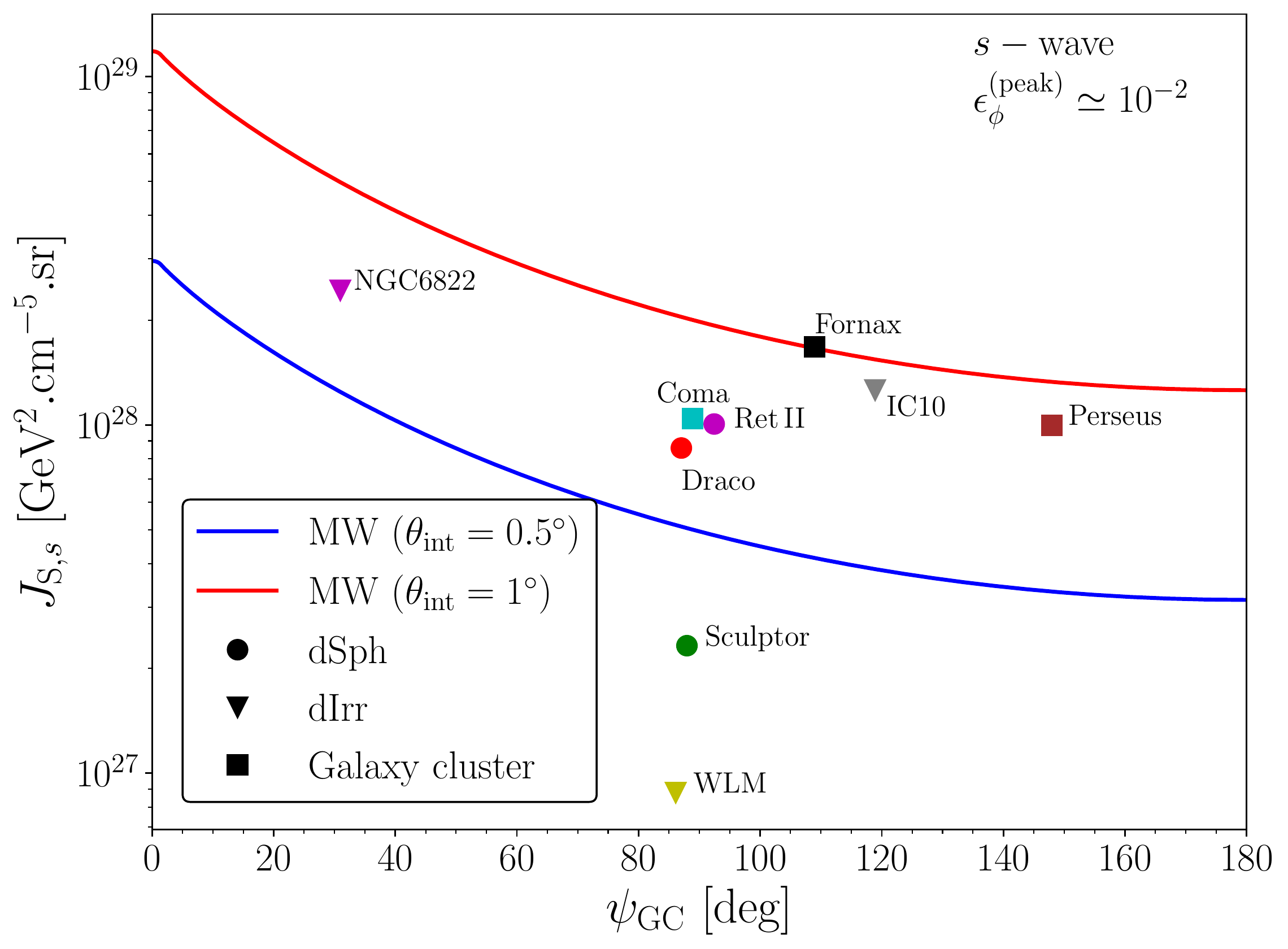}
\includegraphics[width=0.49\linewidth]{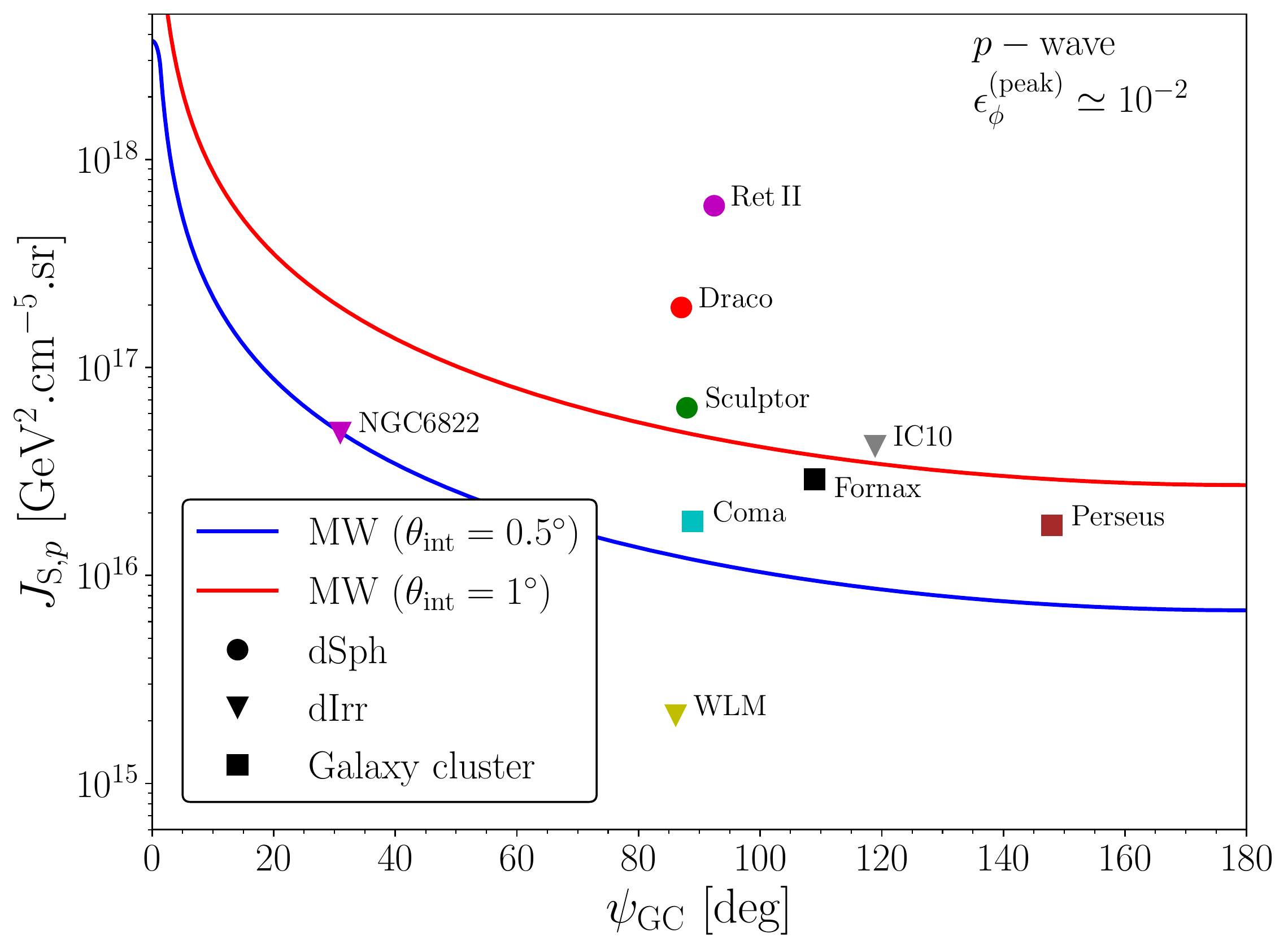}
\caption{Comparison of generalised $J$-factors from all our targets (symbols) against the MW DM foreground (blue and red lines, depending on the integration angle) as a function of the angle $\Phi_0$ (away from the Galactic centre) for $s$-wave (left panels) and $p$-wave (right panels); the top panels are for $\epsilon_\phi\gg1$, i.e., no Sommerfeld enhancement, the middle panels are for $\epsilon_\phi\simeq 10^{-2}$ in the saturation regime off-resonance and the bottom panels are on-resonance also at $\epsilon_\phi\simeq 10^{-2}$ . $J_{\rm S}$ values for all dSphs and dIrrs are calculated for an integration angle $\theta_{\rm int}=0.5^\circ$ while clusters are integrated up to the virial radius ($\theta_{\rm int}=2.2^\circ/1.1^\circ/1.2^\circ$ for Fornax/Coma/Perseus). See text for details. }
\label{fig:comparison_mw_foregrounds}
\end{figure}

The various panels of \citefig{fig:comparison_mw_foregrounds} illustrate that: (i) in most regimes of $s$-wave annihilation (left panels), dSphs remain the most promising targets, whereas dIrr and galaxy cluster signals, $\sim 10$ times lower, remain interesting and complementary targets; (ii) in some regimes of $p$-wave annihilation (e.g., top right panel), an inversion of the ranking is observed and galaxy clusters become the most promising targets, while in some other cases (bottom right panel), dIrrs can shine as bright as some dSphs. In addition, it is interesting to note that for this value of $\theta_{\rm int} = 0.5^{\circ}$, most targets in most regimes outshine the MW DM foreground. Yet, the signal contrast between the former and the latter is significantly smaller for the $p$-wave case: in the no-Sommerfeld enhancement regime (top right panel), the MW is even above all target signals but for the two brightest galaxy clusters. The non-trivial dependence of this contrast with the integration region ($\theta_{\rm int}$) is further discussed in \citeapp{app:MWcontrast}. Overall, this variety of scenarios illustrate that devising the optimal signal region to search for DM signals (or to set limits on DM candidates) is not a trivial task and needs to be studied in detail in a case-by-case basis, as it is sensitive not only to both the individual target and target class but also to the specific particle physics model considered. Such a search strategy should also need to account for the Galactic DM foreground; for completeness, it should also account for the extragalactic DM background (see, e.g.,~\cite{2015JCAP...09..008F}).

\paragraph{Relevance to particle physics.} 
The results discussed here are relevant for particle-physics models involving mediators much lighter than the DM particle. Such a hierarchy appears in minimal setups such as Minimal Dark Matter \cite{CirelliEtAl2005,CirelliEtAl2007,GarciaCelyEtAl2015} but also in models tailored to explain astrophysical observation such as the intense gamma-ray emission observed in the Galactic centre by \textit{Fermi}-LAT \cite{ChoquetteEtAl2016} or the PAMELA/AMS-02 positron flux \cite{FinkbeinerEtAl2010,LiuEtAl2013,DingEtAl2021}. It is in general an expected feature of most models involving multi-TeV WIMPs \cite{Hisano2003,HisanoEtAl2005}, for example in minimal models built from electroweak $n$-uplets \cite{BottaroEtAl2022}. In this case, DM annihilation can proceed through $s$-wave processes. There are also other models based on dark-sector extensions of the standard model (with a dark sector possibly secluded from the visible one), in which DM particles can in principle have masses down to tens of GeV \cite{HambyeEtAl2020a,DasEtAl2017}. For example, for fermionic DM endowed with dark scalar (self-)interactions, $p$-wave annihilation is a natural outcome, providing a phenomenological setup that can therefore be tested through indirect searches from the targets we have investigated here. Our results, however, cannot always be {\it directly} applied to any specific particle-physics model, as one would first need to map that model onto our simplified parameter space (which only contains three parameters: $\alpha_{\rm D}$, $m_\chi$ and $m_\phi$). Our results still provide decent estimates in particular when there is only one light mediator at play for the self-interactions. In turn, they only provide order-of-magnitude estimates in the presence of more complex dark sectors, involving for instance several mediators, provided one can identify a very few dominant interaction(s). Irrespective of any mapping to specific models, we still expect our main result (\ie, very large subhalo boost factors and a model-dependent target hierarchy) to hold for any model featuring a similar velocity dependence of the annihilation cross-section.

\section{Summary and conclusions}
\label{sec:conclusion}

In this work, we have performed a comprehensive study of various classes of astrophysical targets (almost always discussed separately in the literature) for velocity-dependent WIMP annihilations. In particular, we computed their astrophysical $J$-factors entering in the computation of the annihilation flux and ranked them under different velocity-dependent scenarios. The main novelties of our generalised $J$-factor calculations are the following: (i) in addition to the `standard' $s$-wave annihilation case, we also considered Sommerfeld-enhanced $p$-wave annihilation; (ii) for all classes of considered targets --- dSphs, dIrrs, galaxy clusters, and even the MW --- , we self-consistently derived the phase-space distribution function (PSDF) from host halo DM density profiles, accounting for annihilation boost factors from DM substructures (whose PSDFs were also self-consistently derived). It must be noted that very few studies carried out the calculation of the boost for dSphs in this velocity-dependent context, almost none did for galaxy clusters, and this is the first work where this was discussed for dIrrs.

The two most important and probably surprising results, obtained from a case study focusing on a few selected targets among the many available dSphs, dIrrs, and galaxy clusters, are the following. First, substructure boost factors can reach several orders of magnitude on-resonance for $s$-wave annihilation, and also in the Coulomb regime for both $s$- and $p$- wave annihilation; these large boost values are even present for dSphs. All these results are supported and cross-checked by analytical formulae derived (in various regime of the Sommerfeld enhancement) in a companion paper \cite{CompanionPaper}. Second, the standard hierarchy of the most promising classes of targets for indirect DM searches (where, typically, dSphs rank first) can be drastically modified in the presence of both velocity-dependent annihilation and substructure boost. The most striking case is for $s$-wave on resonances and for $p$-wave in the no-Sommerfeld enhancement regime, where galaxy clusters can outshine all other classes of targets. This is a robust result: only in the no-Sommerfeld-enhancement regime we found that uncertainties in the modelling of the DM distribution may significantly impact the resulting ranking between dSphs, dIrrs, and galaxy clusters. We find that uncertainties related to velocity anisotropies are less important than those related to mass modelling.
We stress that the modelling of tidal interactions experienced by subhalos, which is usually a source of large uncertainties, can be safely ignored here. This is because the boost factor is essentially set by subhalos in the outer regions of the targets which are not subject to strong tidal fields.

In this analysis, we have also inspected the spatial morphology of the velocity-dependent annihilation signal from various targets. We showed that subhalos could significantly enlarge the angular size of the signal (up to 10 times for galaxy clusters in the Sommerfeld-enhancement regime). This provides both prospects to identify the underlying particle physics model (if DM-induced $\gamma$-rays are observed in several target classes), but it also brings complications in doing so owing to possible degeneracies between the particle physics parameter space and the many still uncertain DM distribution properties (subhalos, velocity). We have also investigated the signal contrast between the considered targets and the MW DM foreground that is present along the line of sight. The self-consistent calculation of the MW signal showed that the presence of subhalos strongly boosts the signal in both $s$-wave and $p$-wave annihilation (except in the no-Sommerfeld-enhancement regime): accounting for these subhalos is critical on-resonance in all directions (Galactic centre or anticentre), and is also critical especially towards the anticentre in the Coulomb regime.
Though we found a significant number of our targets to exhibit fluxes above the MW foreground, their specific contrast possesses a non-trivial dependence on the signal integration angle and on the considered particle-physics model. Indeed, in some of the studied velocity-dependent regimes and integration angles, some of our targets appear well below the level of the diffuse Galactic DM signal, this way very likely complicating a potential $\gamma$-ray data analysis (typically focused and optimized for point-like sources). 
On the other hand, a calculation of the extragalactic diffuse DM component, similarly to what we did for the Galactic signal, may prove necessary in some specific regimes for which targets outshine all the other ones and the MW foreground. 
Overall, our results show that the analysis strategy (targets and signal regions) should probably be adapted specifically for each of these different particle-physics regimes, in order to optimally search for (or to set constraints on) DM with $\gamma$-ray telescopes. These refinements would complement the toolbox of existing strategies advocated and already explored in the literature to best track down DM signals in $\gamma$ rays (e.g., stacking of a large number of dSphs or galaxy clusters; joint analyses between \textit{Fermi}-LAT and ground-based instruments; optimized search of extended sources; etc.). For velocity-dependent annihilation cross sections, and given the hierarchy dependence of the ranking of targets on the particle physics model, a combined analysis of DM signals from different classes of targets may actually be the most optimal way to provide the most robust and consistent constraints on DM candidates.

This study is especially relevant for most models involving mediators much lighter than the DM particle, from minimal extensions to the Standard Model to complex dark sectors. This is especially important as indirect searches are moving toward multi-TeV DM masses with the advent of CTA. However, rather than covering specific examples, we have studied a simplified setup which allowed us to derive results that should qualitatively apply to a wide range of models.

In a forthcoming effort, we plan to generalise our calculations to a larger list of astrophysical targets and confront our predictions to existing $\gamma$-ray data, also focusing on developing and applying the best data analysis strategy. This should allow us to provide the most robust, up-to-date, and competitive DM limits on generic velocity-dependent annihilation models and associated DM particle candidates.

\acknowledgments
We acknowledge financial support by the CNRS-INSU programs PNHE and PNCG, the {\em GaDaMa} ANR project (ANR-18-CE31-0006), the European Union's Horizon 2020 research and innovation program under the Marie Sk\l{}odowska-Curie grant agreement N$^{\rm o}$~860881-HIDDen. MS acknowlegdes support from Université Savoie Mont Blanc and the PNHE through the AO INSU 2019, grant ``DMSubG" (PI: F. Calore).  GF acknowledges support of the ARC program of the Federation Wallonie-Bruxelles and of the Excellence of Science (EoS) project No. 30820817 - be.h “The H boson gateway to physics beyond the Standard Model”. JPR work is supported by grant SEV-2016-0597-17-2 funded by MCIN/AEI/10.13039/501100011033 and ``ESF Investing in your future''. MASC was also supported by the {\it Atracci\'on de Talento} contracts no. 2016-T1/TIC-1542 and 2020-5A/TIC-19725 granted by the Comunidad de Madrid in Spain. The work of JPR and MASC was additionally supported by the grants PGC2018-095161-B-I00 and CEX2020-001007-S, both funded by MCIN/AEI/10.13039/501100011033 and by ``ERDF A way of making Europe''.

\appendix

\section{Uncertainties related to phase-space modelling}
\label{app:phase_space_modelling}

We complement \citesec{sec:uncertainties_phase_space_smooth} with a discussion of the uncertainty on the generalised $J$-factors coming from the phase-space model itself. The phase space of the DM in our selected targets is modelled assuming equilibrium and spherical symmetry of the halo as well as isotropy of velocities. Under these assumptions, the PSDF of DM particles is given by the well-known Eddington formula
\cite{Eddington1916,BinneyTremaine2008}:
\begin{equation}
\label{Eddington_formula_app}
f(\mathcal{E}) = \dfrac{1}{\sqrt{8}\pi^{2}} \left[ \dfrac{1}{\sqrt{\mathcal{E}}} \left( \dfrac{\mathrm{d}\rho_\chi}{\mathrm{d}\Psi} \right)_{\Psi=0} + \int_{0}^{\mathcal{E}} \! \dfrac{\mathrm{d}^{2}\rho_\chi}{\mathrm{d}\Psi^{2}} \, \dfrac{\mathrm{d}\Psi}{\sqrt{\mathcal{E} - \Psi}}    \right]\,, 
\end{equation}
where $\Psi(r) = \Phi(R_{\mathrm{max}}) - \Phi(r)$ is the (positive-defined) gravitational potential and $\mathcal{E}=\Psi-v^2/2$ the energy. The choice of the radial boundary $R_{\rm max}$ can have a large impact on the phase space in the outer parts of the halo \cite{LacroixEtAl2018}, however the annihilation signal is essentially set by the central region thus we can take $R_{\rm max}\rightarrow\infty$. 

The anisotropy of the DM velocity distribution is, for all intents and purposes, unconstrained in all the gravitational systems we consider. It is therefore important to go beyond the minimal assumption of isotropy and explore the associated uncertainty on predictions of DM-induced $\gamma$-ray fluxes.  In general, the anisotropy of a given component in a gravitational system is quantified by the parameter \cite{Binney1980}
\begin{equation}
\label{eq:beta}
\beta(r) = 1 - \frac{\sigma_{\theta}^2 + \sigma_{\phi}^2}{2\sigma_{r}^2}\,,
\end{equation}
with $\sigma_{r}$, $\sigma_{\theta}$, and $\sigma_{\phi}$ the velocity dispersions in spherical coordinates. We consider the following ansatz for an anisotropic PSDF --- this ansatz makes it possible to obtain a semi-analytic extension of the Eddington formula to an anisotropy profile defined by three parameters, namely the asymptotic values $\beta_{0}$ and $\beta_{\infty}$ at the centre and the outskirts of the galaxy, respectively, and a characteristic angular momentum $L_{0}$ that sets the transition radius between both regimes \cite{WojtakEtAl2008} ---:
\begin{equation}
\label{eq:df_anis}
F(\mathcal{E},L) = f_{\mathcal{E}}(\mathcal{E}) \left( 1 + \dfrac{L^2}{2L_{0}^2} \right)^{-\beta_{\infty}+\beta_{0}} L^{-2\beta_0}\,.
\end{equation}
Constant anisotropy models, for which $\beta(r) = \beta_{0} = \beta_{\infty}$ and 
\begin{equation}
F(\mathcal{E},L) \equiv f_{\beta_{0}}(\mathcal{E},L) = f_{\mathcal{E}}(\mathcal{E}) L^{-2\beta_{0}}\,,
\label{eq:df_constant_beta}
\end{equation}
are a subset of the varying anisotropy models of Eq.~\eqref{eq:df_anis}, and have been extensively discussed in the literature (e.g.~\cite{Henon1973,KentEtAl1982,Cuddeford1991,EvansEtAl2005,BinneyTremaine2008}). 

For all the targets in our sample, for $s$-wave annihilation the relative differences between anisotropic phase-space models with constant negative anisotropy or radius-dependent positive anisotropy and the isotropic (Eddington) result are shown in Table~\ref{tab:PSDF_uncertainties} for the Coulomb regime and the resonances. The value of $J_{\mathrm{S},s}$ for negative (positive) anisotropy is systematically larger (smaller) than the isotropic result by a few tens of \%. 

\begin{table}[t!]
\begin{center}
\begin{tabular}{|c|c|c|c|c|}
\hline
\multirow{2}{*}{Target class} & \multicolumn{2}{c|}{$\dfrac{J_{\mathrm{S},s}^{\beta=-0.5}-J_{\mathrm{S},s}^{\beta=0}}{J_{\mathrm{S},s}^{\beta=0}}$}   & \multicolumn{2}{c|}{$\dfrac{J_{\mathrm{S},s}^{\beta(r)}-J_{\mathrm{S},s}^{\beta=0}}{J_{\mathrm{S},s}^{\beta=0}}$} \\   
\cline{2-5}
&  Coulomb  & Resonances     &  Coulomb  & Resonances \\
\hline
dSphs & $+30\%$ & $+100\%$ & $-10\%$ & $-20\%$ \\
\hline
dIrrs & $+10\%$ & $+40\%$ & -- & -- \\
 \hline
Clusters & $+20\%$ & $+50\%$ & $-10\%$ & $-30\%$ \\
 \hline
\end{tabular}
\caption{\label{tab:PSDF_uncertainties}Summary table of the relative difference between generalised $J$-factors obtained with anisotropic PSDFs and the benchmark Eddington (isotropic) case for each class of targets --- both in the Coulomb regime and on top of resonances in the saturation regime. Sommerfeld enhancement factors used to derive the $J$-factors were computed for $\alpha_{\rm D} = 10^{-2}$.}
\end{center}
\end{table}

For $p$-wave annihilation, the dependence of $J_{\mathrm{S},p}$ on velocity is the same as for $s$-wave in the Coulomb regime, so the uncertainty from the unknown anisotropy is the same in both cases. However, in the saturation regime the dependence on velocity is different from the $s$-wave case. More specifically, there is no dependence on the velocity distribution at resonances, while in between resonances and in the high-$\epsilon_{\phi}$ regime in which there is no Sommerfeld enhancement --- in both cases $J_{\mathrm{S},p} \propto \left\langle v_{\rm rel}^{2} \right\rangle$ --- the relative difference between anisotropic and isotropic models is $\sim 10-20\%$.

Uncertainties from the phase-space model itself are therefore negligible with respect to uncertainties from mass modelling discussed in \citesec{sec:uncertainties_phase_space_smooth}, and they do not affect in any sizable way either the values of the generalised $J$-factors or the ranking of targets. However, as discussed in Sec.~\ref{sec:subhalo_boost_and_classification}, the DM substructure boost has a much more significant impact on generalised $J$-factors, and is the main source of theoretical uncertainties.

\section{The SL17 subhalo population model}
\label{app:SL17}

This section gives a short introduction on the SL17 subhalo population model \cite{StrefEtAl2017}, which was built to address DM searches in general (including the search for subhalos themselves), as flexible as possible to be applied to a diversity of DM candidates and to easily account for changes in relevant cosmological parameters---see also \citerefs{Huetten2019,StrefEtAl2019,FacchinettiEtAl2020,FacchinettiEtAl2022}. This model is analytical in its formulation, but semi-analytical in practice (numerical integrations, iterations, or interpolations are necessary). Assuming a global host halo profile $\rho_{\rm host}$, the model predicts how DM distributes itself between a smooth component and a subhalo component, depending on the host halo profile and its baryonic component. To allow for fast semi-analytical calculations, the model is based on three main approximations: (i) spherically symmetric DM components, (ii) circular orbits for subhalos (i.e. positions are defined by radial distances to the host's centre), (iii) subhalos are independent from each other. The model also includes gravitational tides as sourced by the different components of the host structure (including baryons), which induce a spatial dependence of the subhalo properties and makes the subhalo population model specific to all sorts of host halos (in particular to hosts constrained by observational data). The obtained subhalo distribution is not a fit extrapolated from cosmological simulations, and can consistently cover an arbitrary subhalo mass range. Still, the model allows to qualitatively recover and understand results of cosmological simulations, like the non-trivial spatial distribution of subhalos and the spatial evolution of their structural properties (e.g. ``antibiased'' spatial distributions, spatial dependence of the concentration and mass functions, etc. \cite{DiemandEtAl2004,DiemandEtAl2007a,SpringelEtAl2008,PieriEtAl2011,Benson2017,MolineEtAl2017}). Other analytical or semi-analytical approaches to subhalo population models developed in a broader range of contexts can be found in e.g. \citerefs{BerezinskyEtAl2003,GiocoliEtAl2008,JiangEtAl2016,AndoEtAl2019,IshiyamaEtAl2020a}. Monte-Carlo codes can also be used, e.g. \citerefs{CharbonnierEtAl2012,Benson2012}.

The SL17 model depicts the subhalo population of an arbitrary host halo through probability density functions (PDFs). The subhalo population is bound to be a component of the host halo, consistently with \citeeq{eq:rhohost}. Assuming that subhalos are all independent, the full population can be described with three parameters only: their virial (cosmological) mass $m=M_{\rm 200}$ and concentration $c=c_{\rm 200}$, and their distance to the centre of the host $R$. If subhalos were hard spheres, they would simply track the overall DM potential in the host, as ``particles'' do in $N$-body simulations. Thus, their population could be described by a separable parametric phase-space PDF, because the mass and concentration distributions would not depend on position. This is the starting point of the model, which assumes that the host halo builds up from the aggregation of hard spheres endowed with position, mass and concentration PDFs. The global initial PDF can be written as 
\ben
\label{eq:psub_ini}
\frac{\dd^5{\cal P}_{\rm sub}^{\rm ini}(m,c,R)}{\dd^3\vec{R} \,\dd m \,\dd c} = \frac{\dd^3{\cal P}_V(R)}{\dd^3\vec{R}}\times\frac{\dd {\cal P}_m(m)}{\dd m}\times\frac{\dd {\cal P}_c(c,m)}{\dd c}\,,
\een
i.e. a product of separable PDFs. The number of subhalos before tidal stripping is turned on can be predicted from first principles in a given cosmological framework, as will be explained below. Then, when tidal effects are plugged in, part of the DM initially confined into subhalos is redistributed as a smooth component. This tides intimately depend on the detailed distributions of the various components of the host halo. This induces a mass loss for subhalos, whose efficiency is position-dependent. The spatial dependence of tidal losses translates into a spatial dependence in the subhalo mass function. When tidal effects become disruptive, this turns into a selection in concentration space (more concentrated objects are more resilient to tides), which also implies a spatial dependence in the concentration function. Tidal effects are generically more efficient in the central parts of the host, where the subhalo number density strongly flattens up to almost full depletion close to the centre, depending on disruption criteria. Eventually, the final global PDF of subhalos is fully intricate and not separable anymore, due to mass losses and disruption. It can formally be written as
\ben
\label{eq:psub_fin}
\frac{\dd^5{\cal P}_{\rm sub}(m,c,R)}{\dd^3\vec{R} \,\dd m \,\dd c} = \frac{1}{K_{\rm tidal}}\,\frac{\dd^3{\cal P}_V(R)}{\dd^3\vec{R}}\times\frac{\dd {\cal P}_m(m)}{\dd m}\times\frac{\dd {\cal P}_c(c,m)}{\dd c}\times {\cal T}(m,c,R)\,,
\een
where $K_{\rm tidal}\leqslant 1$ ensures the correct normalisation of the global PDF, and ${\cal T}(m,c,R)$ symbolically encodes tidal stripping and disruption. It typically assigns a tidal radius $r_{\rm t}(m,c,R)$ to a subhalo of virial mass $m$, concentration $c$, and position $R$, given the properties of the host halo components; or it moves the subhalo DM to the smooth component if disrupted.

There is no well-defined way to decide whether a subhalo should be tidally disrupted. It could actually be that a tiny core survives for ever if dense enough and no central collisions with stars occur, simply due to adiabatic protection \cite{Weinberg1994,GnedinEtAl1999}. A practical criterion can still be inspired from studies of cosmological simulations \cite{HayashiEtAl2003}, where it was found that fixing a lower threshold to $x_{\rm t}=r_{\rm t}/r_{\rm s}$, where $r_{\rm s}$ is the scale radius of the structure, was a way to efficiently capturing tidal disruption. We define this threshold as $\epsilon_{\rm t}$. Initially found around $\sim 1$ \cite{HayashiEtAl2003}, it was realized more recently that numerical artifacts could strongly bias these early estimates \cite{vandenBoschEtAl2018a,GreenEtAl2021}, and that one could expect values for $\epsilon_{\rm t}$ as low as $0.01$ or even less, which is consistent with the argument given just above. For definiteness, we use two types of tidal disruption criteria:
\ben
\epsilon_{\rm t} =
\begin{cases}
  1\;\;\;\;\;\; & \text{(fragile subhalos)}\\
  0.01\;\;\;\;\;\; & \text{(resilient subhalos).}
\end{cases}
\een

We can now give a few details about the PDFs introduced above (see \citerefs{StrefEtAl2017,FacchinettiEtAl2022} for an exhaustive presentation). For the initial spatial PDF, we simply assume $\dd{\cal P}_V/dV = \rho_{\rm host}(R)/M_{\rm host}$, where $M_{\rm host}$ is the total DM mass of the host. Note that the final spatial distribution strongly departs from the initial one after tidal effects are activated. For the concentration PDF, we use a log-normal distribution with $\sigma_c^{\rm dec}=0.14$ \cite{BullockEtAl2001b,MaccioEtAl2008,PradaEtAl2012,DuttonEtAl2014} centreed about the mass-relation concentration $c(m)$ given in \citeref{Sanchez-CondeEtAl2014}.

In order to avoid a calibration of the mass function on simulations, as was initially done in the SL17 model to determine the total number of substructures, we have implemented a cosmological mass function from first principles instead. This allows us to potentially change the cosmological parameters or the primordial power spectrum of density fluctuations. The procedure follows previous studies \cite{GiocoliEtAl2008,Benson2012,JiangEtAl2016,vandenBoschEtAl2016,HanEtAl2016b,Benson2017}, which extracted the subhalo cosmological mass function from merger-tree algorithms \cite{SomervilleEtAl1999a,ParkinsonEtAl2008}, built upon the excursion set theory of structure formation \cite{BondEtAl1991a,LaceyEtAl1993,ColeEtAl2000}. We reproduce the same procedure with the merger tree introduced in \citeref{ParkinsonEtAl2008} and we recover that the mass function can be well fitted by \cite{GiocoliEtAl2008,JiangEtAl2016}
\ben
\label{eq:mass_function}
\frac{ \dd N(m,M_{\rm host})}{\dd  m} = \frac{1}{M_{\rm host}} \left[ \sum_{i =1,2}\gamma_i\left(\frac{m}{M_{\rm host}}\right)^{-\alpha_i}  \right] \exp\left\{ -\beta \left(\frac{m}{M_{\rm host}}\right)^{\zeta} \right\}\,.
\een
We find $\gamma_1 = 0.014$, $\gamma_2 = 0.41$, $\alpha_1 = 1.965$, $\alpha_2 = 0.57$, $\beta = 20$, $\zeta = 3.4$, roughly independent of the cosmology and of the host mass---still, we used the cosmological parameters from the latest Planck analysis \cite{PlanckCollab2020}. Contrarily to previous works (only interested in large masses) in this fit we also constrain the low mass part of the spectrum \cite{Facchinetti2021,FacchinettiEtAl2022}. Note that the above mass function is close to a power law in mass $\propto m^{-\alpha}$ with a spectral index $\alpha\simeq \alpha_1$.

The total number of subhalos {\em before tidal stripping} can be determined by integrating the mass function in the specified subhalo mass range, $[m_{\rm min},m_{\rm max}]$
\ben
N_{\rm tot}^{\rm ini} =   \int_{m_{\rm min}}^{m_{\rm max}}\dd m\, \frac{ \dd N(m,M_{\rm host})}{\dd  m } \, ,
\een
such that the total number of subhalos, after tidal disruption effects are plugged, is given by
\ben
N_{\rm tot} = K_{\rm tidal}\times N_{\rm tot}^{\rm ini} \,,
\een
where $K_{\rm tidal}\leqslant 1$ is the normalisation constant introduced in \citeeq{eq:psub_fin}. In practice, we use $m_{\rm max}=10^{-2}M_{\rm host}$, and $m_{\rm min}=10^{-10}\Msun$, unless specified otherwise.

Finally, note that the SL17 model can be used with any assumption for the inner subhalo profiles. In this paper, we use NFW profiles for subhalos, whose properties are completely specified by the virial mass and concentration.

\section{Numerical calculation of the subhalo boost}
\label{app:subhalo_boost}

We present some technical details related to the calculation of the quantity
\begin{eqnarray}
\underline{\rho}^2_{\rm S,tot} = \underline{\rho}^2_{\rm S,sub} + \rho_{\rm S,sm}^2 + 2\,\rho_{\rm S,sm}\,\rho_{\rm sub}.
\end{eqnarray}
First note that the smooth contribution in the presence of Sommerfed enhancement is
\begin{eqnarray}
\rho_{\rm S,sm}(r) = \rho_{\rm sm}(r)\,\left<\overline{\cal S}\right>(r)
\end{eqnarray}
where $\left<\right>$ is the average over the relative velocity distribution of the host halo as defined in Eq.~(\ref{eq:average_O_vrel}). This, combined with the average subhalo density defined in Eq.~(\ref{eq:defrhosub}), also enables the calculation of the cross term $2\,\rho_{\rm S,sm}\,\rho_{\rm sub}$.

To compute the contribution of subhalos, we need to evaluate 
\ben
\xi_{\rm S,t}(m,c,r) = \int_{x\leqslant x_{\rm t}(r,m,c)}\mathrm{d}^3\vec{x}\,\left(\frac{\rho(x) }{\rho_{\circledast}}\right)^2\left<\overline{{\cal S}}\left(\dfrac{v_{\rm rel}}{2}\right)\right>(x)
\een
where the average is now taken over the velocity distribution of the subhalo with parameters $(m,c,R)$.
We then need to perform the average over $m$ and $c$ to get $\underline{\rho^2_{\rm S,sub}}$ as shown in Eq.~(\ref{eq:rho2sub}).
This last step turns out to be very computationally expensive, because an integral over $\vec{v}_{\rm rel}$ has to be performed for each subhalo mass $m$ and concentration $c$.
To speed up the calculation, we instead rely on the following approximations:
\ben
\begin{array}{lc}
\left<\overline{{\cal S}}\left(\dfrac{v_{\rm rel}}{2}\right)\right>(x) \simeq \overline{{\cal S}}\left(\dfrac{\left<v_{\rm rel}^{-2}\right>^{-1/2}(x)}{2}\right) \quad&\quad \mathrm{for}\ s \mathrm{-wave}, \\
\left<\overline{{\cal S}}\left(\dfrac{v_{\rm rel}}{2}\right)\right>(x) \simeq \overline{{\cal S}}\left(\dfrac{\left<v_{\rm rel}^2\right>^{1/2}(x)}{2}\right) \quad&\quad \mathrm{for}\ p\mathrm{-wave}.
\end{array}
\een
These approximated expressions are much faster to compute, because the velocity moments have a simple scaling with the subhalo structural parameters $\left<v_{\rm rel}^p\right>\propto (\rho_{0}\,r_{\rm s}^2)^{p/2}$.
We compute the PSDF $F^{\rm sub}_{\rm rel}$ and the velocity moments using the Eddington inversion method outlined in App.~\ref{app:phase_space_modelling}. We have checked that our approximations introduce an error of at most 30$\%$ for some specific values of the subhalo mass and the $\epsilon_\phi$ parameter, and that the accuracy is better than 10$\%$ in most of the parameter space.

\section{Dependence of $J_{\rm S}^{\rm target}/J_{\rm S}^{\rm MW}$ on $\theta_{\rm int}$ and $\epsilon_\phi$}
\label{app:MWcontrast}

In this paper, we chose not to assess the impact of the MW signal on the detectability of the targets considered. Nevertheless, we briefly illustrate in this Appendix what would be the optimal integration angle $\theta_{\rm int}$ so as to have $J_{\rm S}^{\rm target}/J_{\rm S}^{\rm MW} >1$, that is a favorable contrast between the target and the MW signals. We also illustrate whether this optimal angle depends or not on the Sommerfeld parameter $\epsilon_\phi$.

\begin{figure}[t!]
\centering
\includegraphics[width=0.49\linewidth]{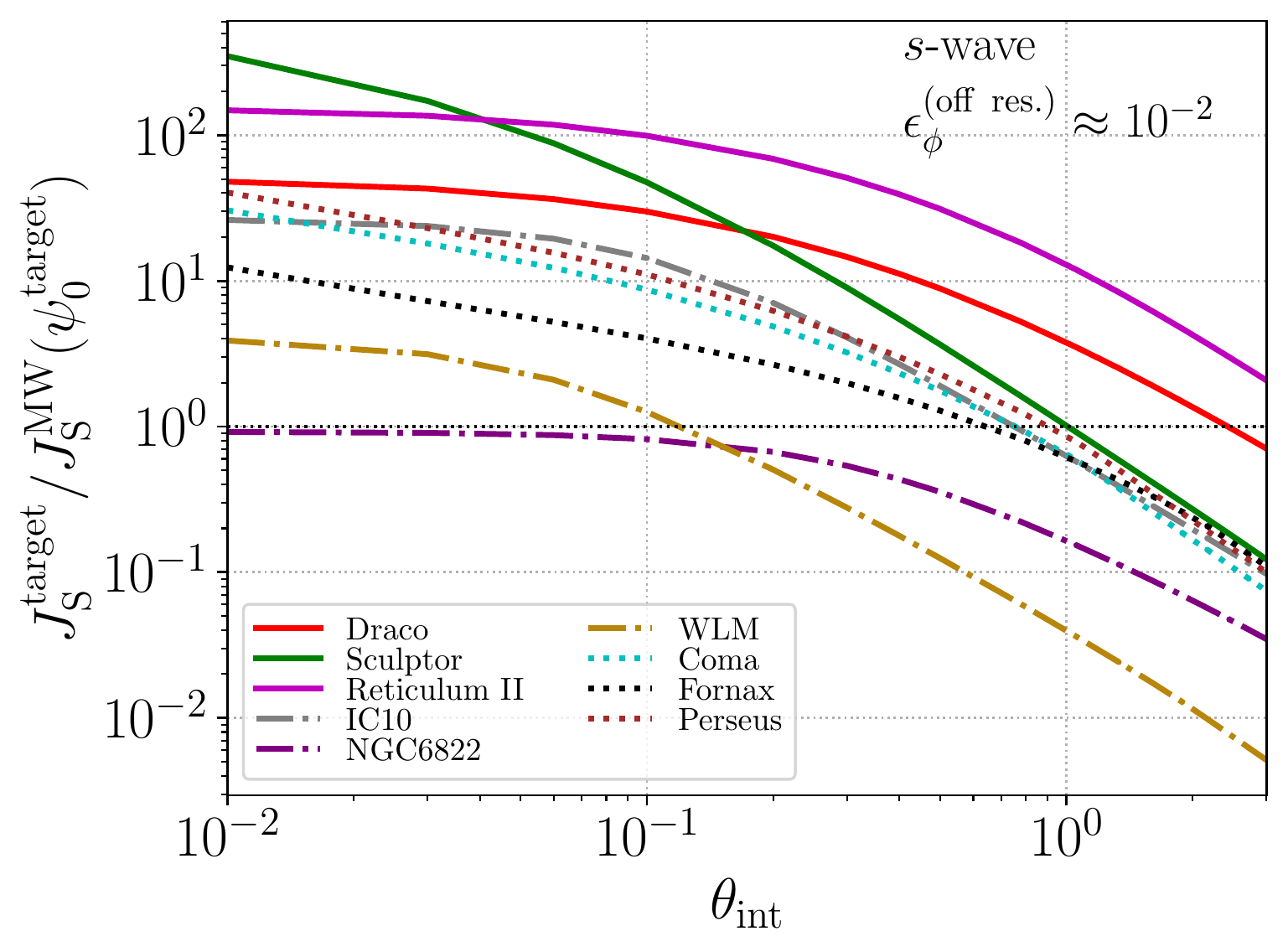} \hfill
\includegraphics[width=0.49\linewidth]{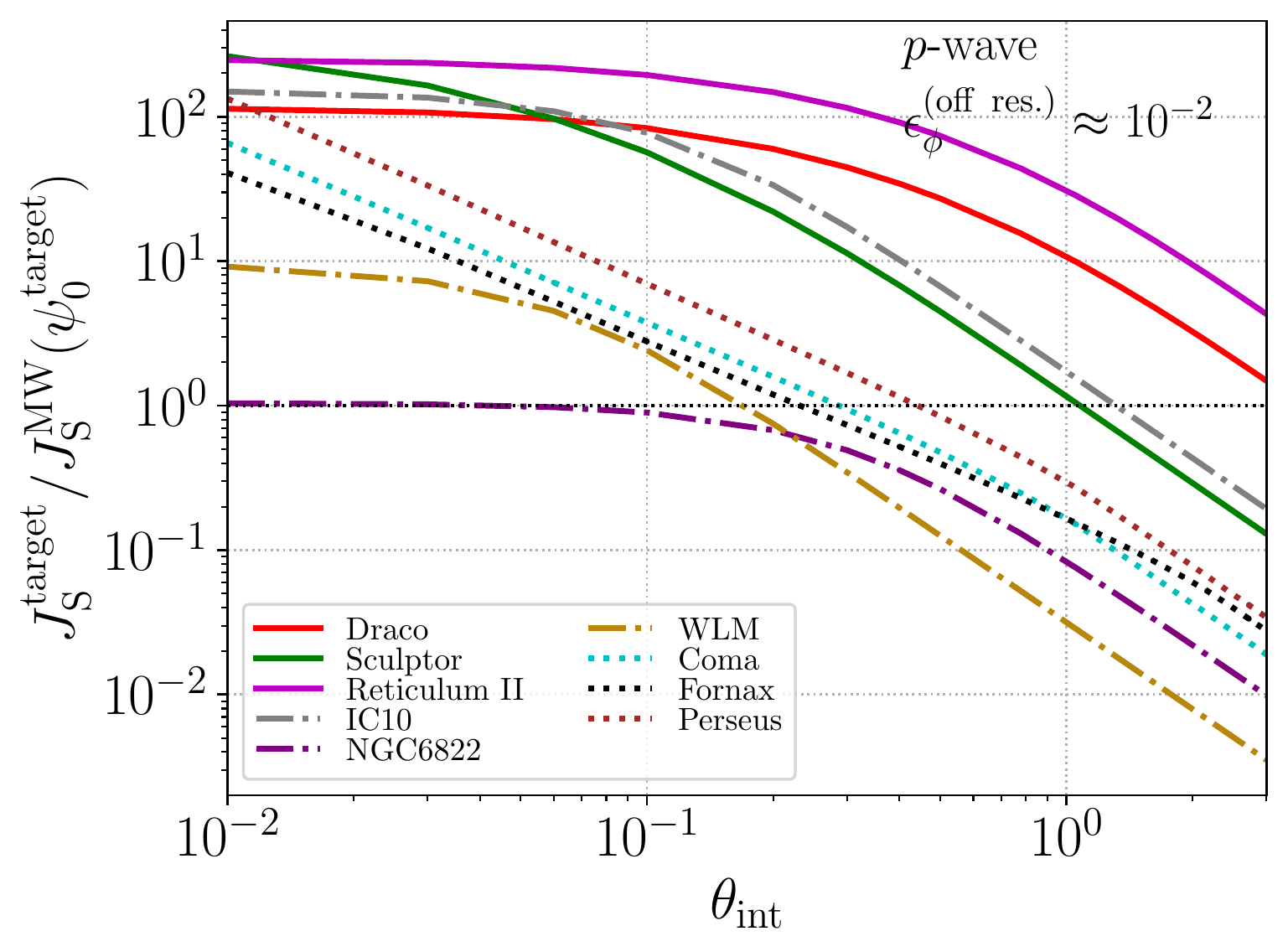} \hfill
\includegraphics[width=0.49\linewidth]{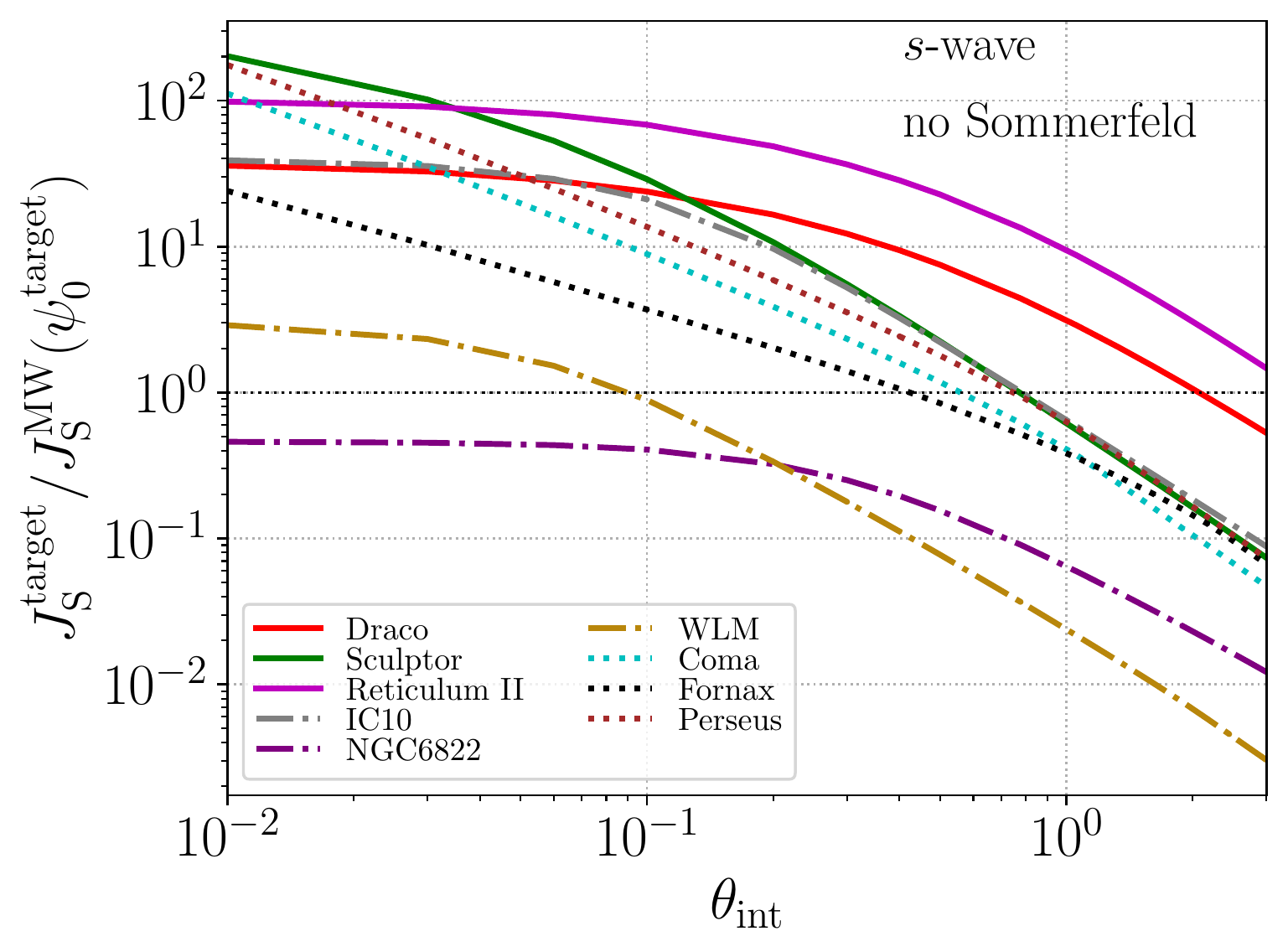} \hfill
\includegraphics[width=0.49\linewidth]{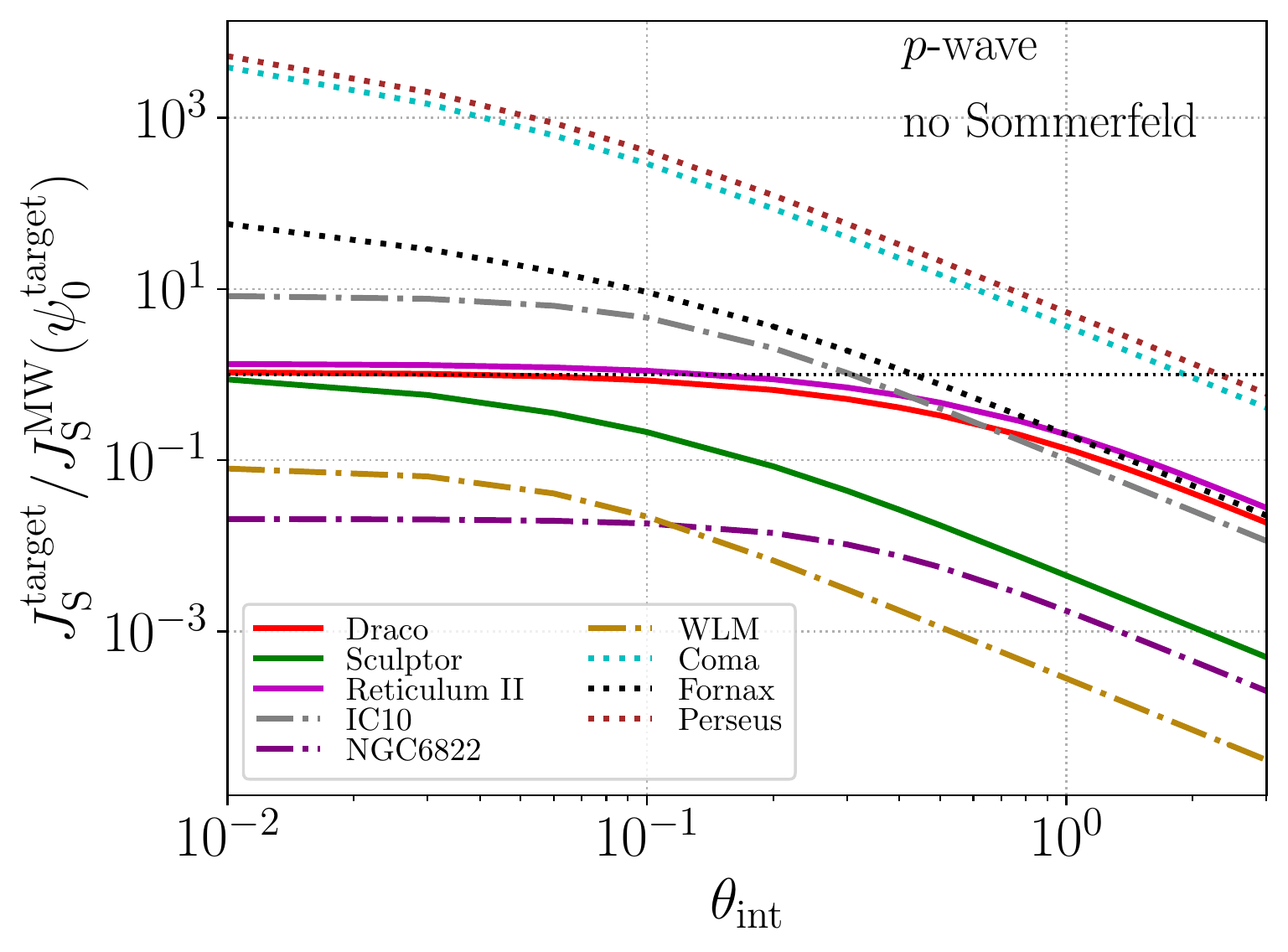}
\caption{Generalised $J$-factor contrast between the targets (color-coded lines) and the MW, $J_{\rm S}^{\rm target}/J_{\rm S}^{\rm MW}$, as a function of the integration angle $\theta_{\rm int}$ around the target position. The three line styles correspond to the three family of targets (solid lines for dSphs, dash-dotted lines for dIrrs, and dotted lines for galaxy clusters). The left (resp. right) panels show the results for $s$-wave (resp. $p$-wave) annihilations, with or without Sommerfeld enhancement (top and bottom panels respectively); the value of the Sommerfeld parameter is fixed to $\epsilon_\phi\approx10^{-2}$ and is chosen off resonance. The black dash-dot-dotted line separates the region between a favorable $(>1)$ and unfavorable ($<1$) contrast. See text for discussion.}
\label{fig:J_S_ratios_vs_fov}
\end{figure}
We start in \citefig{fig:J_S_ratios_vs_fov} with the contrast (between the target and the MW) as a function of the integration angle $\theta_{\rm int}$. All curves for all configurations show the same behavior, i.e., a decreasing contrast with a growing $\theta_{\rm int}$. The most favorable contrast is observed for small integration angles, because the diffuse MW DM foreground $J_{\rm S}^{\rm MW}\propto\theta_{\rm int}^2$, while most of the signal remains located in the central regions of the target so that $J_{\rm S}^{\rm target}$ is independent of $\theta_{\rm int}$ (if $\theta_{\rm int}$ not too small). The steepness or smoothness of the decreases observed are a non-trivial combination of the different structural parameters and relative importance of the substructures (that dominate the signal at large radii). The ordering of the curves (from larger to smaller contrast) follows the ranking established in the main text: at $\theta_{\rm int}=0.5^\circ$, for most configurations shown in the different panels, dSphs (solid lines) reach signal contrasts as high as $\sim 10-50$, but only $\sim 1-5$ for galaxy clusters (dotted lines), whereas dIrrs (dash-dotted lines) are below the MW signal, except for IC10 (dash-dotted grey line). The main difference is for $p$-wave annihilation without Sommerfeld enhancement (bottom right panel), where only galaxy clusters outshine the MW.

\begin{figure}[t!]
\centering
\includegraphics[width=0.49\linewidth]{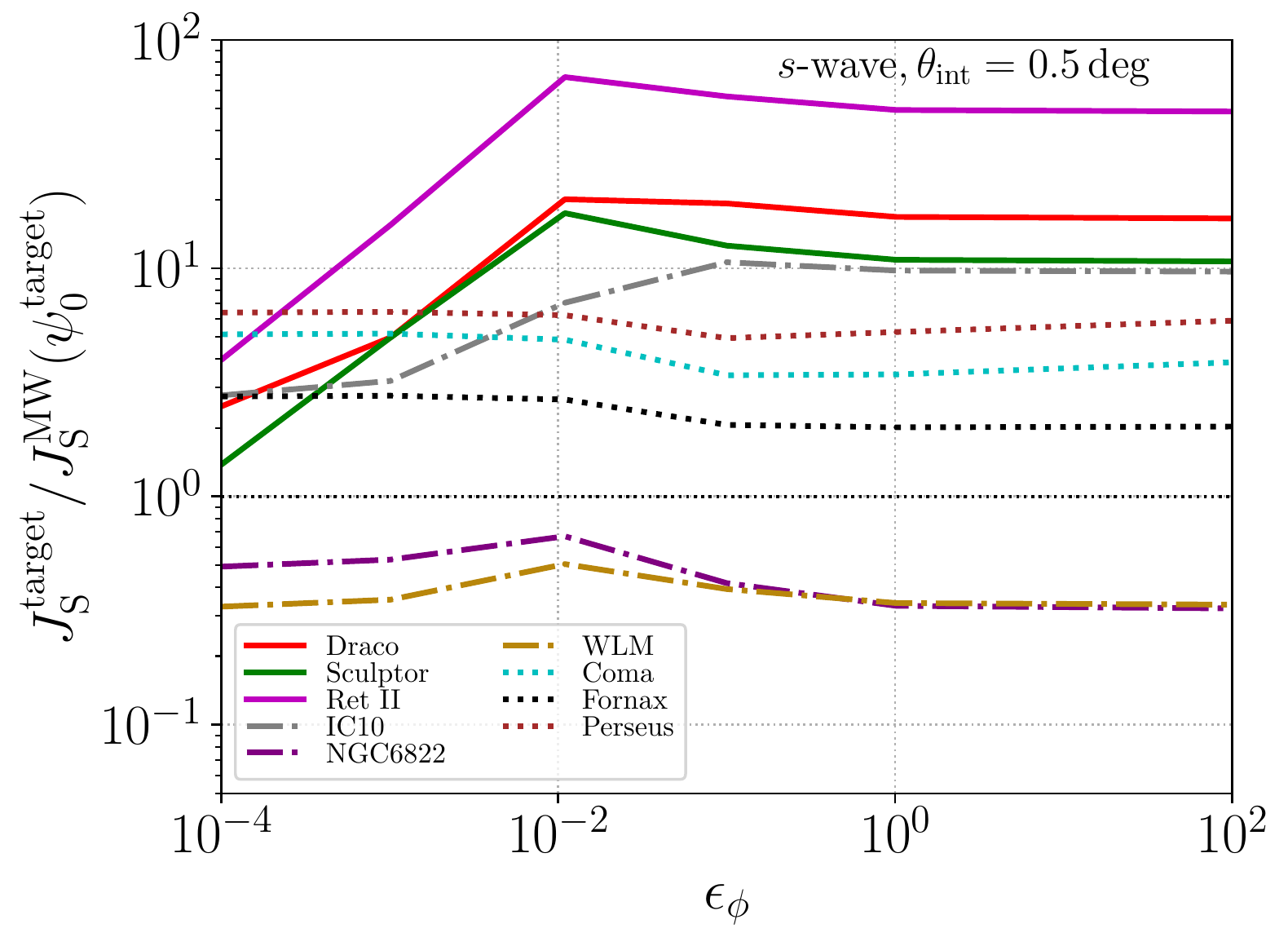} \hfill
\includegraphics[width=0.49\linewidth]{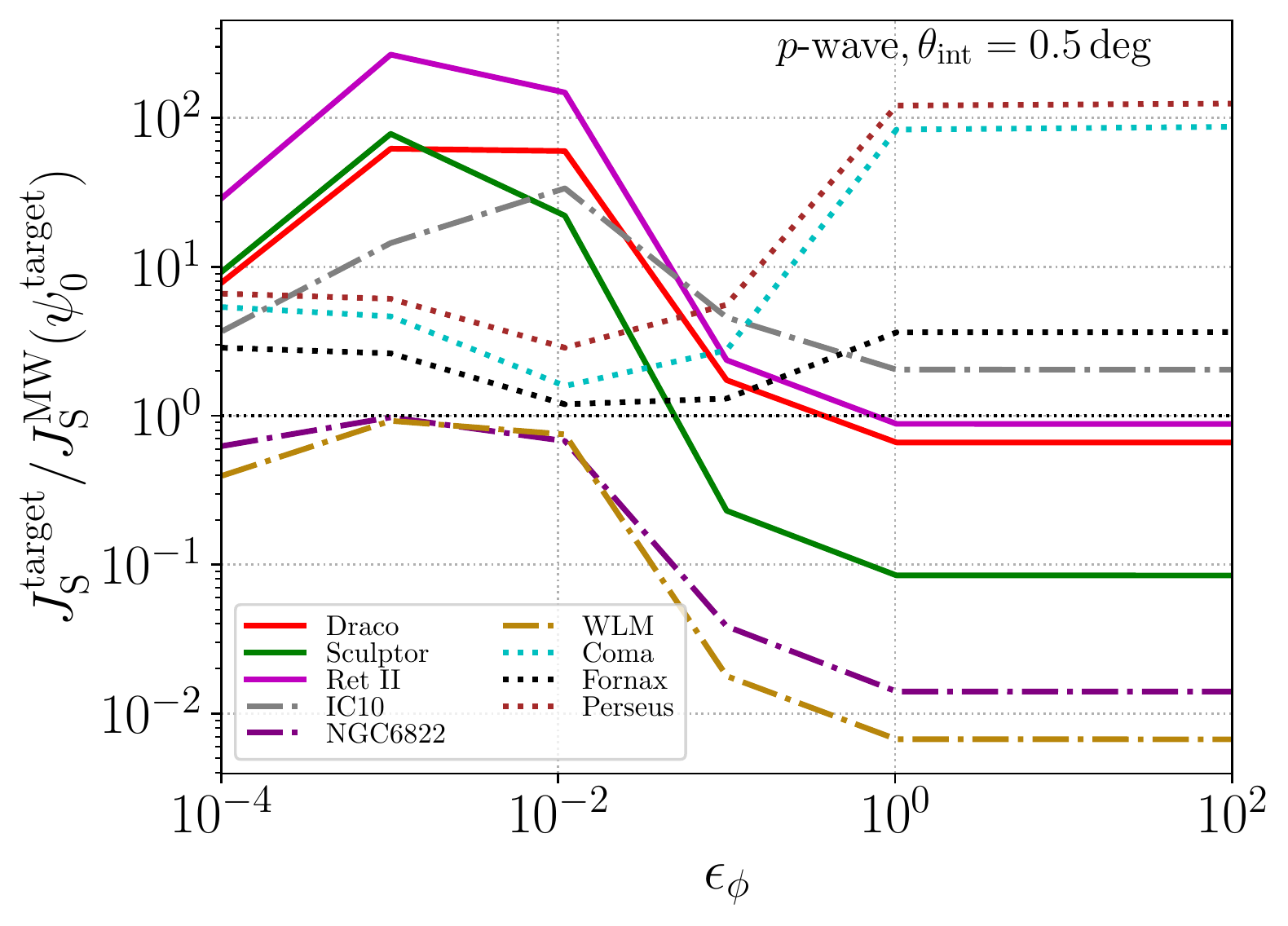}
\caption{Generalised $J$-factor contrast between the targets (color-coded lines) and the MW, $J_{\rm S}^{\rm target}/J_{\rm S}^{\rm MW}$, as a function of the Sommerfeld parameter $\epsilon_\phi$. The three line styles correspond to the three family of targets (solid lines for dSphs, dash-dotted lines for dIrrs, and dotted lines for galaxy clusters). We show  results for $s$-wave (left panel) and $p$-wave (right panel) at a fixed integration angle $\theta_{\rm int}=0.5^\circ$. The black dash-dot-dotted line separates the region between a favorable $(>1)$ and unfavorable ($<1$) contrast. See text for discussion.}
\label{fig:J_S_ratios_vs_eps_phi}
\end{figure}
To know whether these conclusions are generic or not, we show again $J_{\rm S}^{\rm target}/J_{\rm S}^{\rm MW}$ in \citefig{fig:J_S_ratios_vs_eps_phi}, but now as a function of $\epsilon_\phi$ (for $\theta_{\rm int}=0.5^\circ$). We see a strong and complicated dependence that depends on the targets considered and whether $s$-wave (left panel) or $p$-wave (right panel) annihilations are considered. The most dramatic dependence is observed for dSphs (solid lines), for which we can go from favorable ($>1$) to unfavorable ($<1$) contrasts; for the $p$-wave case in particular (right panel), the contrast for Sculptor goes from $\sim 10$ (for $\epsilon_\phi\sim 10^{-3}$) to $\sim 10^{-2}$ (for $\epsilon_\phi>1$).

These results illustrate the fact that devising an optimal data analysis for the various targets considered is not simple. Beside the usual considerations about the instrument characteristics and astrophysical backgrounds, the above figures stress that the optimal signal region is both target- and `particle physics model'-dependent---at least if we wish to consistently analyse the DM signal of some targets, which can possibly lay well below the MW own DM signal \cite{FacchinettiEtAl2020}. This should be kept in mind when searching for DM signals or setting limits on DM candidates from non-detection.

\bibliographystyle{JHEP.bst}
\bibliography{biblio.bib}

\providecommand{\href}[2]{#2}\begingroup\raggedright\begin{thebibliography}{100}

\bibitem{LavalleEtAl2012}
J.~{Lavalle} and P.~{Salati}, \emph{{Dark matter indirect signatures}},
  \href{https://doi.org/10.1016/j.crhy.2012.05.001}{\emph{Comptes Rendus
  Physique} {\bfseries 13} (July, 2012) 740--782},
  [\href{https://arxiv.org/abs/1205.1004}{{\ttfamily 1205.1004}}].

\bibitem{BringmannEtAl2012c}
T.~{Bringmann} and C.~{Weniger}, \emph{{Gamma ray signals from dark matter:
  Concepts, status and prospects}},
  \href{https://doi.org/10.1016/j.dark.2012.10.005}{\emph{Physics of the Dark
  Universe} {\bfseries 1} (Nov., 2012) 194--217},
  [\href{https://arxiv.org/abs/1208.5481}{{\ttfamily 1208.5481}}].

\bibitem{Gaskins2016}
J.~M. {Gaskins}, \emph{{A review of indirect searches for particle dark
  matter}},
  \href{https://doi.org/10.1080/00107514.2016.1175160}{\emph{Contemporary
  Physics} {\bfseries 57} (Oct., 2016) 496--525},
  [\href{https://arxiv.org/abs/1604.00014}{{\ttfamily 1604.00014}}].

\bibitem{DiMauroEtAl2021}
M.~Di~Mauro and M.~W. Winkler, \emph{{Multimessenger constraints on the dark
  matter interpretation of the Fermi-LAT Galactic center excess}},
  \href{https://doi.org/10.1103/PhysRevD.103.123005}{\emph{Phys. Rev. D}
  {\bfseries 103} (2021) 123005},
  [\href{https://arxiv.org/abs/2101.11027}{{\ttfamily 2101.11027}}].

\bibitem{LAT_projections}
E.~Charles, M.~S{\'{a}}nchez-Conde, B.~Anderson, R.~Caputo, A.~Cuoco, M.~D.
  Mauro et~al., \emph{{Sensitivity projections for dark matter searches with
  the Fermi large area telescope}},
  \href{https://doi.org/10.1016/j.physrep.2016.05.001}{\emph{Physics Reports}
  {\bfseries 636} (2016) 1--46}.

\bibitem{fermi_instrument_paper}
{The Fermi-LAT Collaboration} and W.~B. Atwood, \emph{{The Large Area Telescope
  on the Fermi Gamma-ray Space Telescope Mission}},
  \href{https://doi.org/10.1088/0004-637X/697/2/1071}{\emph{APJ} (2009) },
  [\href{https://arxiv.org/abs/0902.1089v1}{{\ttfamily 0902.1089v1}}].

\bibitem{Leane2018}
R.~K. Leane, T.~R. Slatyer, J.~F. Beacom and K.~C.~Y. Ng, \emph{{GeV-scale
  thermal WIMPs: Not even slightly ruled out}},
  \href{https://doi.org/10.1103/PhysRevD.98.023016}{\emph{Phys. Rev.}
  {\bfseries D98} (2018) 023016},
  [\href{https://arxiv.org/abs/1805.10305}{{\ttfamily 1805.10305}}].

\bibitem{2021arXiv211101198D}
M.~{Doro}, M.~A. {S{\'a}nchez-Conde} and M.~{H{\"u}tten}, \emph{{Fundamental
  Physics Searches with IACTs}}, {\emph{arXiv e-prints} (Nov., 2021)
  arXiv:2111.01198}, [\href{https://arxiv.org/abs/2111.01198}{{\ttfamily
  2111.01198}}].

\bibitem{2013APh....43....3A}
{CTA Consortium collaboration}, \emph{{Introducing the CTA concept}},
  \href{https://doi.org/10.1016/j.astropartphys.2013.01.007}{\emph{Astropart.
  Phys.} {\bfseries 43} (Mar., 2013) 3--18}.

\bibitem{CTA_GC_2020}
{The CTA Consortium}, \emph{{Sensitivity of the Cherenkov Telescope Array to a
  dark matter signal from the Galactic centre}},
  \href{https://doi.org/10.1088/1475-7516/2021/01/057}{\emph{Journal of
  Cosmology and Astroparticle Physics} {\bfseries 2021} (Jan, 2021) 057–057}.

\bibitem{AbdallahEtAl2015}
J.~{Abdallah}, H.~{Araujo}, A.~{Arbey}, A.~{Ashkenazi}, A.~{Belyaev},
  J.~{Berger} et~al., \emph{{Simplified Models for Dark Matter Searches at the
  LHC}}, {\emph{ArXiv e-prints} (June, 2015) },
  [\href{https://arxiv.org/abs/1506.03116}{{\ttfamily 1506.03116}}].

\bibitem{ZhaoEtAl2016}
Y.~Zhao, X.-J. Bi, H.-Y. Jia, P.-F. Yin and F.-R. Zhu, \emph{Constraint on the
  velocity dependent dark matter annihilation cross section from fermi-lat
  observations of dwarf galaxies},
  \href{https://doi.org/10.1103/PhysRevD.93.083513}{\emph{\prd} {\bfseries 93}
  (Apr., 2016) 083513}, [\href{https://arxiv.org/abs/1601.02181}{{\ttfamily
  1601.02181}}].

\bibitem{ZhaoEtAl2018}
Y.~{Zhao}, X.-J. {Bi}, P.-F. {Yin} and X.~{Zhang}, \emph{{Constraint on the
  velocity dependent dark matter annihilation cross section from gamma-ray and
  kinematic observations of ultrafaint dwarf galaxies}},
  \href{https://doi.org/10.1103/PhysRevD.97.063013}{\emph{\prd} {\bfseries 97}
  (Mar., 2018) 063013}, [\href{https://arxiv.org/abs/1711.04696}{{\ttfamily
  1711.04696}}].

\bibitem{BoddyEtAl2020}
K.~K. {Boddy}, J.~{Kumar}, A.~B. {Pace}, J.~{Runburg} and L.~E. {Strigari},
  \emph{{Effective J -factors for Milky Way dwarf spheroidal galaxies with
  velocity-dependent annihilation}},
  \href{https://doi.org/10.1103/PhysRevD.102.023029}{\emph{\prd} {\bfseries
  102} (July, 2020) 023029},
  [\href{https://arxiv.org/abs/1909.13197}{{\ttfamily 1909.13197}}].

\bibitem{Arkani-HamedEtAl2009}
N.~{Arkani-Hamed}, D.~P. {Finkbeiner}, T.~R. {Slatyer} and N.~{Weiner},
  \emph{{A theory of dark matter}},
  \href{https://doi.org/10.1103/PhysRevD.79.015014}{\emph{\prd} {\bfseries 79}
  (Jan., 2009) 015014}, [\href{https://arxiv.org/abs/0810.0713}{{\ttfamily
  0810.0713}}].

\bibitem{PospelovEtAl2008}
M.~Pospelov and A.~Ritz, \emph{{Astrophysical Signatures of Secluded Dark
  Matter}}, \href{https://doi.org/10.1016/j.physletb.2008.12.012}{\emph{Phys.
  Lett.} {\bfseries B671} (2009) 391--397},
  [\href{https://arxiv.org/abs/0810.1502}{{\ttfamily 0810.1502}}].

\bibitem{Sommerfeld1931}
A.~{Sommerfeld}, \emph{{{\"U}ber die Beugung und Bremsung der Elektronen}},
  \href{https://doi.org/10.1002/andp.19314030302}{\emph{Annalen der Physik}
  {\bfseries 403} (Jan., 1931) 257--330}.

\bibitem{Hisano2003}
J.~Hisano, S.~Matsumoto and M.~M. Nojiri, \emph{{Explosive dark matter
  annihilation}},
  \href{https://doi.org/10.1103/PhysRevLett.92.031303}{\emph{Phys. Rev. Lett.}
  {\bfseries 92} (2004) 031303},
  [\href{https://arxiv.org/abs/hep-ph/0307216}{{\ttfamily hep-ph/0307216}}].

\bibitem{HisanoEtAl2005}
J.~{Hisano}, S.~{Matsumoto}, M.~M. {Nojiri} and O.~{Saito},
  \emph{{Nonperturbative effect on dark matter annihilation and gamma ray
  signature from the galactic center}},
  \href{https://doi.org/10.1103/PhysRevD.71.063528}{\emph{\prd} {\bfseries 71}
  (Mar., 2005) 063528}, [\href{https://arxiv.org/abs/hep-ph/0412403}{{\ttfamily
  hep-ph/0412403}}].

\bibitem{Profumo2005}
S.~Profumo, \emph{{TeV gamma-rays and the largest masses and annihilation cross
  sections of neutralino dark matter}},
  \href{https://doi.org/10.1103/PhysRevD.72.103521}{\emph{Phys. Rev.}
  {\bfseries D72} (2005) 103521},
  [\href{https://arxiv.org/abs/astro-ph/0508628}{{\ttfamily
  astro-ph/0508628}}].

\bibitem{CirelliEtAl2007}
M.~Cirelli, A.~Strumia and M.~Tamburini, \emph{{Cosmology and Astrophysics of
  Minimal Dark Matter}},
  \href{https://doi.org/10.1016/j.nuclphysb.2007.07.023}{\emph{Nucl. Phys.}
  {\bfseries B787} (2007) 152--175},
  [\href{https://arxiv.org/abs/0706.4071}{{\ttfamily 0706.4071}}].

\bibitem{MarchRussellEtAl2008}
J.~March-Russell, S.~M. West, D.~Cumberbatch and D.~Hooper, \emph{{Heavy Dark
  Matter Through the Higgs Portal}},
  \href{https://doi.org/10.1088/1126-6708/2008/07/058}{\emph{JHEP} {\bfseries
  07} (2008) 058}, [\href{https://arxiv.org/abs/0801.3440}{{\ttfamily
  0801.3440}}].

\bibitem{SilkEtAl1993}
J.~{Silk} and A.~{Stebbins}, \emph{{Clumpy cold dark matter}},
  \href{https://doi.org/10.1086/172846}{\emph{\apj} {\bfseries 411} (July,
  1993) 439--449}.

\bibitem{BergstroemEtAl1999a}
L.~{Bergstr{\"o}m}, J.~{Edsj{\"o}}, P.~{Gondolo} and P.~{Ullio}, \emph{{Clumpy
  neutralino dark matter}},
  \href{https://doi.org/10.1103/PhysRevD.59.043506}{\emph{\prd} {\bfseries 59}
  (Feb., 1999) 043506},
  [\href{https://arxiv.org/abs/astro-ph/9806072}{{\ttfamily
  astro-ph/9806072}}].

\bibitem{CalcaneoRoldanEtAl2000}
C.~Calcaneo-Roldan and B.~Moore, \emph{{The Surface brightness of dark matter:
  Unique signatures of neutralino annihilation in the galactic halo}},
  \href{https://doi.org/10.1103/PhysRevD.62.123005}{\emph{Phys. Rev. D}
  {\bfseries 62} (2000) 123005},
  [\href{https://arxiv.org/abs/astro-ph/0010056}{{\ttfamily
  astro-ph/0010056}}].

\bibitem{BerezinskyEtAl2003}
V.~Berezinsky, V.~Dokuchaev and Y.~Eroshenko, \emph{{Small - scale clumps in
  the galactic halo and dark matter annihilation}},
  \href{https://doi.org/10.1103/PhysRevD.68.103003}{\emph{Phys. Rev.}
  {\bfseries D68} (2003) 103003},
  [\href{https://arxiv.org/abs/astro-ph/0301551}{{\ttfamily
  astro-ph/0301551}}].

\bibitem{StoehrEtAl2003}
F.~Stoehr, S.~D.~M. White, V.~Springel, G.~Tormen and N.~Yoshida, \emph{{Dark
  matter annihilation in the halo of the Milky Way}},
  \href{https://doi.org/10.1046/j.1365-2966.2003.07052.x}{\emph{Mon. Not. Roy.
  Astron. Soc.} {\bfseries 345} (2003) 1313},
  [\href{https://arxiv.org/abs/astro-ph/0307026}{{\ttfamily
  astro-ph/0307026}}].

\bibitem{LavalleEtAl2007}
J.~{Lavalle}, J.~{Pochon}, P.~{Salati} and R.~{Taillet}, \emph{{Clumpiness of
  dark matter and the positron annihilation signal}},
  \href{https://doi.org/10.1051/0004-6361:20065312}{\emph{\aap} {\bfseries 462}
  (Feb., 2007) 827--840},
  [\href{https://arxiv.org/abs/arXiv:astro-ph/0603796}{{\ttfamily
  arXiv:astro-ph/0603796}}].

\bibitem{KuhlenEtAl2008}
M.~{Kuhlen}, J.~{Diemand} and P.~{Madau}, \emph{{The Dark Matter Annihilation
  Signal from Galactic Substructure: Predictions for GLAST}},
  \href{https://doi.org/10.1086/590337}{\emph{\apj} {\bfseries 686} (Oct.,
  2008) 262--278}, [\href{https://arxiv.org/abs/0805.4416}{{\ttfamily
  0805.4416}}].

\bibitem{PieriEtAl2011}
L.~Pieri, J.~Lavalle, G.~Bertone and E.~Branchini, \emph{{Implications of
  High-Resolution Simulations on Indirect Dark Matter Searches}},
  \href{https://doi.org/10.1103/PhysRevD.83.023518}{\emph{\prd} {\bfseries 83}
  (Jan., 2011) 023518}, [\href{https://arxiv.org/abs/0908.0195}{{\ttfamily
  0908.0195}}].

\bibitem{NezriEtAl2012}
E.~{Nezri}, R.~{White}, C.~{Combet}, J.~A. {Hinton}, D.~{Maurin} and
  E.~{Pointecouteau}, \emph{{{\ensuremath{\gamma}} -rays from annihilating dark
  matter in galaxy clusters: stacking versus single source analysis}},
  \href{https://doi.org/10.1111/j.1365-2966.2012.21484.x}{\emph{\mnras}
  {\bfseries 425} (Sept., 2012) 477--489},
  [\href{https://arxiv.org/abs/1203.1165}{{\ttfamily 1203.1165}}].

\bibitem{Sanchez-CondeEtAl2014}
M.~A. {S{\'a}nchez-Conde} and F.~{Prada}, \emph{{The flattening of the
  concentration-mass relation towards low halo masses and its implications for
  the annihilation signal boost}},
  \href{https://doi.org/10.1093/mnras/stu1014}{\emph{\mnras} {\bfseries 442}
  (Aug., 2014) 2271--2277}, [\href{https://arxiv.org/abs/1312.1729}{{\ttfamily
  1312.1729}}].

\bibitem{BonnivardEtAl2016}
V.~{Bonnivard}, M.~{H{\"u}tten}, E.~{Nezri}, A.~{Charbonnier}, C.~{Combet} and
  D.~{Maurin}, \emph{{CLUMPY: Jeans analysis, {\ensuremath{\gamma}}-ray and
  {\ensuremath{\nu}} fluxes from dark matter (sub-)structures}},
  \href{https://doi.org/10.1016/j.cpc.2015.11.012}{\emph{Computer Physics
  Communications} {\bfseries 200} (Mar., 2016) 336--349},
  [\href{https://arxiv.org/abs/1506.07628}{{\ttfamily 1506.07628}}].

\bibitem{MolineEtAl2017}
{\'A}.~{Molin{\'e}}, M.~A. {S{\'a}nchez-Conde}, S.~{Palomares-Ruiz} and
  F.~{Prada}, \emph{{Characterization of subhalo structural properties and
  implications for dark matter annihilation signals}},
  \href{https://doi.org/10.1093/mnras/stx026}{\emph{\mnras} {\bfseries 466}
  (Apr., 2017) 4974--4990}, [\href{https://arxiv.org/abs/1603.04057}{{\ttfamily
  1603.04057}}].

\bibitem{StrefEtAl2017}
M.~{Stref} and J.~{Lavalle}, \emph{{Modeling dark matter subhalos in a
  constrained galaxy: Global mass and boosted annihilation profiles}},
  \href{https://doi.org/10.1103/PhysRevD.95.063003}{\emph{\prd} {\bfseries 95}
  (Mar., 2017) 063003}, [\href{https://arxiv.org/abs/1610.02233}{{\ttfamily
  1610.02233}}].

\bibitem{LattanziAndSilk2009}
M.~{Lattanzi} and J.~{Silk}, \emph{{Can the WIMP annihilation boost factor be
  boosted by the Sommerfeld enhancement?}},
  \href{https://doi.org/10.1103/PhysRevD.79.083523}{\emph{\prd} {\bfseries 79}
  (Apr., 2009) 083523}, [\href{https://arxiv.org/abs/0812.0360}{{\ttfamily
  0812.0360}}].

\bibitem{Bovy2009}
J.~Bovy, \emph{Substructure boosts to dark matter annihilation from sommerfeld
  enhancement}, \href{https://doi.org/10.1103/PhysRevD.79.083539}{\emph{\prd}
  {\bfseries 79} (Apr., 2009) 083539},
  [\href{https://arxiv.org/abs/0903.0413}{{\ttfamily 0903.0413}}].

\bibitem{KuhlenMadauSilk2009}
M.~{Kuhlen}, P.~{Madau} and J.~{Silk}, \emph{{Exploring Dark Matter with Milky
  Way Substructure}},
  \href{https://doi.org/10.1126/science.1174881}{\emph{Science} {\bfseries 325}
  (Aug., 2009) 970}, [\href{https://arxiv.org/abs/0907.0005}{{\ttfamily
  0907.0005}}].

\bibitem{KamionkowskiEtAl2010}
M.~{Kamionkowski}, S.~M. {Koushiappas} and M.~{Kuhlen}, \emph{{Galactic
  substructure and dark-matter annihilation in the Milky Way halo}},
  \href{https://doi.org/10.1103/PhysRevD.81.043532}{\emph{\prd} {\bfseries 81}
  (Feb., 2010) 043532}, [\href{https://arxiv.org/abs/1001.3144}{{\ttfamily
  1001.3144}}].

\bibitem{RobertsonAndZentner2009}
B.~E. {Robertson} and A.~R. {Zentner}, \emph{{Dark matter annihilation rates
  with velocity-dependent annihilation cross sections}},
  \href{https://doi.org/10.1103/PhysRevD.79.083525}{\emph{\prd} {\bfseries 79}
  (Apr., 2009) 083525}, [\href{https://arxiv.org/abs/0902.0362}{{\ttfamily
  0902.0362}}].

\bibitem{EssigEtAl2010}
R.~{Essig}, N.~{Sehgal}, L.~E. {Strigari}, M.~{Geha} and J.~D. {Simon},
  \emph{{Indirect dark matter detection limits from the ultrafaint Milky Way
  satellite Segue 1}},
  \href{https://doi.org/10.1103/PhysRevD.82.123503}{\emph{\prd} {\bfseries 82}
  (Dec., 2010) 123503}, [\href{https://arxiv.org/abs/1007.4199}{{\ttfamily
  1007.4199}}].

\bibitem{FerrerEtAl2013}
F.~{Ferrer} and D.~R. {Hunter}, \emph{{The impact of the phase-space density on
  the indirect detection of dark matter}},
  \href{https://doi.org/10.1088/1475-7516/2013/09/005}{\emph{\jcap} {\bfseries
  9} (Sept., 2013) 5}, [\href{https://arxiv.org/abs/1306.6586}{{\ttfamily
  1306.6586}}].

\bibitem{BoddyEtAl2017}
K.~K. Boddy, J.~Kumar, L.~E. Strigari and M.-Y. Wang, \emph{Sommerfeld-enhanced
  j -factors for dwarf spheroidal galaxies},
  \href{https://doi.org/10.1103/PhysRevD.95.123008}{\emph{\prd} {\bfseries 95}
  (June, 2017) 123008}, [\href{https://arxiv.org/abs/1702.00408}{{\ttfamily
  1702.00408}}].

\bibitem{BergstromEtAl2018}
S.~{Bergstr{\"o}m}, R.~{Catena}, A.~{Chiappo}, J.~{Conrad}, B.~{Eurenius},
  M.~{Eriksson} et~al., \emph{{J -factors for self-interacting dark matter in
  20 dwarf spheroidal galaxies}},
  \href{https://doi.org/10.1103/PhysRevD.98.043017}{\emph{\prd} {\bfseries 98}
  (Aug., 2018) 043017}, [\href{https://arxiv.org/abs/1712.03188}{{\ttfamily
  1712.03188}}].

\bibitem{PetacEtAl2018}
M.~{Peta{\v{c}}}, P.~{Ullio} and M.~{Valli}, \emph{{On velocity-dependent dark
  matter annihilations in dwarf satellites}},
  \href{https://doi.org/10.1088/1475-7516/2018/12/039}{\emph{\jcap} {\bfseries
  2018} (Dec., 2018) 039}, [\href{https://arxiv.org/abs/1804.05052}{{\ttfamily
  1804.05052}}].

\bibitem{AndoAndIshiwata2021}
S.~Ando and K.~Ishiwata, \emph{{Sommerfeld-enhanced dark matter searches with
  dwarf spheroidal galaxies}},
  \href{https://doi.org/10.1103/PhysRevD.104.023016}{\emph{Phys. Rev. D}
  {\bfseries 104} (2021) 023016},
  [\href{https://arxiv.org/abs/2103.01446}{{\ttfamily 2103.01446}}].

\bibitem{BaxterEtAl2021}
E.~J. Baxter, J.~Kumar, A.~B. Pace and J.~Runburg, \emph{{Prospects for
  measuring dark matter microphysics with observations of dwarf spheroidal
  galaxies}}, \href{https://doi.org/10.1088/1475-7516/2021/07/030}{\emph{JCAP}
  {\bfseries 07} (2021) 030},
  [\href{https://arxiv.org/abs/2103.11646}{{\ttfamily 2103.11646}}].

\bibitem{BoddyEtAl2018}
K.~K. Boddy, J.~Kumar and L.~E. Strigari, \emph{The effective j-factor of the
  galactic center for velocity-dependent dark matter annihilation},
  \href{https://doi.org/10.1103/PhysRevD.98.063012}{\emph{\prd} {\bfseries 98}
  (May, 2018) 063012}, [\href{https://arxiv.org/abs/1805.08379}{{\ttfamily
  1805.08379}}].

\bibitem{JohnsonEtAl2019}
C.~{Johnson}, R.~{Caputo}, C.~{Karwin}, S.~{Murgia}, S.~{Ritz}, J.~{Shelton}
  et~al., \emph{{Search for gamma-ray emission from p -wave dark matter
  annihilation in the Galactic Center}},
  \href{https://doi.org/10.1103/PhysRevD.99.103007}{\emph{\prd} {\bfseries 99}
  (May, 2019) 103007}, [\href{https://arxiv.org/abs/1904.06261}{{\ttfamily
  1904.06261}}].

\bibitem{Petac2020}
M.~{Peta{\v{c}}}, \emph{{Equilibrium axisymmetric halo model for the Milky Way
  and its implications for direct and indirect dark matter searches}},
  \href{https://doi.org/10.1103/PhysRevD.102.123028}{\emph{\prd} {\bfseries
  102} (Dec., 2020) 123028},
  [\href{https://arxiv.org/abs/2008.11172}{{\ttfamily 2008.11172}}].

\bibitem{BoardEtAl2021}
E.~{Board}, N.~{Bozorgnia}, L.~E. {Strigari}, R.~J.~J. {Grand}, A.~{Fattahi},
  C.~S. {Frenk} et~al., \emph{{Velocity-dependent J-factors for annihilation
  radiation from cosmological simulations}},
  \href{https://doi.org/10.1088/1475-7516/2021/04/070}{\emph{\jcap} {\bfseries
  2021} (Apr., 2021) 070}, [\href{https://arxiv.org/abs/2101.06284}{{\ttfamily
  2101.06284}}].

\bibitem{LuEtAl2018}
B.-Q. {Lu}, Y.-L. {Wu}, W.-H. {Zhang} and Y.-F. {Zhou}, \emph{{Constraints on
  the Sommerfeld- enhanced dark matter annihilation from the gamma rays of
  subhalos and dwarf galaxies}},
  \href{https://doi.org/10.1088/1475-7516/2018/04/035}{\emph{\jcap} {\bfseries
  2018} (Apr., 2018) 035}, [\href{https://arxiv.org/abs/1711.00749}{{\ttfamily
  1711.00749}}].

\bibitem{RunburgEtAl2021}
J.~{Runburg}, E.~J. {Baxter} and J.~{Kumar}, \emph{{Constraining Dark Matter
  Microphysics with the Annihilation Signal from Subhalos}}, {\emph{arXiv
  e-prints} (June, 2021) arXiv:2106.10399},
  [\href{https://arxiv.org/abs/2106.10399}{{\ttfamily 2106.10399}}].

\bibitem{AbramowskiEtAl2012}
A.~{Abramowski}, F.~{Acero}, F.~{Aharonian}, A.~G. {Akhperjanian}, G.~{Anton},
  A.~{Balzer} et~al., \emph{{Search for Dark Matter Annihilation Signals from
  the Fornax Galaxy Cluster with H.E.S.S.}},
  \href{https://doi.org/10.1088/0004-637X/750/2/123}{\emph{\apj} {\bfseries
  750} (May, 2012) 123}, [\href{https://arxiv.org/abs/1202.5494}{{\ttfamily
  1202.5494}}].

\bibitem{CampbellEtAl2010}
S.~{Campbell}, B.~{Dutta} and E.~{Komatsu}, \emph{{Effects of
  velocity-dependent dark matter annihilation on the energy spectrum of the
  extragalactic gamma-ray background}},
  \href{https://doi.org/10.1103/PhysRevD.82.095007}{\emph{\prd} {\bfseries 82}
  (Nov., 2010) 095007}, [\href{https://arxiv.org/abs/1009.3530}{{\ttfamily
  1009.3530}}].

\bibitem{CampbellAndDutta2011}
S.~{Campbell} and B.~{Dutta}, \emph{{Effects of p-wave annihilation on the
  angular power spectrum of extragalactic gamma-rays from dark matter
  annihilation}}, \href{https://doi.org/10.1103/PhysRevD.84.075004}{\emph{\prd}
  {\bfseries 84} (Oct., 2011) 075004},
  [\href{https://arxiv.org/abs/1106.4621}{{\ttfamily 1106.4621}}].

\bibitem{ZavalaEtAl2009}
J.~Zavala, M.~Vogelsberger and S.~D.~M. White, \emph{{Relic density and CMB
  constraints on dark matter annihilation with Sommerfeld enhancement}},
  \href{https://doi.org/10.1103/PhysRevD.81.083502}{\emph{Phys. Rev. D}
  {\bfseries 81} (2010) 083502},
  [\href{https://arxiv.org/abs/0910.5221}{{\ttfamily 0910.5221}}].

\bibitem{HannestadEt2010}
S.~Hannestad and T.~Tram, \emph{{Sommerfeld Enhancement of DM Annihilation:
  Resonance Structure, Freeze-Out and CMB Spectral Bound}},
  \href{https://doi.org/10.1088/1475-7516/2011/01/016}{\emph{JCAP} {\bfseries
  01} (2011) 016}, [\href{https://arxiv.org/abs/1008.1511}{{\ttfamily
  1008.1511}}].

\bibitem{HisanoEtAl2011}
J.~Hisano, M.~Kawasaki, K.~Kohri, T.~Moroi, K.~Nakayama and T.~Sekiguchi,
  \emph{{Cosmological constraints on dark matter models with velocity-dependent
  annihilation cross section}},
  \href{https://doi.org/10.1103/PhysRevD.83.123511}{\emph{Phys. Rev. D}
  {\bfseries 83} (2011) 123511},
  [\href{https://arxiv.org/abs/1102.4658}{{\ttfamily 1102.4658}}].

\bibitem{CompanionPaper}
G.~Facchinetti, M.~Stref, T.~Lacroix, J.~Lavalle, J.~P\'erez-Romero, D.~Maurin
  et~al., \emph{{Analytical insight into dark matter subhalo boost factors for
  Sommerfeld-enhanced $s$- and $p$-wave $\gamma$-ray signals}},
  \href{https://arxiv.org/abs/2203.16491}{{\ttfamily 2203.16491}}.

\bibitem{Cassel2010}
S.~{Cassel}, \emph{{Sommerfeld factor for arbitrary partial wave processes}},
  \href{https://doi.org/10.1088/0954-3899/37/10/105009}{\emph{Journal of
  Physics G Nuclear Physics} {\bfseries 37} (Oct., 2010) 105009},
  [\href{https://arxiv.org/abs/0903.5307}{{\ttfamily 0903.5307}}].

\bibitem{Iengo2009}
R.~{Iengo}, \emph{{Sommerfeld enhancement: general results from field theory
  diagrams}},
  \href{https://doi.org/10.1088/1126-6708/2009/05/024}{\emph{Journal of High
  Energy Physics} {\bfseries 2009} (May, 2009) 024},
  [\href{https://arxiv.org/abs/0902.0688}{{\ttfamily 0902.0688}}].

\bibitem{Slatyer2010}
T.~R. {Slatyer}, \emph{{The Sommerfeld enhancement for dark matter with an
  excited state}},
  \href{https://doi.org/10.1088/1475-7516/2010/02/028}{\emph{\jcap} {\bfseries
  2010} (Feb., 2010) 028}, [\href{https://arxiv.org/abs/0910.5713}{{\ttfamily
  0910.5713}}].

\bibitem{FengEtAl2010}
J.~L. {Feng}, M.~{Kaplinghat} and H.-B. {Yu}, \emph{{Sommerfeld enhancements
  for thermal relic dark matter}},
  \href{https://doi.org/10.1103/PhysRevD.82.083525}{\emph{\prd} {\bfseries 82}
  (Oct., 2010) 083525}, [\href{https://arxiv.org/abs/1005.4678}{{\ttfamily
  1005.4678}}].

\bibitem{BlumEtAl2016}
K.~{Blum}, R.~{Sato} and T.~R. {Slatyer}, \emph{{Self-consistent calculation of
  the Sommerfeld enhancement}},
  \href{https://doi.org/10.1088/1475-7516/2016/06/021}{\emph{\jcap} {\bfseries
  2016} (June, 2016) 021}, [\href{https://arxiv.org/abs/1603.01383}{{\ttfamily
  1603.01383}}].

\bibitem{PetrakiEtAl2015}
K.~Petraki, M.~Postma and M.~Wiechers, \emph{{Dark-matter bound states from
  Feynman diagrams}},
  \href{https://doi.org/10.1007/JHEP06(2015)128}{\emph{JHEP} {\bfseries 06}
  (2015) 128}, [\href{https://arxiv.org/abs/1505.00109}{{\ttfamily
  1505.00109}}].

\bibitem{PetrakiEtAl2016}
K.~Petraki, M.~Postma and J.~de~Vries, \emph{{Radiative bound-state-formation
  cross-sections for dark matter interacting via a Yukawa potential}},
  \href{https://doi.org/10.1007/JHEP04(2017)077}{\emph{JHEP} {\bfseries 04}
  (2017) 077}, [\href{https://arxiv.org/abs/1611.01394}{{\ttfamily
  1611.01394}}].

\bibitem{Eddington1916}
A.~S. {Eddington}, \emph{{The distribution of stars in globular clusters}},
  \href{https://doi.org/10.1093/mnras/76.7.572}{\emph{\mnras} {\bfseries 76}
  (May, 1916) 572--585}.

\bibitem{BinneyTremaine2008}
J.~{Binney} and S.~{Tremaine}, \emph{{Galactic Dynamics: Second Edition}}.
\newblock Princeton University Press, 2008.

\bibitem{LacroixEtAl2018}
T.~Lacroix, M.~Stref and J.~Lavalle, \emph{{Anatomy of Eddington-like inversion
  methods in the context of dark matter searches}},
  \href{https://doi.org/10.1088/1475-7516/2018/09/040}{\emph{JCAP} {\bfseries
  1809} (2018) 040}, [\href{https://arxiv.org/abs/1805.02403}{{\ttfamily
  1805.02403}}].

\bibitem{Huetten2019}
M.~H{\"u}tten, M.~Stref, C.~Combet, J.~Lavalle and D.~Maurin,
  \emph{{$\gamma$}-ray and {$\nu$} searches for dark-matter subhalos in the
  milky way with a baryonic potential},
  \href{https://doi.org/10.3390/galaxies7020060}{\emph{Galaxies} {\bfseries 7}
  (May, 2019) 60}, [\href{https://arxiv.org/abs/1904.10935}{{\ttfamily
  1904.10935}}].

\bibitem{StrefEtAl2019}
M.~Stref, T.~Lacroix and J.~Lavalle, \emph{Remnants of galactic subhalos and
  their impact on indirect dark-matter searches},
  \href{https://doi.org/10.3390/galaxies7020065}{\emph{Galaxies} {\bfseries 7}
  (June, 2019) 65}, [\href{https://arxiv.org/abs/1905.02008}{{\ttfamily
  1905.02008}}].

\bibitem{FacchinettiEtAl2020}
G.~{Facchinetti}, J.~{Lavalle} and M.~{Stref}, \emph{{Statistics for dark
  matter subhalo searches in gamma rays from a kinematically constrained
  population model. I: Fermi-LAT-like telescopes}}, {\emph{arXiv e-prints}
  (July, 2020) arXiv:2007.10392},
  [\href{https://arxiv.org/abs/2007.10392}{{\ttfamily 2007.10392}}].

\bibitem{FacchinettiEtAl2022}
G.~Facchinetti, M.~Stref and J.~Lavalle, \emph{Tidal stripping of dark matter
  subhalos by baryons from analytical perspectives: disk shocking and
  encounters with stars}, {\emph{arXiv e-prints} (Jan., 2022)
  arXiv:2201.09788}, [\href{https://arxiv.org/abs/2201.09788}{{\ttfamily
  2201.09788}}].

\bibitem{Bringmann2009}
T.~Bringmann, \emph{{Particle Models and the Small-Scale Structure of Dark
  Matter}}, \href{https://doi.org/10.1088/1367-2630/11/10/105027}{\emph{New J.
  Phys.} {\bfseries 11} (2009) 105027},
  [\href{https://arxiv.org/abs/0903.0189}{{\ttfamily 0903.0189}}].

\bibitem{Lake1990}
G.~{Lake}, \emph{{Detectability of gamma-rays from clumps of dark matter}},
  \href{https://doi.org/10.1038/346039a0}{\emph{\nat} {\bfseries 346} (July,
  1990) 39--40}.

\bibitem{EvansEtAl2004}
N.~W. {Evans}, F.~{Ferrer} and S.~{Sarkar}, \emph{{A travel guide to the dark
  matter annihilation signal}},
  \href{https://doi.org/10.1103/PhysRevD.69.123501}{\emph{\prd} {\bfseries 69}
  (June, 2004) 123501},
  [\href{https://arxiv.org/abs/astro-ph/0311145}{{\ttfamily
  astro-ph/0311145}}].

\bibitem{MAGIC2016}
{\scshape MAGIC, Fermi-LAT} collaboration, M.~L. Ahnen et~al., \emph{{Limits to
  Dark Matter Annihilation Cross-Section from a Combined Analysis of MAGIC and
  Fermi-LAT Observations of Dwarf Satellite Galaxies}},
  \href{https://doi.org/10.1088/1475-7516/2016/02/039}{\emph{JCAP} {\bfseries
  02} (2016) 039}, [\href{https://arxiv.org/abs/1601.06590}{{\ttfamily
  1601.06590}}].

\bibitem{AlbertEtAl2017}
A.~{Albert}, B.~{Anderson}, K.~{Bechtol}, A.~{Drlica-Wagner}, M.~{Meyer},
  M.~{S{\'a}nchez-Conde} et~al., \emph{{Searching for Dark Matter Annihilation
  in Recently Discovered Milky Way Satellites with Fermi-Lat}},
  \href{https://doi.org/10.3847/1538-4357/834/2/110}{\emph{\apj} {\bfseries
  834} (Jan., 2017) 110}, [\href{https://arxiv.org/abs/1611.03184}{{\ttfamily
  1611.03184}}].

\bibitem{HoofEtAl2020}
S.~{Hoof}, A.~{Geringer-Sameth} and R.~{Trotta}, \emph{{A global analysis of
  dark matter signals from 27 dwarf spheroidal galaxies using 11 years of
  Fermi-LAT observations}},
  \href{https://doi.org/10.1088/1475-7516/2020/02/012}{\emph{\jcap} {\bfseries
  2020} (Feb., 2020) 012}, [\href{https://arxiv.org/abs/1812.06986}{{\ttfamily
  1812.06986}}].

\bibitem{Strigari2018}
L.~E. {Strigari}, \emph{{Dark matter in dwarf spheroidal galaxies and indirect
  detection: a review}},
  \href{https://doi.org/10.1088/1361-6633/aaae16}{\emph{Reports on Progress in
  Physics} {\bfseries 81} (May, 2018) 056901},
  [\href{https://arxiv.org/abs/1805.05883}{{\ttfamily 1805.05883}}].

\bibitem{BechtolEtAl2015}
K.~{Bechtol}, A.~{Drlica-Wagner}, E.~{Balbinot}, A.~{Pieres}, J.~D. {Simon},
  B.~{Yanny} et~al., \emph{{Eight New Milky Way Companions Discovered in
  First-year Dark Energy Survey Data}},
  \href{https://doi.org/10.1088/0004-637X/807/1/50}{\emph{\apj} {\bfseries 807}
  (July, 2015) 50}, [\href{https://arxiv.org/abs/1503.02584}{{\ttfamily
  1503.02584}}].

\bibitem{KoposovEtAl2015}
S.~E. {Koposov}, V.~{Belokurov}, G.~{Torrealba} and N.~W. {Evans},
  \emph{{Beasts of the Southern Wild: Discovery of Nine Ultra Faint Satellites
  in the Vicinity of the Magellanic Clouds.}},
  \href{https://doi.org/10.1088/0004-637X/805/2/130}{\emph{\apj} {\bfseries
  805} (June, 2015) 130}, [\href{https://arxiv.org/abs/1503.02079}{{\ttfamily
  1503.02079}}].

\bibitem{Drlica-Wagner2015}
A.~{Drlica-Wagner}, K.~{Bechtol}, E.~S. {Rykoff}, E.~{Luque}, A.~{Queiroz},
  Y.~Y. {Mao} et~al., \emph{{Eight Ultra-faint Galaxy Candidates Discovered in
  Year Two of the Dark Energy Survey}},
  \href{https://doi.org/10.1088/0004-637X/813/2/109}{\emph{\apj} {\bfseries
  813} (Nov., 2015) 109}, [\href{https://arxiv.org/abs/1508.03622}{{\ttfamily
  1508.03622}}].

\bibitem{LaevensEtAl2015a}
B.~P.~M. {Laevens}, N.~F. {Martin}, R.~A. {Ibata}, H.-W. {Rix}, E.~J.
  {Bernard}, E.~F. {Bell} et~al., \emph{{A New Faint Milky Way Satellite
  Discovered in the Pan-STARRS1 3{\ensuremath{\pi}} Survey}},
  \href{https://doi.org/10.1088/2041-8205/802/2/L18}{\emph{\apjl} {\bfseries
  802} (Apr., 2015) L18}, [\href{https://arxiv.org/abs/1503.05554}{{\ttfamily
  1503.05554}}].

\bibitem{LaevensEtAl2015b}
B.~P.~M. {Laevens}, N.~F. {Martin}, E.~J. {Bernard}, E.~F. {Schlafly},
  B.~{Sesar}, H.-W. {Rix} et~al., \emph{{Sagittarius II, Draco II and Laevens
  3: Three New Milky Way Satellites Discovered in the Pan-STARRS 1
  3{\ensuremath{\pi}} Survey}},
  \href{https://doi.org/10.1088/0004-637X/813/1/44}{\emph{\apj} {\bfseries 813}
  (Nov., 2015) 44}, [\href{https://arxiv.org/abs/1507.07564}{{\ttfamily
  1507.07564}}].

\bibitem{HommaEtAl2016}
D.~{Homma}, M.~{Chiba}, S.~{Okamoto}, Y.~{Komiyama}, M.~{Tanaka}, M.~{Tanaka}
  et~al., \emph{{A New Milky Way Satellite Discovered in the Subaru/Hyper
  Suprime-Cam Survey}},
  \href{https://doi.org/10.3847/0004-637X/832/1/21}{\emph{\apj} {\bfseries 832}
  (Nov., 2016) 21}, [\href{https://arxiv.org/abs/1609.04346}{{\ttfamily
  1609.04346}}].

\bibitem{HommaEtAl2018}
D.~{Homma}, M.~{Chiba}, S.~{Okamoto}, Y.~{Komiyama}, M.~{Tanaka}, M.~{Tanaka}
  et~al., \emph{{Searches for new Milky Way satellites from the first two years
  of data of the Subaru/Hyper Suprime-Cam survey: Discovery of Cetus III}},
  \href{https://doi.org/10.1093/pasj/psx050}{\emph{\pasj} {\bfseries 70} (Jan.,
  2018) S18}, [\href{https://arxiv.org/abs/1704.05977}{{\ttfamily
  1704.05977}}].

\bibitem{HommaEtAl2019}
D.~{Homma}, M.~{Chiba}, Y.~{Komiyama}, M.~{Tanaka}, S.~{Okamoto}, M.~{Tanaka}
  et~al., \emph{{Bo{\"o}tes. IV. A new Milky Way satellite discovered in the
  Subaru Hyper Suprime-Cam Survey and implications for the missing satellite
  problem}}, \href{https://doi.org/10.1093/pasj/psz076}{\emph{\pasj} {\bfseries
  71} (Oct., 2019) 94}, [\href{https://arxiv.org/abs/1906.07332}{{\ttfamily
  1906.07332}}].

\bibitem{Torrealba2019}
G.~{Torrealba}, V.~{Belokurov}, S.~E. {Koposov}, T.~S. {Li}, M.~G. {Walker},
  J.~L. {Sanders} et~al., \emph{{The hidden giant: discovery of an enormous
  Galactic dwarf satellite in Gaia DR2}},
  \href{https://doi.org/10.1093/mnras/stz1624}{\emph{\mnras} {\bfseries 488}
  (Sept., 2019) 2743--2766},
  [\href{https://arxiv.org/abs/1811.04082}{{\ttfamily 1811.04082}}].

\bibitem{Drlica-Wagner2020}
A.~{Drlica-Wagner}, K.~{Bechtol}, S.~{Mau}, M.~{McNanna}, E.~O. {Nadler}, A.~B.
  {Pace} et~al., \emph{{Milky Way Satellite Census. I. The Observational
  Selection Function for Milky Way Satellites in DES Y3 and Pan-STARRS DR1}},
  \href{https://doi.org/10.3847/1538-4357/ab7eb9}{\emph{\apj} {\bfseries 893}
  (Apr., 2020) 47}, [\href{https://arxiv.org/abs/1912.03302}{{\ttfamily
  1912.03302}}].

\bibitem{Mutlu-Pakdil2021}
B.~{Mutlu-Pakdil}, D.~J. {Sand}, D.~{Crnojevi{\'c}}, A.~{Drlica-Wagner},
  N.~{Caldwell}, P.~{Guhathakurta} et~al., \emph{{Resolved Dwarf Galaxy
  Searches within 5 Mpc with the Vera Rubin Observatory and Subaru Hyper
  Suprime-Cam}}, \href{https://doi.org/10.3847/1538-4357/ac0db8}{\emph{\apj}
  {\bfseries 918} (Sept., 2021) 88},
  [\href{https://arxiv.org/abs/2105.01658}{{\ttfamily 2105.01658}}].

\bibitem{StrigariEtAl2007}
L.~E. {Strigari}, S.~M. {Koushiappas}, J.~S. {Bullock} and M.~{Kaplinghat},
  \emph{{Precise constraints on the dark matter content of MilkyWay dwarf
  galaxies for gamma-ray experiments}},
  \href{https://doi.org/10.1103/PhysRevD.75.083526}{\emph{\prd} {\bfseries 75}
  (Apr., 2007) 083526},
  [\href{https://arxiv.org/abs/astro-ph/0611925}{{\ttfamily
  astro-ph/0611925}}].

\bibitem{MartinezEtAl2009}
G.~D. {Martinez}, J.~S. {Bullock}, M.~{Kaplinghat}, L.~E. {Strigari} and
  R.~{Trotta}, \emph{{Indirect Dark Matter detection from Dwarf satellites:
  joint expectations from astrophysics and supersymmetry}},
  \href{https://doi.org/10.1088/1475-7516/2009/06/014}{\emph{\jcap} {\bfseries
  2009} (June, 2009) 014}, [\href{https://arxiv.org/abs/0902.4715}{{\ttfamily
  0902.4715}}].

\bibitem{CharbonnierEtAl2011}
A.~{Charbonnier}, C.~{Combet}, M.~{Daniel}, S.~{Funk}, J.~A. {Hinton},
  D.~{Maurin} et~al., \emph{{Dark matter profiles and annihilation in dwarf
  spheroidal galaxies: prospectives for present and future
  {\ensuremath{\gamma}}-ray observatories - I. The classical dwarf spheroidal
  galaxies}},
  \href{https://doi.org/10.1111/j.1365-2966.2011.19387.x}{\emph{\mnras}
  {\bfseries 418} (Dec., 2011) 1526--1556},
  [\href{https://arxiv.org/abs/1104.0412}{{\ttfamily 1104.0412}}].

\bibitem{Geringer-SamethEtAl2015}
A.~{Geringer-Sameth}, S.~M. {Koushiappas} and M.~{Walker}, \emph{{Dwarf Galaxy
  Annihilation and Decay Emission Profiles for Dark Matter Experiments}},
  \href{https://doi.org/10.1088/0004-637X/801/2/74}{\emph{\apj} {\bfseries 801}
  (Mar., 2015) 74}, [\href{https://arxiv.org/abs/1408.0002}{{\ttfamily
  1408.0002}}].

\bibitem{BonnivardEtAl2015b}
V.~{Bonnivard}, C.~{Combet}, M.~{Daniel}, S.~{Funk}, A.~{Geringer-Sameth},
  J.~A. {Hinton} et~al., \emph{{Dark matter annihilation and decay in dwarf
  spheroidal galaxies: the classical and ultrafaint dSphs}},
  \href{https://doi.org/10.1093/mnras/stv1601}{\emph{\mnras} {\bfseries 453}
  (Oct., 2015) 849--867}, [\href{https://arxiv.org/abs/1504.02048}{{\ttfamily
  1504.02048}}].

\bibitem{EvansEtAl2016}
N.~W. {Evans}, J.~L. {Sanders} and A.~{Geringer-Sameth}, \emph{{Simple
  J-factors and D-factors for indirect dark matter detection}},
  \href{https://doi.org/10.1103/PhysRevD.93.103512}{\emph{\prd} {\bfseries 93}
  (May, 2016) 103512}, [\href{https://arxiv.org/abs/1604.05599}{{\ttfamily
  1604.05599}}].

\bibitem{SandersEtAl2016}
J.~L. {Sanders} and J.~{Binney}, \emph{{A review of action estimation methods
  for galactic dynamics}},
  \href{https://doi.org/10.1093/mnras/stw106}{\emph{\mnras} {\bfseries 457}
  (Apr., 2016) 2107--2121}, [\href{https://arxiv.org/abs/1511.08213}{{\ttfamily
  1511.08213}}].

\bibitem{PaceAndStrigari2019}
A.~B. {Pace} and L.~E. {Strigari}, \emph{{Scaling relations for dark matter
  annihilation and decay profiles in dwarf spheroidal galaxies}},
  \href{https://doi.org/10.1093/mnras/sty2839}{\emph{\mnras} {\bfseries 482}
  (Jan., 2019) 3480--3496}, [\href{https://arxiv.org/abs/1802.06811}{{\ttfamily
  1802.06811}}].

\bibitem{ChiappoEtAl2019}
A.~{Chiappo}, J.~{Cohen-Tanugi}, J.~{Conrad} and L.~E. {Strigari}, \emph{{Dwarf
  spheroidal J-factor likelihoods for generalized NFW profiles}},
  \href{https://doi.org/10.1093/mnras/stz1871}{\emph{\mnras} {\bfseries 488}
  (Sept., 2019) 2616--2628},
  [\href{https://arxiv.org/abs/1810.09917}{{\ttfamily 1810.09917}}].

\bibitem{AlvarezEtAl2020}
A.~{Alvarez}, F.~{Calore}, A.~{Genina}, J.~{Read}, P.~D. {Serpico} and
  B.~{Zaldivar}, \emph{{Dark matter constraints from dwarf galaxies with
  data-driven J-factors}},
  \href{https://doi.org/10.1088/1475-7516/2020/09/004}{\emph{\jcap} {\bfseries
  2020} (Sept., 2020) 004}, [\href{https://arxiv.org/abs/2002.01229}{{\ttfamily
  2002.01229}}].

\bibitem{MunozEtAl2018}
R.~R. {Mu{\~n}oz}, P.~{C{\^o}t{\'e}}, F.~A. {Santana}, M.~{Geha}, J.~D.
  {Simon}, G.~A. {Oyarz{\'u}n} et~al., \emph{{A MegaCam Survey of Outer Halo
  Satellites. III. Photometric and Structural Parameters}},
  \href{https://doi.org/10.3847/1538-4357/aac16b}{\emph{\apj} {\bfseries 860}
  (June, 2018) 66}, [\href{https://arxiv.org/abs/1806.06891}{{\ttfamily
  1806.06891}}].

\bibitem{Simon2019}
J.~D. {Simon}, \emph{{The Faintest Dwarf Galaxies}},
  \href{https://doi.org/10.1146/annurev-astro-091918-104453}{\emph{\araa}
  {\bfseries 57} (Aug., 2019) 375--415},
  [\href{https://arxiv.org/abs/1901.05465}{{\ttfamily 1901.05465}}].

\bibitem{SimonEtAl2020}
J.~D. {Simon}, T.~S. {Li}, D.~{Erkal}, A.~B. {Pace}, A.~{Drlica-Wagner}, D.~J.
  {James} et~al., \emph{{Birds of a Feather? Magellan/IMACS Spectroscopy of the
  Ultra-faint Satellites Grus II, Tucana IV, and Tucana V}},
  \href{https://doi.org/10.3847/1538-4357/ab7ccb}{\emph{\apj} {\bfseries 892}
  (Apr., 2020) 137}, [\href{https://arxiv.org/abs/1911.08493}{{\ttfamily
  1911.08493}}].

\bibitem{JenkinsEtAl2021}
S.~{Jenkins}, T.~S. {Li}, A.~B. {Pace}, A.~P. {Ji}, S.~E. {Koposov} and
  B.~{Mutlu-Pakdil}, \emph{{VLT Spectroscopy of Ultra-Faint Dwarf Galaxies. 1.
  Bo\{{\"o}\}tes I, Leo IV, Leo V}}, {\emph{arXiv e-prints} (Dec., 2020)
  arXiv:2101.00013}, [\href{https://arxiv.org/abs/2101.00013}{{\ttfamily
  2101.00013}}].

\bibitem{BonnivardEtAl2015c}
V.~{Bonnivard}, C.~{Combet}, D.~{Maurin}, A.~{Geringer-Sameth}, S.~M.
  {Koushiappas}, M.~G. {Walker} et~al., \emph{{Dark Matter Annihilation and
  Decay Profiles for the Reticulum II Dwarf Spheroidal Galaxy}},
  \href{https://doi.org/10.1088/2041-8205/808/2/L36}{\emph{\apjl} {\bfseries
  808} (Aug., 2015) L36}, [\href{https://arxiv.org/abs/1504.03309}{{\ttfamily
  1504.03309}}].

\bibitem{HuettenEtAl2019}
M.~{H{\"u}tten}, C.~{Combet} and D.~{Maurin}, \emph{{CLUMPY v3:
  {\ensuremath{\gamma}}-ray and {\ensuremath{\nu}} signals from dark matter at
  all scales}}, \href{https://doi.org/10.1016/j.cpc.2018.10.001}{\emph{Computer
  Physics Communications} {\bfseries 235} (Feb., 2019) 336--345},
  [\href{https://arxiv.org/abs/1806.08639}{{\ttfamily 1806.08639}}].

\bibitem{BonnivardEtAl2015a}
V.~{Bonnivard}, C.~{Combet}, D.~{Maurin} and M.~G. {Walker}, \emph{{Spherical
  Jeans analysis for dark matter indirect detection in dwarf spheroidal
  galaxies - impact of physical parameters and triaxiality}},
  \href{https://doi.org/10.1093/mnras/stu2296}{\emph{\mnras} {\bfseries 446}
  (Jan., 2015) 3002--3021}, [\href{https://arxiv.org/abs/1407.7822}{{\ttfamily
  1407.7822}}].

\bibitem{Mateo1998}
M.~L. {Mateo}, \emph{{Dwarf Galaxies of the Local Group}},
  \href{https://doi.org/10.1146/annurev.astro.36.1.435}{\emph{\araa} {\bfseries
  36} (Jan., 1998) 435--506},
  [\href{https://arxiv.org/abs/astro-ph/9810070}{{\ttfamily
  astro-ph/9810070}}].

\bibitem{McConnachie2012}
A.~W. {McConnachie}, \emph{{The Observed Properties of Dwarf Galaxies in and
  around the Local Group}},
  \href{https://doi.org/10.1088/0004-6256/144/1/4}{\emph{\aj} {\bfseries 144}
  (July, 2012) 4}, [\href{https://arxiv.org/abs/1204.1562}{{\ttfamily
  1204.1562}}].

\bibitem{Oh:2015xoa}
S.-H. Oh et~al., \emph{{High-resolution mass models of dwarf galaxies from
  LITTLE THINGS}},
  \href{https://doi.org/10.1088/0004-6256/149/6/180}{\emph{Astron. J.}
  {\bfseries 149} (2015) 180},
  [\href{https://arxiv.org/abs/1502.01281}{{\ttfamily 1502.01281}}].

\bibitem{oh1}
S.-H. Oh, W.~J.~G. de~Blok, E.~Brinks, F.~Walter and R.~C. Kennicutt, Jr,
  \emph{{Dark and luminous matter in THINGS dwarf galaxies}},
  \href{https://doi.org/10.1088/0004-6256/141/6/193}{\emph{Astron. J.}
  {\bfseries 141} (2011) 193},
  [\href{https://arxiv.org/abs/1011.0899}{{\ttfamily 1011.0899}}].

\bibitem{Gentile:2006hv}
G.~Gentile, P.~Salucci, U.~Klein and G.~L. Granato, \emph{{NGC 3741: Dark halo
  profile from the most extended rotation curve}},
  \href{https://doi.org/10.1111/j.1365-2966.2006.11283.x}{\emph{Mon. Not. Roy.
  Astron. Soc.} {\bfseries 375} (2007) 199--212},
  [\href{https://arxiv.org/abs/astro-ph/0611355}{{\ttfamily
  astro-ph/0611355}}].

\bibitem{Winter:2016wmy}
M.~Winter, G.~Zaharijas, K.~Bechtol and J.~Vandenbroucke, \emph{{Estimating the
  GeV Emission of Millisecond Pulsars in Dwarf Spheroidal Galaxies}},
  \href{https://doi.org/10.3847/2041-8205/832/1/L6}{\emph{Astrophys. J. Lett.}
  {\bfseries 832} (2016) L6},
  [\href{https://arxiv.org/abs/1607.06390}{{\ttfamily 1607.06390}}].

\bibitem{Gammaldi:2017mio}
V.~Gammaldi, E.~Karukes and P.~Salucci, \emph{{Theoretical predictions for dark
  matter detection in dwarf irregular galaxies with gamma rays}},
  \href{https://doi.org/10.1103/PhysRevD.98.083008}{\emph{Phys. Rev. D}
  {\bfseries 98} (2018) 083008},
  [\href{https://arxiv.org/abs/1706.01843}{{\ttfamily 1706.01843}}].

\bibitem{2021PhRvD.104h3026G}
V.~{Gammaldi}, J.~{P{\'e}rez-Romero}, J.~{Coronado-Bl{\'a}zquez}, M.~{Di
  Mauro}, E.~V. {Karukes}, M.~A. {S{\'a}nchez-Conde} et~al., \emph{{Dark matter
  search in dwarf irregular galaxies with the Fermi Large Area Telescope}},
  \href{https://doi.org/10.1103/PhysRevD.104.083026}{\emph{\prd} {\bfseries
  104} (Oct., 2021) 083026},
  [\href{https://arxiv.org/abs/2109.11291}{{\ttfamily 2109.11291}}].

\bibitem{Salucci:2007tm}
P.~Salucci, A.~Lapi, C.~Tonini, G.~Gentile, I.~Yegorova and U.~Klein,
  \emph{{The Universal Rotation Curve of Spiral Galaxies. 2. The Dark Matter
  Distribution out to the Virial Radius}},
  \href{https://doi.org/10.1111/j.1365-2966.2007.11696.x}{\emph{Mon. Not. Roy.
  Astron. Soc.} {\bfseries 378} (2007) 41--47},
  [\href{https://arxiv.org/abs/astro-ph/0703115}{{\ttfamily
  astro-ph/0703115}}].

\bibitem{Navarro:1995iw}
J.~F. Navarro, C.~S. Frenk and S.~D.~M. White, \emph{{The Structure of cold
  dark matter halos}}, \href{https://doi.org/10.1086/177173}{\emph{Astrophys.
  J.} {\bfseries 462} (1996) 563--575},
  [\href{https://arxiv.org/abs/astro-ph/9508025}{{\ttfamily
  astro-ph/9508025}}].

\bibitem{NavarroEtAl1997}
J.~F. {Navarro}, C.~S. {Frenk} and S.~D.~M. {White}, \emph{{A Universal Density
  Profile from Hierarchical Clustering}},
  \href{https://doi.org/10.1086/304888}{\emph{\apj} {\bfseries 490} (Dec.,
  1997) 493--508}, [\href{https://arxiv.org/abs/astro-ph/9611107}{{\ttfamily
  astro-ph/9611107}}].

\bibitem{Einasto1965}
J.~{Einasto}, \emph{{On the Construction of a Composite Model for the Galaxy
  and on the Determination of the System of Galactic Parameters}}, {\emph{Trudy
  Astrofizicheskogo Instituta Alma-Ata} {\bfseries 5} (1965) 87--100}.

\bibitem{Gomez-Vargas:2013bea}
G.~A. G\'omez-Vargas, M.~A. S\'anchez-Conde, J.-H. Huh, M.~Peir\'o, F.~Prada,
  A.~Morselli et~al., \emph{{Constraints on WIMP annihilation for contracted
  dark matterin the inner Galaxy with the Fermi-LAT}},
  \href{https://doi.org/10.1088/1475-7516/2013/10/029}{\emph{JCAP} {\bfseries
  10} (2013) 029}, [\href{https://arxiv.org/abs/1308.3515}{{\ttfamily
  1308.3515}}].

\bibitem{Schaller:2014uwa}
M.~Schaller, C.~S. Frenk, R.~G. Bower, T.~Theuns, A.~Jenkins, J.~Schaye et~al.,
  \emph{{Baryon effects on the internal structure of \ensuremath{\Lambda}CDM
  haloes in the EAGLE simulations}},
  \href{https://doi.org/10.1093/mnras/stv1067}{\emph{Mon. Not. Roy. Astron.
  Soc.} {\bfseries 451} (2015) 1247--1267},
  [\href{https://arxiv.org/abs/1409.8617}{{\ttfamily 1409.8617}}].

\bibitem{2017MNRAS.472.2153P}
S.~{Peirani}, Y.~{Dubois}, M.~{Volonteri}, J.~{Devriendt}, K.~{Bundy},
  J.~{Silk} et~al., \emph{{Density profile of dark matter haloes and galaxies
  in the HORIZON-AGN simulation: the impact of AGN feedback}},
  \href{https://doi.org/10.1093/mnras/stx2099}{\emph{\mnras} {\bfseries 472}
  (Dec., 2017) 2153--2169}, [\href{https://arxiv.org/abs/1611.09922}{{\ttfamily
  1611.09922}}].

\bibitem{Bose:2018oaj}
S.~Bose et~al., \emph{{No cores in dark matter-dominated dwarf galaxies with
  bursty star formation histories}},
  \href{https://doi.org/10.1093/mnras/stz1168}{\emph{Mon. Not. Roy. Astron.
  Soc.} {\bfseries 486} (2019) 4790--4804},
  [\href{https://arxiv.org/abs/1810.03635}{{\ttfamily 1810.03635}}].

\bibitem{2019MNRAS.488.2387B}
A.~{Ben{\'\i}tez-Llambay}, C.~S. {Frenk}, A.~D. {Ludlow} and J.~F. {Navarro},
  \emph{{Baryon-induced dark matter cores in the EAGLE simulations}},
  \href{https://doi.org/10.1093/mnras/stz1890}{\emph{\mnras} {\bfseries 488}
  (Sept., 2019) 2387--2404},
  [\href{https://arxiv.org/abs/1810.04186}{{\ttfamily 1810.04186}}].

\bibitem{Burkert:1995yz}
A.~Burkert, \emph{{The Structure of dark matter halos in dwarf galaxies}},
  \href{https://doi.org/10.1086/309560}{\emph{IAU Symp.} {\bfseries 171} (1996)
  175}, [\href{https://arxiv.org/abs/astro-ph/9504041}{{\ttfamily
  astro-ph/9504041}}].

\bibitem{RevModPhys.77.207}
G.~M. Voit, \emph{Tracing cosmic evolution with clusters of galaxies},
  \href{https://doi.org/10.1103/RevModPhys.77.207}{\emph{Rev. Mod. Phys.}
  {\bfseries 77} (Apr, 2005) 207--258}.

\bibitem{vanWeeren:2019vxy}
R.~J. van Weeren, F.~de~Gasperin, H.~Akamatsu, M.~Br\"uggen, L.~Feretti,
  H.~Kang et~al., \emph{{Diffuse Radio Emission from Galaxy Clusters}},
  \href{https://doi.org/10.1007/s11214-019-0584-z}{\emph{Space Sci. Rev.}
  {\bfseries 215} (2019) 16},
  [\href{https://arxiv.org/abs/1901.04496}{{\ttfamily 1901.04496}}].

\bibitem{Ackermann:2013iaq}
{\scshape Fermi-LAT} collaboration, M.~Ackermann et~al., \emph{{Search for
  cosmic-ray induced gamma-ray emission in Galaxy Clusters}},
  \href{https://doi.org/10.1088/0004-637X/787/1/18}{\emph{Astrophys. J.}
  {\bfseries 787} (2014) 18},
  [\href{https://arxiv.org/abs/1308.5654}{{\ttfamily 1308.5654}}].

\bibitem{Colavincenzo:2019jtj}
M.~Colavincenzo, X.~Tan, S.~Ammazzalorso, S.~Camera, M.~Regis, J.-Q. Xia
  et~al., \emph{{Searching for gamma-ray emission from galaxy clusters at low
  redshift}}, \href{https://doi.org/10.1093/mnras/stz3263}{\emph{Mon. Not. Roy.
  Astron. Soc.} {\bfseries 491} (2020) 3225--3244},
  [\href{https://arxiv.org/abs/1907.05264}{{\ttfamily 1907.05264}}].

\bibitem{AckermannEtAl2016}
M.~{Ackermann}, M.~{Ajello}, A.~{Albert}, W.~B. {Atwood}, L.~{Baldini},
  J.~{Ballet} et~al., \emph{{Search for Gamma-Ray Emission from the Coma
  Cluster with Six Years of Fermi-LAT Data}},
  \href{https://doi.org/10.3847/0004-637X/819/2/149}{\emph{\apj} {\bfseries
  819} (Mar., 2016) 149}, [\href{https://arxiv.org/abs/1507.08995}{{\ttfamily
  1507.08995}}].

\bibitem{XiEtAl2018}
S.-Q. {Xi}, X.-Y. {Wang}, Y.-F. {Liang}, F.-K. {Peng}, R.-Z. {Yang} and R.-Y.
  {Liu}, \emph{{Detection of gamma-ray emission from the Coma cluster with
  Fermi Large Area Telescope and tentative evidence for an extended spatial
  structure}}, \href{https://doi.org/10.1103/PhysRevD.98.063006}{\emph{\prd}
  {\bfseries 98} (Sept., 2018) 063006},
  [\href{https://arxiv.org/abs/1709.08319}{{\ttfamily 1709.08319}}].

\bibitem{AdamEtAl2021}
R.~{Adam}, H.~{Goksu}, S.~{Brown}, L.~{Rudnick} and C.~{Ferrari},
  \emph{{{\ensuremath{\gamma}}-ray detection toward the Coma cluster with
  Fermi-LAT: Implications for the cosmic ray content in the hadronic
  scenario}}, \href{https://doi.org/10.1051/0004-6361/202039660}{\emph{\aap}
  {\bfseries 648} (Apr., 2021) A60},
  [\href{https://arxiv.org/abs/2102.02251}{{\ttfamily 2102.02251}}].

\bibitem{BaghmanyanEtAl2021}
V.~{Baghmanyan}, D.~{Zargaryan}, F.~{Aharonian}, R.~{Yang}, S.~{Casanova} and
  J.~{Mackey}, \emph{{Detailed study of extended gamma-ray morphology in the
  vicinity of the Coma cluster with Fermi-LAT}}, {\emph{arXiv e-prints} (Oct.,
  2021) arXiv:2110.00309}, [\href{https://arxiv.org/abs/2110.00309}{{\ttfamily
  2110.00309}}].

\bibitem{Blasi:2007pm}
P.~Blasi, S.~Gabici and G.~Brunetti, \emph{{Gamma rays from clusters of
  galaxies}}, \href{https://doi.org/10.1142/S0217751X0703529X}{\emph{Int. J.
  Mod. Phys. A} {\bfseries 22} (2007) 681--706},
  [\href{https://arxiv.org/abs/astro-ph/0701545}{{\ttfamily
  astro-ph/0701545}}].

\bibitem{2010MNRAS.409..449P}
A.~{Pinzke} and C.~{Pfrommer}, \emph{{Simulating the {\ensuremath{\gamma}}-ray
  emission from galaxy clusters: a universal cosmic ray spectrum and spatial
  distribution}},
  \href{https://doi.org/10.1111/j.1365-2966.2010.17328.x}{\emph{\mnras}
  {\bfseries 409} (Dec., 2010) 449--480},
  [\href{https://arxiv.org/abs/1001.5023}{{\ttfamily 1001.5023}}].

\bibitem{JeltemaEtAl2009}
T.~E. {Jeltema}, J.~{Kehayias} and S.~{Profumo}, \emph{{Gamma rays from
  clusters and groups of galaxies: Cosmic rays versus dark matter}},
  \href{https://doi.org/10.1103/PhysRevD.80.023005}{\emph{\prd} {\bfseries 80}
  (July, 2009) 023005}, [\href{https://arxiv.org/abs/0812.0597}{{\ttfamily
  0812.0597}}].

\bibitem{2010JCAP...05..025A}
M.~{Ackermann}, M.~{Ajello}, A.~{Allafort}, L.~{Baldini}, J.~{Ballet},
  G.~{Barbiellini} et~al., \emph{{Constraints on dark matter annihilation in
  clusters of galaxies with the Fermi large area telescope}},
  \href{https://doi.org/10.1088/1475-7516/2010/05/025}{\emph{\jcap} {\bfseries
  2010} (May, 2010) 025}, [\href{https://arxiv.org/abs/1002.2239}{{\ttfamily
  1002.2239}}].

\bibitem{PinzkeEtAl2011}
A.~{Pinzke}, C.~{Pfrommer} and L.~{Bergstr{\"o}m}, \emph{{Prospects of
  detecting gamma-ray emission from galaxy clusters: Cosmic rays and dark
  matter annihilations}},
  \href{https://doi.org/10.1103/PhysRevD.84.123509}{\emph{\prd} {\bfseries 84}
  (Dec., 2011) 123509}, [\href{https://arxiv.org/abs/1105.3240}{{\ttfamily
  1105.3240}}].

\bibitem{CombetEtAl2012}
C.~{Combet}, D.~{Maurin}, E.~{Nezri}, E.~{Pointecouteau}, J.~A. {Hinton} and
  R.~{White}, \emph{{Decaying dark matter: Stacking analysis of galaxy clusters
  to improve on current limits}},
  \href{https://doi.org/10.1103/PhysRevD.85.063517}{\emph{\prd} {\bfseries 85}
  (Mar., 2012) 063517}, [\href{https://arxiv.org/abs/1203.1164}{{\ttfamily
  1203.1164}}].

\bibitem{2012JCAP...07..017A}
S.~{Ando} and D.~{Nagai}, \emph{{Fermi-LAT constraints on dark matter
  annihilation cross section from observations of the Fornax cluster}},
  \href{https://doi.org/10.1088/1475-7516/2012/07/017}{\emph{\jcap} {\bfseries
  2012} (July, 2012) 017}, [\href{https://arxiv.org/abs/1201.0753}{{\ttfamily
  1201.0753}}].

\bibitem{Ackermann:2015fdi}
{\scshape Fermi-LAT} collaboration, M.~Ackermann et~al., \emph{{Search for
  extended gamma-ray emission from the Virgo galaxy cluster with Fermi-LAT}},
  \href{https://doi.org/10.1088/0004-637X/812/2/159}{\emph{Astrophys. J.}
  {\bfseries 812} (2015) 159},
  [\href{https://arxiv.org/abs/1510.00004}{{\ttfamily 1510.00004}}].

\bibitem{Acciari:2018sjn}
{\scshape MAGIC} collaboration, V.~A. Acciari et~al., \emph{{Constraining Dark
  Matter lifetime with a deep gamma-ray survey of the Perseus Galaxy Cluster
  with MAGIC}}, \href{https://doi.org/10.1016/j.dark.2018.08.002}{\emph{Phys.
  Dark Univ.} {\bfseries 22} (2018) 38--47},
  [\href{https://arxiv.org/abs/1806.11063}{{\ttfamily 1806.11063}}].

\bibitem{MaurinEtAl2012}
D.~{Maurin}, C.~{Combet}, E.~{Nezri} and E.~{Pointecouteau},
  \emph{{Disentangling cosmic-ray and dark-matter induced
  {\ensuremath{\gamma}}-rays in galaxy clusters}},
  \href{https://doi.org/10.1051/0004-6361/201218986}{\emph{\aap} {\bfseries
  547} (Nov., 2012) A16}, [\href{https://arxiv.org/abs/1203.1166}{{\ttfamily
  1203.1166}}].

\bibitem{2011JCAP...12..011S}
M.~A. {S{\'a}nchez-Conde}, M.~{Cannoni}, F.~{Zandanel}, M.~E. {G{\'o}mez} and
  F.~{Prada}, \emph{{Dark matter searches with Cherenkov telescopes: nearby
  dwarf galaxies or local galaxy clusters?}},
  \href{https://doi.org/10.1088/1475-7516/2011/12/011}{\emph{\jcap} {\bfseries
  2011} (Dec., 2011) 011}, [\href{https://arxiv.org/abs/1104.3530}{{\ttfamily
  1104.3530}}].

\bibitem{2002ApJ...567..716R}
T.~H. {Reiprich} and H.~{B{\"o}hringer}, \emph{{The Mass Function of an X-Ray
  Flux-limited Sample of Galaxy Clusters}},
  \href{https://doi.org/10.1086/338753}{\emph{\apj} {\bfseries 567} (Mar.,
  2002) 716--740}, [\href{https://arxiv.org/abs/astro-ph/0111285}{{\ttfamily
  astro-ph/0111285}}].

\bibitem{2011A&A...534A.109P}
R.~{Piffaretti}, M.~{Arnaud}, G.~W. {Pratt}, E.~{Pointecouteau} and J.~B.
  {Melin}, \emph{{The MCXC: a meta-catalogue of x-ray detected clusters of
  galaxies}}, \href{https://doi.org/10.1051/0004-6361/201015377}{\emph{\aap}
  {\bfseries 534} (Oct., 2011) A109},
  [\href{https://arxiv.org/abs/1007.1916}{{\ttfamily 1007.1916}}].

\bibitem{Snowden:2007jg}
S.~L. Snowden, R.~M. Mushotzky, K.~D. Kuntz and D.~S. Davis, \emph{{A Catalog
  of Galaxy Clusters Observed by XMM-Newton}},
  \href{https://doi.org/10.1051/0004-6361:20077930}{\emph{Astron. Astrophys.}
  {\bfseries 478} (2008) 615--658},
  [\href{https://arxiv.org/abs/0710.2241}{{\ttfamily 0710.2241}}].

\bibitem{Vikhlinin:2008cd}
A.~Vikhlinin et~al., \emph{{Chandra Cluster Cosmology Project II: Samples and
  X-ray Data Reduction}},
  \href{https://doi.org/10.1088/0004-637X/692/2/1033}{\emph{Astrophys. J.}
  {\bfseries 692} (2009) 1033--1059},
  [\href{https://arxiv.org/abs/0805.2207}{{\ttfamily 0805.2207}}].

\bibitem{Schellenberger:2017wdw}
G.~Schellenberger and T.~H. Reiprich, \emph{{HICOSMO \textendash{} cosmology
  with a complete sample of galaxy clusters \textendash{} I. Data analysis,
  sample selection and luminosity\textendash{}mass scaling relation}},
  \href{https://doi.org/10.1093/mnras/stx1022}{\emph{Mon. Not. Roy. Astron.
  Soc.} {\bfseries 469} (2017) 3738--3761},
  [\href{https://arxiv.org/abs/1705.05842}{{\ttfamily 1705.05842}}].

\bibitem{Pratt:2019cnf}
G.~W. Pratt, M.~Arnaud, A.~Biviano, D.~Eckert, S.~Ettori, D.~Nagai et~al.,
  \emph{{The galaxy cluster mass scale and its impact on cosmological
  constraints from the cluster population}},
  \href{https://doi.org/10.1007/s11214-019-0591-0}{\emph{Space Sci. Rev.}
  {\bfseries 215} (2019) 25},
  [\href{https://arxiv.org/abs/1902.10837}{{\ttfamily 1902.10837}}].

\bibitem{2017A&A...599A.138Z}
Y.-Y. {Zhang}, T.~H. {Reiprich}, P.~{Schneider}, N.~{Clerc}, A.~{Merloni},
  A.~{Schwope} et~al., \emph{{HIFLUGCS: X-ray luminosity-dynamical mass
  relation and its implications for mass calibrations with the SPIDERS and
  4MOST surveys}},
  \href{https://doi.org/10.1051/0004-6361/201628971}{\emph{\aap} {\bfseries
  599} (Mar., 2017) A138}, [\href{https://arxiv.org/abs/1608.06585}{{\ttfamily
  1608.06585}}].

\bibitem{2016A&A...594A..27P}
{Planck Collaboration}, P.~A.~R. {Ade}, N.~{Aghanim}, M.~{Arnaud},
  M.~{Ashdown}, J.~{Aumont} et~al., \emph{{Planck 2015 results. XXVII. The
  second Planck catalogue of Sunyaev-Zeldovich sources}},
  \href{https://doi.org/10.1051/0004-6361/201525823}{\emph{\aap} {\bfseries
  594} (Sept., 2016) A27}, [\href{https://arxiv.org/abs/1502.01598}{{\ttfamily
  1502.01598}}].

\bibitem{Chen:2007sz}
Y.~Chen, T.~H. Reiprich, H.~Bohringer, Y.~Ikebe and Y.~Y. Zhang,
  \emph{{Statistics of X-ray observables for the cooling-core and non-cooling
  core galaxy clusters}},
  \href{https://doi.org/10.1051/0004-6361:20066471}{\emph{Astron. Astrophys.}
  {\bfseries 466} (2007) 805},
  [\href{https://arxiv.org/abs/astro-ph/0702482}{{\ttfamily
  astro-ph/0702482}}].

\bibitem{Adam:2020atc}
R.~Adam, H.~Goksu, A.~Leingartner-Goth, S.~Ettori, R.~Gnatyk, B.~Hnatyk et~al.,
  \emph{{MINOT: Modeling the intracluster medium (non-)thermal content and
  observable prediction tools}},
  \href{https://doi.org/10.1051/0004-6361/202039091}{\emph{Astron. Astrophys.}
  {\bfseries 644} (2020) A70},
  [\href{https://arxiv.org/abs/2009.05373}{{\ttfamily 2009.05373}}].

\bibitem{KuhlenEtAl2009}
M.~Kuhlen and D.~Malyshev, \emph{Atic, pamela, hess, and fermi data and nearby
  dark matter subhalos},
  \href{https://doi.org/10.1103/PhysRevD.79.123517}{\emph{\prd} {\bfseries 79}
  (June, 2009) 123517}, [\href{https://arxiv.org/abs/0904.3378}{{\ttfamily
  0904.3378}}].

\bibitem{PieriEtAl2009}
L.~{Pieri}, M.~{Lattanzi} and J.~{Silk}, \emph{{Constraining the dark matter
  annihilation cross-section with Cherenkov telescope observations of dwarf
  galaxies}},
  \href{https://doi.org/10.1111/j.1365-2966.2009.15388.x}{\emph{\mnras}
  {\bfseries 399} (Nov., 2009) 2033--2040},
  [\href{https://arxiv.org/abs/0902.4330}{{\ttfamily 0902.4330}}].

\bibitem{SlatyerEtAl2012}
T.~R. {Slatyer}, N.~{Toro} and N.~{Weiner}, \emph{{Sommerfeld-enhanced
  annihilation in dark matter substructure: Consequences for constraints on
  cosmic-ray excesses}},
  \href{https://doi.org/10.1103/PhysRevD.86.083534}{\emph{\prd} {\bfseries 86}
  (Oct., 2012) 083534}, [\href{https://arxiv.org/abs/1107.3546}{{\ttfamily
  1107.3546}}].

\bibitem{ZavalaEtAl2014a}
J.~{Zavala} and N.~{Afshordi}, \emph{{Clustering in the phase space of dark
  matter haloes - II. Stable clustering and dark matter annihilation}},
  \href{https://doi.org/10.1093/mnras/stu506}{\emph{\mnras} {\bfseries 441}
  (June, 2014) 1329--1339}, [\href{https://arxiv.org/abs/1311.3296}{{\ttfamily
  1311.3296}}].

\bibitem{BoddyEtAl2019}
K.~K. {Boddy}, J.~{Kumar}, J.~{Runburg} and L.~E. {Strigari}, \emph{{Angular
  distribution of gamma-ray emission from velocity-dependent dark matter
  annihilation in subhalos}},
  \href{https://doi.org/10.1103/PhysRevD.100.063019}{\emph{\prd} {\bfseries
  100} (Sept., 2019) 063019},
  [\href{https://arxiv.org/abs/1905.03431}{{\ttfamily 1905.03431}}].

\bibitem{PiccirilloEtAl2022}
E.~Piccirillo, K.~Blanchette, N.~Bozorgnia, L.~E. Strigari, C.~S. Frenk,
  R.~J.~J. Grand et~al., \emph{{Velocity-dependent annihilation radiation from
  dark matter subhalos in cosmological simulations}},
  \href{https://arxiv.org/abs/2203.08853}{{\ttfamily 2203.08853}}.

\bibitem{Palomares-RuizEtAl2010}
S.~{Palomares-Ruiz} and J.~M. {Siegal-Gaskins}, \emph{{Annihilation vs. decay:
  constraining dark matter properties from a gamma-ray detection}},
  \href{https://doi.org/10.1088/1475-7516/2010/07/023}{\emph{\jcap} {\bfseries
  2010} (July, 2010) 023}, [\href{https://arxiv.org/abs/1003.1142}{{\ttfamily
  1003.1142}}].

\bibitem{2021arXiv211009653B}
B.~{Boucher}, J.~{Kumar}, V.~B. {Le} and J.~{Runburg}, \emph{{$J$-factors for
  Velocity-dependent Dark Matter}}, {\emph{arXiv e-prints} (Oct., 2021)
  arXiv:2110.09653}, [\href{https://arxiv.org/abs/2110.09653}{{\ttfamily
  2110.09653}}].

\bibitem{2015JCAP...09..008F}
{Fermi LAT Collaboration}, \emph{{Limits on dark matter annihilation signals
  from the Fermi LAT 4-year measurement of the isotropic gamma-ray
  background}},
  \href{https://doi.org/10.1088/1475-7516/2015/09/008}{\emph{\jcap} {\bfseries
  2015} (Sept., 2015) 008}, [\href{https://arxiv.org/abs/1501.05464}{{\ttfamily
  1501.05464}}].

\bibitem{McMillan2017}
P.~J. {McMillan}, \emph{{The mass distribution and gravitational potential of
  the Milky Way}}, \href{https://doi.org/10.1093/mnras/stw2759}{\emph{\mnras}
  {\bfseries 465} (Feb., 2017) 76--94},
  [\href{https://arxiv.org/abs/1608.00971}{{\ttfamily 1608.00971}}].

\bibitem{CirelliEtAl2005}
M.~Cirelli, N.~Fornengo and A.~Strumia, \emph{{Minimal dark matter}},
  \href{https://doi.org/10.1016/j.nuclphysb.2006.07.012}{\emph{Nucl. Phys. B}
  {\bfseries 753} (2006) 178--194},
  [\href{https://arxiv.org/abs/hep-ph/0512090}{{\ttfamily hep-ph/0512090}}].

\bibitem{GarciaCelyEtAl2015}
C.~{Garcia-Cely}, A.~{Ibarra}, A.~S. {Lamperstorfer} and M.~H.~G. {Tytgat},
  \emph{{Gamma-rays from Heavy Minimal Dark Matter}},
  \href{https://doi.org/10.1088/1475-7516/2015/10/058}{\emph{\jcap} {\bfseries
  2015} (Oct., 2015) 058}, [\href{https://arxiv.org/abs/1507.05536}{{\ttfamily
  1507.05536}}].

\bibitem{ChoquetteEtAl2016}
J.~{Choquette}, J.~M. {Cline} and J.~M. {Cornell}, \emph{{p -wave annihilating
  dark matter from a decaying predecessor and the Galactic Center excess}},
  \href{https://doi.org/10.1103/PhysRevD.94.015018}{\emph{\prd} {\bfseries 94}
  (July, 2016) 015018}, [\href{https://arxiv.org/abs/1604.01039}{{\ttfamily
  1604.01039}}].

\bibitem{FinkbeinerEtAl2010}
D.~P. Finkbeiner, L.~Goodenough, T.~R. Slatyer, M.~Vogelsberger and N.~Weiner,
  \emph{{Consistent Scenarios for Cosmic-Ray Excesses from Sommerfeld-Enhanced
  Dark Matter Annihilation}},
  \href{https://doi.org/10.1088/1475-7516/2011/05/002}{\emph{JCAP} {\bfseries
  05} (2011) 002}, [\href{https://arxiv.org/abs/1011.3082}{{\ttfamily
  1011.3082}}].

\bibitem{LiuEtAl2013}
Z.-P. Liu, Y.-L. Wu and Y.-F. Zhou, \emph{{Sommerfeld enhancements with vector,
  scalar and pseudoscalar force-carriers}},
  \href{https://doi.org/10.1103/PhysRevD.88.096008}{\emph{Phys. Rev. D}
  {\bfseries 88} (2013) 096008},
  [\href{https://arxiv.org/abs/1305.5438}{{\ttfamily 1305.5438}}].

\bibitem{DingEtAl2021}
Y.-C. Ding, Y.-L. Ku, C.-C. Wei and Y.-F. Zhou, \emph{{Consistent explanation
  for the cosmic-ray positron excess in p-wave Sommerfeld-enhanced dark matter
  annihilation}},
  \href{https://doi.org/10.1088/1475-7516/2021/09/005}{\emph{JCAP} {\bfseries
  09} (2021) 005}, [\href{https://arxiv.org/abs/2104.14881}{{\ttfamily
  2104.14881}}].

\bibitem{BottaroEtAl2022}
S.~Bottaro, D.~Buttazzo, M.~Costa, R.~Franceschini, P.~Panci, D.~Redigolo
  et~al., \emph{Closing the window on wimp dark matter},
  \href{https://doi.org/10.1140/epjc/s10052-021-09917-9}{\emph{\epjc}
  {\bfseries 82} (2022) 31},
  [\href{https://arxiv.org/abs/2107.09688}{{\ttfamily 2107.09688}}].

\bibitem{HambyeEtAl2020a}
T.~Hambye and L.~Vanderheyden, \emph{Minimal self-interacting dark matter
  models with light mediator},
  \href{https://doi.org/10.1088/1475-7516/2020/05/001}{\emph{\jcap} {\bfseries
  2020} (May, 2020) 001}, [\href{https://arxiv.org/abs/1912.11708}{{\ttfamily
  1912.11708}}].

\bibitem{DasEtAl2017}
A.~Das and B.~Dasgupta, \emph{Selection rule for enhanced dark matter
  annihilation},
  \href{https://doi.org/10.1103/PhysRevLett.118.251101}{\emph{\prl} {\bfseries
  118} (June, 2017) 251101},
  [\href{https://arxiv.org/abs/1611.04606}{{\ttfamily 1611.04606}}].

\bibitem{Binney1980}
J.~{Binney}, \emph{{The radius-dependence of velocity dispersion in elliptical
  galaxies}}, \href{https://doi.org/10.1093/mnras/190.4.873}{\emph{\mnras}
  {\bfseries 190} (Mar., 1980) 873--880}.

\bibitem{WojtakEtAl2008}
R.~{Wojtak}, E.~L. {{\L}okas}, G.~A. {Mamon}, S.~{Gottl{\"o}ber}, A.~{Klypin}
  and Y.~{Hoffman}, \emph{{The distribution function of dark matter in massive
  haloes}},
  \href{https://doi.org/10.1111/j.1365-2966.2008.13441.x}{\emph{\mnras}
  {\bfseries 388} (Aug., 2008) 815--828},
  [\href{https://arxiv.org/abs/0802.0429}{{\ttfamily 0802.0429}}].

\bibitem{Henon1973}
M.~{Henon}, \emph{{Numerical Experiments on the Stability of Spherical Stellar
  Systems}}, {\emph{\aap} {\bfseries 24} (Apr., 1973) 229}.

\bibitem{KentEtAl1982}
S.~M. Kent and J.~E. Gunn, \emph{The dynamics of rich clusters of galaxies. i -
  the coma cluster}, \href{https://doi.org/10.1086/113178}{\emph{\aj}
  {\bfseries 87} (July, 1982) 945--971}.

\bibitem{Cuddeford1991}
P.~{Cuddeford}, \emph{{An analytic inversion for anisotropic spherical
  galaxies}}, \href{https://doi.org/10.1093/mnras/253.3.414}{\emph{\mnras}
  {\bfseries 253} (Dec., 1991) 414--426}.

\bibitem{EvansEtAl2005}
N.~W. Evans and J.~H. An, \emph{{Distribution function of the dark matter}},
  \href{https://doi.org/10.1103/PhysRevD.73.023524}{\emph{Phys. Rev.}
  {\bfseries D73} (2006) 023524},
  [\href{https://arxiv.org/abs/astro-ph/0511687}{{\ttfamily
  astro-ph/0511687}}].

\bibitem{DiemandEtAl2004}
J.~{Diemand}, B.~{Moore} and J.~{Stadel}, \emph{{Velocity and spatial biases in
  cold dark matter subhalo distributions}},
  \href{https://doi.org/10.1111/j.1365-2966.2004.07940.x}{\emph{\mnras}
  {\bfseries 352} (Aug., 2004) 535--546},
  [\href{https://arxiv.org/abs/astro-ph/0402160}{{\ttfamily
  astro-ph/0402160}}].

\bibitem{DiemandEtAl2007a}
J.~{Diemand}, M.~{Kuhlen} and P.~{Madau}, \emph{{Formation and Evolution of
  Galaxy Dark Matter Halos and Their Substructure}},
  \href{https://doi.org/10.1086/520573}{\emph{\apj} {\bfseries 667} (Oct.,
  2007) 859--877}, [\href{https://arxiv.org/abs/astro-ph/0703337}{{\ttfamily
  astro-ph/0703337}}].

\bibitem{SpringelEtAl2008}
V.~{Springel}, J.~{Wang}, M.~{Vogelsberger}, A.~{Ludlow}, A.~{Jenkins},
  A.~{Helmi} et~al., \emph{{The Aquarius Project: the subhaloes of galactic
  haloes}},
  \href{https://doi.org/10.1111/j.1365-2966.2008.14066.x}{\emph{\mnras}
  {\bfseries 391} (Dec., 2008) 1685--1711},
  [\href{https://arxiv.org/abs/0809.0898}{{\ttfamily 0809.0898}}].

\bibitem{Benson2017}
A.~J. Benson, \emph{The mass function of unprocessed dark matter haloes and
  merger tree branching rates},
  \href{https://doi.org/10.1093/mnras/stx343}{\emph{\mnras} {\bfseries 467}
  (May, 2017) 3454--3466}, [\href{https://arxiv.org/abs/1610.01057}{{\ttfamily
  1610.01057}}].

\bibitem{GiocoliEtAl2008}
C.~{Giocoli}, L.~{Pieri} and G.~{Tormen}, \emph{{Analytical approach to subhalo
  population in dark matter haloes}},
  \href{https://doi.org/10.1111/j.1365-2966.2008.13283.x}{\emph{\mnras}
  {\bfseries 387} (June, 2008) 689--697},
  [\href{https://arxiv.org/abs/0712.1476}{{\ttfamily 0712.1476}}].

\bibitem{JiangEtAl2016}
F.~{Jiang} and F.~C. {van den Bosch}, \emph{{Statistics of Dark Matter
  Substructure: I. Model and Universal Fitting Functions}},
  \href{https://doi.org/10.1093/mnras/stw439}{\emph{\mnras} {\bfseries 458}
  (May, 2016) 2848--2869}, [\href{https://arxiv.org/abs/1403.6827}{{\ttfamily
  1403.6827}}].

\bibitem{AndoEtAl2019}
S.~Ando, T.~Ishiyama and N.~Hiroshima, \emph{Halo substructure boosts to the
  signatures of dark matter annihilation},
  \href{https://doi.org/10.3390/galaxies7030068}{\emph{Galaxies} {\bfseries 7}
  (July, 2019) 68}, [\href{https://arxiv.org/abs/1903.11427}{{\ttfamily
  1903.11427}}].

\bibitem{IshiyamaEtAl2020a}
T.~Ishiyama and S.~Ando, \emph{The abundance and structure of subhaloes near
  the free streaming scale and their impact on indirect dark matter searches},
  \href{https://doi.org/10.1093/mnras/staa069}{\emph{\mnras} {\bfseries 492}
  (Mar., 2020) 3662--3671}, [\href{https://arxiv.org/abs/1907.03642}{{\ttfamily
  1907.03642}}].

\bibitem{CharbonnierEtAl2012}
A.~{Charbonnier}, C.~{Combet} and D.~{Maurin}, \emph{{CLUMPY: A code for
  {$\gamma$}-ray signals from dark matter structures}},
  \href{https://doi.org/10.1016/j.cpc.2011.10.017}{\emph{Computer Physics
  Communications} {\bfseries 183} (Mar., 2012) 656--668},
  [\href{https://arxiv.org/abs/1201.4728}{{\ttfamily 1201.4728}}].

\bibitem{Benson2012}
A.~J. Benson, \emph{Galacticus: A semi-analytic model of galaxy formation},
  \href{https://doi.org/10.1016/j.newast.2011.07.004}{\emph{\na} {\bfseries 17}
  (Feb., 2012) 175--197}, [\href{https://arxiv.org/abs/1008.1786}{{\ttfamily
  1008.1786}}].

\bibitem{Weinberg1994}
M.~D. {Weinberg}, \emph{{Adiabatic invariants in stellar dynamics. 1: Basic
  concepts}}, \href{https://doi.org/10.1086/117161}{\emph{\aj} {\bfseries 108}
  (Oct., 1994) 1398--1402},
  [\href{https://arxiv.org/abs/astro-ph/9404015}{{\ttfamily
  astro-ph/9404015}}].

\bibitem{GnedinEtAl1999}
O.~Y. {Gnedin}, L.~{Hernquist} and J.~P. {Ostriker}, \emph{{Tidal Shocking by
  Extended Mass Distributions}},
  \href{https://doi.org/10.1086/306910}{\emph{\apj} {\bfseries 514} (Mar.,
  1999) 109--118}.

\bibitem{HayashiEtAl2003}
E.~{Hayashi}, J.~F. {Navarro}, J.~E. {Taylor}, J.~{Stadel} and T.~{Quinn},
  \emph{{The Structural Evolution of Substructure}},
  \href{https://doi.org/10.1086/345788}{\emph{\apj} {\bfseries 584} (Feb.,
  2003) 541--558}, [\href{https://arxiv.org/abs/astro-ph/0203004}{{\ttfamily
  astro-ph/0203004}}].

\bibitem{vandenBoschEtAl2018a}
F.~C. van~den Bosch, G.~Ogiya, O.~Hahn and A.~Burkert, \emph{Disruption of dark
  matter substructure: fact or fiction?},
  \href{https://doi.org/10.1093/mnras/stx2956}{\emph{\mnras} {\bfseries 474}
  (Mar., 2018) 3043--3066}, [\href{https://arxiv.org/abs/1711.05276}{{\ttfamily
  1711.05276}}].

\bibitem{GreenEtAl2021}
S.~B. Green, F.~C. van~den Bosch and F.~Jiang, \emph{The tidal evolution of
  dark matter substructure - ii. the impact of artificial disruption on subhalo
  mass functions and radial profiles},
  \href{https://doi.org/10.1093/mnras/stab696}{\emph{\mnras} (Mar., 2021) },
  [\href{https://arxiv.org/abs/2103.01227}{{\ttfamily 2103.01227}}].

\bibitem{BullockEtAl2001b}
J.~S. {Bullock}, T.~S. {Kolatt}, Y.~{Sigad}, R.~S. {Somerville}, A.~V.
  {Kravtsov}, A.~A. {Klypin} et~al., \emph{{Profiles of dark haloes: evolution,
  scatter and environment}},
  \href{https://doi.org/10.1046/j.1365-8711.2001.04068.x}{\emph{\mnras}
  {\bfseries 321} (Mar., 2001) 559--575},
  [\href{https://arxiv.org/abs/astro-ph/9908159}{{\ttfamily
  astro-ph/9908159}}].

\bibitem{MaccioEtAl2008}
A.~V. {Macci{\`o}}, A.~A. {Dutton} and F.~C. {van den Bosch},
  \emph{{Concentration, spin and shape of dark matter haloes as a function of
  the cosmological model: WMAP1, WMAP3 and WMAP5 results}},
  \href{https://doi.org/10.1111/j.1365-2966.2008.14029.x}{\emph{\mnras}
  {\bfseries 391} (Dec., 2008) 1940--1954},
  [\href{https://arxiv.org/abs/0805.1926}{{\ttfamily 0805.1926}}].

\bibitem{PradaEtAl2012}
F.~{Prada}, A.~A. {Klypin}, A.~J. {Cuesta}, J.~E. {Betancort-Rijo} and
  J.~{Primack}, \emph{{Halo concentrations in the standard {$\Lambda$} cold
  dark matter cosmology}},
  \href{https://doi.org/10.1111/j.1365-2966.2012.21007.x}{\emph{\mnras}
  {\bfseries 423} (July, 2012) 3018--3030},
  [\href{https://arxiv.org/abs/1104.5130}{{\ttfamily 1104.5130}}].

\bibitem{DuttonEtAl2014}
A.~A. {Dutton} and A.~V. {Macci{\`o}}, \emph{{Cold dark matter haloes in the
  Planck era: evolution of structural parameters for Einasto and NFW
  profiles}}, \href{https://doi.org/10.1093/mnras/stu742}{\emph{\mnras}
  {\bfseries 441} (July, 2014) 3359--3374},
  [\href{https://arxiv.org/abs/1402.7073}{{\ttfamily 1402.7073}}].

\bibitem{vandenBoschEtAl2016}
F.~C. {van den Bosch} and F.~{Jiang}, \emph{{Statistics of Dark Matter
  Substructure: II. Comparison of Model with Simulation Results}},
  \href{https://doi.org/10.1093/mnras/stw440}{\emph{\mnras} {\bfseries 458}
  (May, 2016) 2870--2884}, [\href{https://arxiv.org/abs/1403.6835}{{\ttfamily
  1403.6835}}].

\bibitem{HanEtAl2016b}
J.~{Han}, S.~{Cole}, C.~S. {Frenk} and Y.~{Jing}, \emph{{A unified model for
  the spatial and mass distribution of subhaloes}},
  \href{https://doi.org/10.1093/mnras/stv2900}{\emph{\mnras} {\bfseries 457}
  (Apr., 2016) 1208--1223}, [\href{https://arxiv.org/abs/1509.02175}{{\ttfamily
  1509.02175}}].

\bibitem{SomervilleEtAl1999a}
R.~S. Somerville and T.~S. Kolatt, \emph{How to plant a merger tree},
  \href{https://doi.org/10.1046/j.1365-8711.1999.02154.x}{\emph{\mnras}
  {\bfseries 305} (May, 1999) 1--14},
  [\href{https://arxiv.org/abs/astro-ph/9711080}{{\ttfamily
  astro-ph/9711080}}].

\bibitem{ParkinsonEtAl2008}
H.~Parkinson, S.~Cole and J.~Helly, \emph{Generating dark matter halo merger
  trees}, \href{https://doi.org/10.1111/j.1365-2966.2007.12517.x}{\emph{\mnras}
  {\bfseries 383} (Jan., 2008) 557--564},
  [\href{https://arxiv.org/abs/0708.1382}{{\ttfamily 0708.1382}}].

\bibitem{BondEtAl1991a}
J.~R. {Bond}, S.~{Cole}, G.~{Efstathiou} and N.~{Kaiser}, \emph{{Excursion set
  mass functions for hierarchical Gaussian fluctuations}},
  \href{https://doi.org/10.1086/170520}{\emph{\apj} {\bfseries 379} (Oct.,
  1991) 440--460}.

\bibitem{LaceyEtAl1993}
C.~{Lacey} and S.~{Cole}, \emph{{Merger rates in hierarchical models of galaxy
  formation}}, {\emph{\mnras} {\bfseries 262} (June, 1993) 627--649}.

\bibitem{ColeEtAl2000}
S.~Cole, C.~G. Lacey, C.~M. Baugh and C.~S. Frenk, \emph{Hierarchical galaxy
  formation},
  \href{https://doi.org/10.1046/j.1365-8711.2000.03879.x}{\emph{\mnras}
  {\bfseries 319} (Nov., 2000) 168--204},
  [\href{https://arxiv.org/abs/astro-ph/0007281}{{\ttfamily
  astro-ph/0007281}}].

\bibitem{PlanckCollab2020}
{Planck Collaboration}, \emph{Planck 2018 results. vi. cosmological
  parameters}, \href{https://doi.org/10.1051/0004-6361/201833910}{\emph{\aap}
  {\bfseries 641} (Sept., 2020) A6},
  [\href{https://arxiv.org/abs/1807.06209}{{\ttfamily 1807.06209}}].

\bibitem{Facchinetti2021}
G.~Facchinetti, \emph{Analytical study of particle dark matter structuring on
  small scales and implications for dark matter searches}, Ph.D. thesis,
  University of Montpellier, 2021.

\end{thebibliography}\endgroup



\providecommand{\href}[2]{#2}\begingroup\raggedright\endgroup

\end{document}